\documentclass[iop]{emulateapj}
\usepackage{apjfonts}                   

\usepackage{epsfig,graphicx,latexsym,amsmath,amssymb}
\usepackage[export]{adjustbox}[2011/08/13]

\usepackage{natbib}
\usepackage{hyperref}
\usepackage{mathrsfs}
\usepackage{lastpage}

\usepackage{color}

\bibpunct[,]{(}{)}{;}{a}{}{,}

\begin{document}

\author{
Pau Amaro Seoane\altaffilmark{1,2,3,4}\thanks{e-mail: amaro@riseup.net}
}

\altaffiltext{1}{
Institute of Multidisciplinary Mathematics, Universitat Politècnica de València, València, Spain}
\altaffiltext{2}{Max-Planck-Institute for Extraterrestrial Physics, Garching, Germany}
\altaffiltext{3}{Higgs Centre for Theoretical Physics, Edinburgh, UK}
\altaffiltext{4}{Kavli Institute for Astronomy and Astrophysics, Beijing, China}

\date{\today}

\label{firstpage}

\title{Transient stellar collisions as multimessenger probes:\\
Non-thermal-, gravitational wave emission and the cosmic ladder argument
}

\begin{abstract}

In dense stellar clusters like galactic nuclei and globular clusters stellar
densities are so high that stars might physically collide with each other. In
galactic nuclei the energy and power output can be close, and even exceed, to
those from supernovae events. We address the event rate and the electromagnetic
characteristics of collisions of main sequence stars (MS) and red giants (RG).
We also investigate the case in which the cores form a binary and emit
gravitational waves. In the case of RGs this is particularly interesting
because the cores are degenerate.  We find that MS event rate can be as high as
tens per year, and that of RGs one order of magnitude larger.  The collisions
are powerful enough to mimic supernovae- or tidal disruptions events. We find Zwicky Transient Facility observational data which
seem to exhibit the features we describe. The cores
embedded in the gaseous debris experience a friction force which has an impact
on the chirping mass of the gravitational wave. As a consequence, the two small
cores in principle mimic two supermassive black holes merging.  However, their
evolution in frequency along with the precedent electromagnetic burst and the
ulterior afterglow are efficient tools to reveal the impostors. In the
particular case of RGs, we derive the properties of the degenerate He cores and
their H-burning shells to analyse the formation of the binaries. The merger is
such that it can be misclassified with SN Ia events.  Because the masses and
densities of the cores are so dissimilar in values depending on their
evolutionary stage, the argument about standard candles and cosmic ladder
should be re-evaluated.  

\end{abstract}


\maketitle

\section{Motivation}
\label{sec.motivation}

Dense stellar systems such as globular clusters and galactic nuclei have
stellar densities ranging between a million and a hundred million stars per
cubic parsec.  In them, relative velocities of the order of $\sim
\text{a~few~}10\,\text{km/s}$ in the case of globular clusters and of $\sim$
$100-1000\,\text{km/s}$ in the case of galactic nuclei can be reached
\citep{NeumayerEtAl2020,Spitzer87,BinneyTremaine08}.  In these exceptional
conditions, and unlike anywhere else in the host galaxy, collisional effects
come into play.  With ``collisional'' we mean in general mutual gravitational
deflections which lead to an exchange of energy and angular momentum, but also
in particular genuine contact collisions.  The possibility that collisions
between stars play a fundamental role both in explaining particular
observations and in the global influence of dense stellar systems  has been
studied with dedicated numerical studies
\citep{SS66,DDC87a,Sanders70b,BenzHills1987,DDC87b,DaviesEtAl1991,BenzHills1992,MCD91,LRS93,LRS95,LRS96,1999MNRAS.308..257B,DaviesEtAl1998,BaileyEtAl1999,LombardiEtAl2002,2002ASPC..263....1S,AdamsEtAl2004,TracEtAl2007,DaleEtAl2009,2020ApJ...901...44W,Mastrobuono-BattistiEtAl2021,2021A&A...649A.160V}.

We have chosen to first focus on galactic nuclei. We then address globular
clusters, in which the rates are larger due to the smaller relative velocities
between the stars participating in the collision (which is of the order of the
velocity dispersion).  For galactic nuclei, we first derive the event rate of
these collisions as a function of the host galaxy cusp
(section~\ref{sec.rates}) and analyse analytically the non-thermal properties
of the outcome of such collisions (sections \ref{sec.energy} and
\ref{sec.temperature}). This analysis is performed for both main-sequence stars
and, later, for red giants (section \ref{sec.RedGiants}).  The electromagnetic
analysis reveals that these collisions can mimic over periods of time tidal
disruptions but also Type Ia supernovae \citep{daSilva1993}.  Our analysis is a
dynamical and analytical one, and depends on solely two free parameters which
whose value should be extracted with dedicated numerical simulations.

We extend the analysis to the gravitational radiation phase as emitted by a
subset of these collisions, namely those in which the core survives and forms a
binary (section \ref{sec.GW}). Red giants have a very compact nucleus and can
always withstand the onslaught of the collision. 

We find that the number of gravitational wave sources that form is not
negligible, and leads to the emergence of a type of source that can be
misleading. A source that drastically changes its characteristics within a very
short time. In a matter of months, the binary that forms initially appears to
have a few solar masses to later appear as a supermassive black hole binary.
Similarly, the luminosity distance varies tremendously in that short interval
of time.

Due to the multi-messenger characteristics of this source, the extraction of
information is very interesting and complementary. That is, electromagnetic
data can help us to break various degeneracies in the analysis of gravitational
waves and vice versa. In the particular case of the red giants, the rates are
very high and, because the electromagnetic nature of the process very strongly
depends on the stage of the evolution of the colliding red giants, if these
collisions were confused with supernovae events, which are used as a kind of
standard candles, the ladder argument to calculate cosmological distances would
be in danger of revision. 

Although galactic nuclei are often left out from the supernova searches, it is
often difficult if not impossible to discern the nucleus due to a lack of
resolution. Moreover, collisions happen more frequently in globular clusters,
as mentioned before, which are located off the plane and away from the galactic
nucleus, and are hence not excluded in the searches. However, the low relative
velocities lead to a different kind of phenomenon: stellar pulsations.  In
Sec.~(\ref{sec.globular}) we find that the collisions in globular clusters can
lead to the classical Cepheids pulsation phenomenon. We show that in the
adiabatic, spherical case this is a stable phenomenon, and we calculate the
associated timescale (sections \ref{sec.pulsations} and
\ref{sec.AdiabaticAppr}).  However, ulterior inputs of energy are required if
the vibrational or thermal instability dissipate the oscillations. This
additional inputs of energy can happen if futher collisions take place with the
same companion star in the case of binary formation, or with another star, or
in the case in which internal instabilities lead to them.

The classical pulsation problem has been envisaged as another rung in the
standard candle classification of the cosmological ladder, so that this must be
addressed in more detail than we present here, and will be presented elsewhere.
We discuss the supernovae and pulsating star misclassification in the context
of the cosmological ladder in Sec.~(\ref{sec.Ladder}).

Finally, in Sec.~(\ref{sec.conclusions}) we present a summary of all of the
conclusions from our investigations.

\section{Event rate derivation}
\label{sec.rates}

The quasi-steady solution for how stars distribute around a massive black hole
(MBH) follows an isotropic distribution function in physical space of the form
$\rho(r)\sim R^{-\gamma}$, where $\rho$ is the stellar density $\rho$ and $R$ the radius
\citep{Peebles1972,BW76}.  This mathematical derivation has been corroborated
using numerical techniques
\citep{SM78,MS79,MS80,ST85,FB01a,ASEtAl04,PretoMerrittSpurzem04} and, recently,
a comparison with data from our Galactic Centre yields a very good match
between observations, theory and numerical simulations
\citep{BaumgardtEtAl2018,Gallego-CanoEtAl2018,SchoedelEtAl2018}.

Therefore, we assume a power-law mass distribution for the numerical density of
stars around the MBH, $n_{*}(R) \propto R^{-\gamma}$, with $R$ the radius. Following this, we
can derive that the enclosed stellar mass around the MBH within a given radius is \citep[see e.g.][]{Amaro-Seoane2019}

\begin{equation}
M_{*}(R) = M_{\bullet} \left(\frac{R}{R_{\rm infl}}\right)^{3-\gamma}. 
\end{equation}

\noindent
In this last equation $M_{*}(R)$ is the stellar mass at a radius $R$, $M_{\bullet}$ is the
mass of the MBH, $R_{\rm infl}$ is the influence radius of the MBH (i.e. the radius within which
the potential is dominated by the MBH) and $\gamma$ is the exponent of the power law. Hence, the
total number of stars at that radius is

\begin{equation}
N_{*}(R) = \frac{M_{\bullet}}{m_{*}} \left(\frac{R}{R_{\rm infl}} \right)^{3-\gamma},
\label{eq.Nstar}
\end{equation}

\noindent
where $m_{*}$ is the mass of one star and we are assuming for simplicity that all stars have the
same mass and radius $R_{*}$, so that the stellar mass density at a given radius is $\rho_{*}(R)=m_{*}\,n_{*}(R)$. Therefore, we have that
the numerical density is

\begin{equation}
n_{*}(R) = \frac{3-\gamma}{4\pi} \frac{M_{\bullet}}{m_{*}R_{\rm infl}^3} \left(\frac{R}{R_{\rm infl}}\right)^{-\gamma}, 
\end{equation}

\noindent
since $dN_{*}/dR=4\pi R^2\,n_{*}$.

At the radii of interest, those close to the MBH, within the radius of influence, 
the typical relative velocity between stars

\begin{equation}
V_{\rm rel}(R) = K_v\sqrt{\frac{GM_{\bullet}}{R}} 
\end{equation}

\noindent
is $V_{\rm rel}(R) \geqslant V_{\rm esc}$, with $V_{\rm esc}$ the escape velocity from the stellar surface,

\
\begin{equation}
V_{\rm esc} = \sqrt{\frac{2Gm_{*}}{R_{*}}}, 
\end{equation}

\noindent
and $K_v$ depends on $\gamma$ and is of order unity.

The collision rate for one star can be estimated as

\begin{equation}
\frac{1}{T_{{\rm coll},\,1}(R)} = n_{*}(R)\,V_{\rm rel}(R)\,S, 
\label{eq.T1}
\end{equation}

with $S$ the cross-section, 

\begin{equation}
S = \pi \left( f_{\rm coll} 2R_{*}\right)^2, 
\label{eq.S}
\end{equation}

\noindent
since we are neglecting the gravitational focusing, because $V_{\rm rel}(R)
\geqslant V_{\rm esc}$, so that $S$ can be computed geometrically. In practise
this means that we are looking at a lower-limit case, since the rates could be
slightly enhanced. This is particularly true in globular clusters, where the
relative velocity is lower.  As stated in the introduction, nonetheless, we are
focusing in galactic nuclei, which is a lower-limit case of the general
scenario.  In this equation, $f_{\rm coll}$ defines how deep a collision is. 

Introducing Eq.~(\ref{eq.S}), $n_{*}(R)$ and $V_{\rm rel}(R)$ in Eq.~(\ref{eq.T1}), we have that

\begin{equation}
\frac{1}{T_{{\rm coll},\,1}(R)} = (3-\gamma) K_v f_{\rm coll}^2 \left(\frac{R_{*}}{R_{\rm infl}}\right)^{2}
                                \frac{M_{\bullet}}{m_{*}}\sqrt{\frac{GM_{\bullet}}{R_{\rm infl}^3}} \left(\frac{R}{R_{\rm infl}}\right)^{-(\gamma+1/2)}
\end{equation}

The total collisional rate in the cusp around the MBH is

\begin{equation}
\Gamma_{\rm coll}= \frac{N_{*}}{2}\frac{1}{T_{\rm coll,\,tot}}, 
\label{eq.GammaColl}
\end{equation}

\noindent
since $N_{*}=4\pi R^2 n_{*}$ and we take into account that for a collision we need two stars. Therefore

\begin{equation}
\Gamma_{\rm coll}= 2\pi \int^{R_{\rm max}}_{R_{\rm min}} n_{*}(R) \frac{R^2}{{T_{\rm coll,\,1}(R)}} dR.
\label{eq.GammaIntegral}
\end{equation}

In this integral we choose the maximum radius $R_{\rm max}$ to be the distance within the influence radius at which
$V_{\rm rel}(R)=V_{\rm esc}(R)$, i.e.

\begin{equation}
R_{\rm max}=K_v^{-2}R_{*}\frac{M_{\bullet}}{m_{*}}, 
\end{equation}

\noindent
and the minimum radius $R_{\rm min}$ to be the radius which contains on average one star. From
Eq.~(\ref{eq.Nstar}) we derive that

\begin{equation}
R_{\rm min} = R_{\rm infl} \left(\frac{m_{*}}{M_{\bullet}} \right)^{\frac{1}{3-\gamma}}. 
\label{eq.Rmin}
\end{equation}

We note that the interior mass enclosed in $R_{\rm max}$ is

\begin{align}
M_{*,\,{\rm max}} & =K_v^{-2(3-\gamma)} M_{\bullet} \left(\frac{R_{*}}{R_{\rm infl}}\right)^{3-\gamma} 
                  \left( \frac{M_{\bullet}}{m_{*}} \right)^{3-\gamma} \approx \nonumber \\
                  & M_{\bullet} \left(\frac{\sigma_v^2}{V_{\rm esc}^2}\right) \ll M_{\bullet},
\end{align}

\noindent
where $\sigma_v$ is the velocity dispersion at large distances from the MBH. This last equation means that $R_{\rm max} \ll R_{\rm infl}$.

We can now integrate Eq.~(\ref{eq.GammaIntegral}),

\begin{align}
\Gamma_{\rm coll} & = 2.12\times 10^{-9} \frac{1}{yr} \frac{(3-\gamma)^2}{5-4 \gamma} K_v \left(\frac{f_{\rm coll}}{0.25}\right)^2
                      \left(\frac{R_{*}}{1\,R_{\odot}}\right)^2 \times \nonumber \\
                  &   \left(\frac{M_{\bullet}}{10^6M_{\odot}}\right)^{5/2} \left(\frac{m_{*}}{1\,M_{\odot}}\right)^{-2}
                                                                \left(\frac{R_{\rm infl}}{1\,\textrm{pc}}\right)^{-7/2} \times \nonumber \\
                  &   \left[ A(\gamma)\left(\frac{R_{*}}{R_{\odot}}\right)^{-2\gamma+5/2} \left(\frac{M_{\bullet}}{10^6M_{\odot}}\right)^{-2\gamma+5/2} 
                                                                \left(\frac{R_{\rm infl}}{1\textrm{pc}}\right)^{2\gamma-5/2} \times \nonumber \right. \\
                  &   \left. \left( \frac{m_{*}}{1\,M_{\odot}}  \right)^{2\gamma-5/2} 
                                                                - B(\gamma) \left(\frac{M_{\bullet}}{10^6M_{\odot}}\right)^{\frac{2\gamma-5/2}{3-\gamma}}
                                                                \left(\frac{m_{*}}{1\,M_{\odot}}\right)^{\frac{5/2-2\gamma}{3-\gamma}}  \right],  
\label{eq.GammaColl}
\end{align}

\noindent
where we have defined

\begin{align}
A(\gamma):= &2.25^{-2\gamma +5/2} 10^{4\gamma -5} K_v^{4\gamma -5} \nonumber\\
B(\gamma):= &10^{\frac{12\gamma-15}{3-\gamma}}
\end{align}

We note that, since $R_{\rm min} \ll R_{\rm max}$, the rates are dominated at
short distances from the MBH, so that the first term in the square brackets of
Eq.~(\ref{eq.GammaColl}) can in principle be neglected. 
However, since this could artificially increase the rates, we do not neglect
it. We have normalised $f_{\rm coll}$ to $0.25$ because we are interested in
collisions which lead to a total disruption of the stars. This situation is
achieved when the periastron distance of a gravitational two-body hyperbolic
encounter in the centre-of-mass reference frame $d_{\rm min}$ has the value

\begin{equation}
d_{\rm min}=\left(R_{{\rm half}, 1} + R_{{\rm half}, 2} \right),
\label{eq.dmin}
\end{equation}

\noindent
with $R_{{\rm half}, 1}$ the half-mass radius of the first star participating in the collision
(and $R_{{\rm half}, 1}=R_{{\rm half}, 2}$ since we assume they have the same radius and mass).
Therefore, for a complete disruptive collision $f_{\rm coll}$, a measure of the depth of the impact, as
we explained, is

\begin{equation}
f_{\rm coll} \approx \frac{R_{\rm half}}{R_*}.
\label{eq.fcoll}
\end{equation}

As we can see in e.g. Fig. 4 of \cite{FB05} (and see also their Fig. 9), for
$m_{*}= 1\,M_{\odot}$, $R_{*}= 1\, R_{\odot}$, and then $f_{\rm coll} = 0.25$.
For $m_{*}= 10\,M_{\odot}$, $R_{*}= 6\, R_{\odot}$, and $f_{\rm coll} = 0.2$

As for the influence radius, we use the so-called ``mass-sigma'' correlation
\citep{McConnellEtAl2011,KormendyHo2013,DaviesEtAl2017} for black hole masses in nearby galaxies,

\begin{equation}
\frac{M_{\bullet}}{3\times 10^8\,M_{\odot}} \cong \left(\frac{\sigma}{200\,\textrm{km\,s}^{-1}}\right)^5,
\end{equation}

\noindent
with $\sigma$ the velocity dispersion of the stars. This combined with the definition of the
influence radius which takes into account the overall effect on the motion of a
star by the bulge, including those that have moved away from the MBH, as
introduced by \cite{Peebles1972},

\begin{equation}
R_{\rm infl} = \frac{G\,M_{\bullet}}{\sigma^2},
\end{equation}

\noindent
leads to

\begin{equation}
R_{\rm infl} = 1.05\, \textrm{pc} \times \left(\frac{M_{\bullet}}{10^6\,M_{\odot}}\right)^{0.6}, 
\end{equation}

\noindent
and note that for $M_{\bullet}=4\times 10^6\,M_{\odot}$ such as the one in our
Galactic Centre, $R_{\rm infl} = 2.5\,\textrm{pc}$, which is close to the value
observed of $\sim 3 \textrm{pc}$ \citep{SchoedelEtAl2014,SchoedelEtAl2018}.
Hence, for $M_{\bullet}=10^7\,M_{\odot}$, $R_{\rm infl} = 4.2\,\textrm{pc}$ and
for $M_{\bullet}=10^5\,M_{\odot}$, $R_{\rm infl} = 0.27\,\textrm{pc}$.

For a Bahcall-Wolf power-law \citep{BW76}, $\gamma=7/4$, taking $f_{\rm coll}=0.25$,
and the default values given in Eq.~(\ref{eq.GammaColl}), we obtain that
$\Gamma_{{\rm coll},\,6} \cong 10^{-4}\textrm{yr}^{-1}$ for a Milky-Way-like nucleus, i.e.
with a MBH in this mass range, $M_{\bullet} = 10^6\,M_{\odot}$ (as indicated with the
sub-index 6). 

The calculation of the event rate applies to nuclei hosting MBHs with masses
between $\sim 10^5 - 10^7\,M_{\odot}$ , since for larger MBH masses the
relaxation time would exceed a Hubble time, and for lighter MBH masses the MBH
is in the intermediate-mass regime and hence cannot be envisaged as fixed in
the centre of the potential, but wandering, which renders the calculation much
more complicated.  For $M_{\bullet} = 10^7\,M_{\odot}$, and taking the same
parameters as for the $M_{\bullet} = 10^6\,M_{\odot}$ case but for the influence radius, we obtain that
$\Gamma_{{\rm coll},\,7} \cong 2\times 10^{-4}\textrm{yr}^{-1}$, and for $M_{\bullet} =
10^5\,M_{\odot}$, $\Gamma_{{\rm coll},\,5} \cong 10^{-5} \textrm{yr}^{-1}$.

\begin{figure}
\resizebox{\hsize}{!}
          {\includegraphics[scale=1,clip]{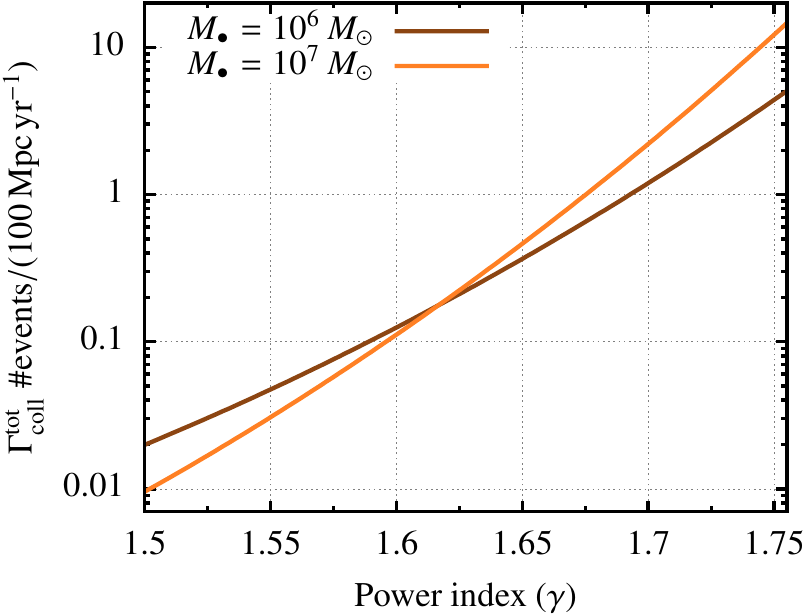}}
\caption
   {
Total amount of events per year in a volume of $100\,\textrm{Mpc}$ of radius for
two different values of MBHs and for typical values of the power index $\gamma$.
We note that $\gamma=1.75$ corresponds to the theoretical expectation of a relaxed
nucleus for a single-mass population \citep{Peebles1972,BW76}. We show lower values
as an illustration for the dependency of $\Gamma_{{\rm coll}}^{\rm tot}$ with $\gamma$,
which is not obvious from Eq.~(\ref{eq.GammaColl}).
At smaller values of $\gamma$, $10^{6}\,M_{\odot}$ is the upper curve and from
$\gamma \sim 1.625$ the situation reverts and the upper one corresponds to $10^{7}\,M_{\odot}$.
   }
\label{fig.G}
\end{figure}

Assuming an observable distance of $100\,\textrm{Mpc}$ for these events, this
translates into an observable volume of $\sim 4.2 \times 10^6 \,
\textrm{Mpc}^3$.  Within this volume, and assuming $10^{-2}$ MBH of $M_{\bullet}=10^6\,M_{\odot}$ per
$\textrm{Mpc}^3$ \citep[see Fig.2 of][]{KellyMerloni2012}, we derive a total of
$4.2\times 10^4$ sources, i.e.  nuclei hosting MBHs with a mass of $M_{\bullet} =
10^6\,M_{\odot}$, so that this multiplied by $\Gamma_{{\rm coll},\,6}$ leads to
a total event rate of $\Gamma_{{\rm coll},\,6}^{\rm tot} \sim 4.2\,\textrm{yr}^{-1}$.
For MBHs with masses of $10^7\,M_{\odot}$, the work of \cite{KellyMerloni2012}
yields $6\times 10^{-3}$ MBH per $\textrm{Mpc}^3$, and hence $\Gamma_{{\rm coll},\,7}^{\rm tot} \sim
5\,\textrm{yr}^{-1}$. For MBHs with masses of $10^5\,M_{\odot}$, and
extrapolating the results of \cite{KellyMerloni2012}, to about $10^{-2}$ MBH
per $\textrm{Mpc}^3$ as well, we have that $\Gamma_{{\rm coll},\,5}^{\rm tot}
\sim 0.42\,\textrm{yr}^{-1}$. Therefore, neglecting the contribution of $10^5\,M_{\odot}$ MBHs,
and for a mass range for the MBH between $[10^6,\,\textrm{a~few\,}10^7]\,M_{\odot}$, we have
a total integrated event rate of $\gtrsim 100 \textrm{\,yr}^{-1}$ in $100\,\textrm{Mpc}$.
In Fig.(\ref{fig.G}) we show $\Gamma_{{\rm coll},\,6}^{\rm tot}$ and $\Gamma_{{\rm coll},\,7}^{\rm tot}$
for various typical values of $\gamma$ in a volume of $100\,\textrm{Mpc}$ of radius.

\section{Energy release}
\label{sec.energy}

During the collision release of nuclear energy is negligible
\citep[see][]{Mathis67,RYBMH89}. Gravitational energy can also be neglected in
the kind of collisions we are considering (very high velocities and $f\sim
0.2$).  We can also neglect radiative transport, since the merging stars are
obviously optically thick while the collision it taking place. During it, the
energy transport by radiation is diffusive.  

In this kind of almost head-on stellar collisions and in our framework of high
relative velocities, the colliding stars merge into a single object surrounded
by a gaseous structure which is approximately spherical \citep[see e.g. the
numerical work of ][]{FB05}.  This gaseous cloud will expand at a speed which
is equivalent to the average relative speeds one observes at galactic centres
harbouring MBHs of masses
$M_{\bullet}=[10^{6},\,\textrm{a~few}\,10^{7}]\,M_{\odot}$.

In this section, we first estimate the timescale for the energy to diffuse from
the centre of the cloud to the surface and the timescale associated for the
cloud to become transparent. Then we calculate the total emission of the
energy and its time dependency, as well as the luminosity.

\subsection{Diffusion of energy: Timescales}
\label{sec.diff_times}

We estimate the associated timescales for a cloud to diffuse energy to the
surface and for it to become fully transparent. We consider it to be
transparent when the mean free path of photons is larger than the radius of the
cloud.

We define the mean free path $l(t)$ (which changes over time) as the average
distance for a photon between two interactions with two electrons at a given
time, so that the time to cover it is $l(t)/c$, with $c$ the speed of light.
Since we are talking about a random-walk process, the average number of steps
of length $l(t)$ for the photon to cover a distance $R(t)$ (the radius of the
cloud, function of time) is

\begin{equation}
N(t) = \left( \frac{R(t)}{l(t)} \right)^2,
\end{equation}

\noindent
because the average of the squared distance is proportional to the time
in a random walk.

\noindent
We define the diffusion time as this number of steps multiplied
by the time to cover the distance between two interactions, so that

\begin{equation}
T_{\rm diff}(t) \cong \frac{N(t)\,l(t)}{c}. 
\end{equation}

We now calculate the mean free path by estimating the probability $P_{\rm
coll}$ that an electron collides with a photon after a distance $x$, 

\begin{equation}
dP_{\rm coll} = S_{\rm eff}\, n\, dx, 
\end{equation}

\noindent
with $S_{\rm eff}$ is the effective area and $n$ the numerical density of electrons.
Hence, the collisional rate for one electron is 

\begin{equation}
\Gamma_{\rm e} = \frac{dP_{\rm coll}}{dt} = S_{\rm eff} \, n \, v, 
\end{equation}

\noindent
with $v$ the relative velocity between the electron and the photon, i.e. $v=c$.
Therefore, the average number of collisions over a distance $x$ is

\begin{equation}
N_{\rm coll} = S_{\rm eff} \, n \, x.
\end{equation}

\noindent
By setting $N_{\rm coll} =1$ in this last equation, we derive the value of $x$,
i.e. the mean free path,

\begin{equation}
l = \frac{1}{S_{\rm eff}\,n}.
\label{eq.l}
\end{equation}

\noindent
Since $n = \rho_{\rm g}/m$, with $m$ the mass of one ``gas particle'' (i.e. the proton mass, since we assume
that we have completly ionised H) per electron,

\begin{equation}
l = \frac{m}{\rho_{\rm g}\,S_{\rm eff}}, 
\end{equation}

\noindent
which allows us to introduce the usual definition of opacity, $\kappa = S_{\rm
eff}/m$.  If we assume that the ionisation degree does not change, then $l
\propto 1/\rho_{\rm g}$, and since $\rho_{\rm g} \simeq M/R(t)^3$, we derive
that $l(t) \propto R(t)^3$. Therefore, there must be a time in which $l(t) >
R(t)$ and the cloud is transparent, $t = t_{\rm transp}$.  If at that moment,
which we denote as $t = t_{\rm transp}$, there is still enough energy in form
of photons in the cloud, they will be able to escape it instantaneously even if
they are located at the centre of the cloud, in a straight line, without
diffusion.  

I.e. if $t$ is the time passed since the formation of the cloud (i.e. right
after the collision), and $T_{\rm diff} \ll t$, then most of the photons are
still trapped in the cloud. Nonetheless, $t$ obviously increases and $T_{\rm
diff}$ varies in time, so that there might be a moment in which $t > T_{\rm
diff}$ before we reach $t = t_{\rm transp}$.  We need to estimate these
timescales. From the previous equations, we have that

\begin{equation}
T_{\rm diff}(t) \simeq \frac{\kappa}{c}\frac{M}{R(t)}. 
\label{eq.Tdiff}
\end{equation}

\noindent
With $\kappa=0.04\,\textrm{m}^2
\textrm{kg}^{-1}$ (a lower bound for an ionised gas due to electron
scattering).
In the right hand side of this last equation everything is constant but for $R(t)$,
which increases, so that $T_{\rm diff}(t)$ decreases with time. 
We can calculate at what time $T_{\rm diff}$ is reached, so as to compare it
with $t = t_{\rm transp}$. An approximation is to set $T_{\rm diff}=t$ in 
Eq.~(\ref{eq.Tdiff}), so that if we approximate the expansion velocity $V_{\rm exp}$ to the
relative velocity, $V_{\rm exp}=10^{4}\,\textrm{km\,s}^{-1}$ (we will elaborate on this
choice later), we have that $t=\sqrt{\kappa M/(V_{\rm exp}c)}$. Hence,

\begin{equation}
T_{\rm diff} \sim 0.16\,\textrm{yrs} \sim 2\,\textrm{months} 
\label{eq.Tdiffmonths}
\end{equation}

\noindent
After reaching this time, approximately half of the total energy contained in
the cloud has been released and the remaining half is still trapped in it. If
we wait two times this amount of time, half of half the initial energy will
still be in the cloud, so that the remaining amount of energy in the cloud goes
as $1/2^{n}$ the initial amount, with $n$ the amount of temporal intervals
corresponding to $T_{\rm diff}$. We note that this assumes that $T_{\rm diff}(t)$ is
the same as $T_{\rm diff}(0)$. This is of course not true but it gives us a first
rough estimate of the initial timescale for half of the energy to be released.
We will improve this approximation in Sec.~(\ref{sec.energypower}).

To calculate at what time $t_{\rm transp}$ is reached, we substitute $R(t)=l(t)$,
so that,

\begin{equation}
R(t) \simeq \sqrt{\kappa \,M}. 
\end{equation}

\noindent
Adopting the same values as before, we find that $t_{\rm transp} \sim 9\,\textrm{yr}$. When the cloud has become
transparent, all of the energy will have been already radiated away via
diffusion.

\subsection{Total emission of energy}
\label{sec.tot_energy}

The total energy involved in the collision $E_{\rm tot}$ is the sum of three
contributions: The binding energy of the stars ($E_{\rm bin}$) which take place
in the collision plus the kinetic energy $E_{\rm kin}$ at infinity. For one of
the stars participating in the collision, these values are

\begin{align}
E_{\rm kin} & = \frac{\mu}{2} V_{\rm rel}^2 \nonumber \\
E_{\rm bin} & = \alpha \frac{G m_*^2}{R_*},
\end{align}

\noindent
with $\mu:= m_{*,\,1}m_{*,\,2}/(m_{*,\,1}+m_{*,\,2})$ the reduced mass,
and $\alpha=3/(5-n)$, with $n=3$ for a Sun-like star \citep[see][for the equation
and value]{Chandra42}. We can
approximate $E_{\rm bin} \approx m_* V_{\rm esc}^2$, so that for the two stars

\begin{equation}
E_{\rm tot} \approx -\left(m_{*,\,1} V_{\rm esc,\,1}^2 + m_{*,\,2} V_{\rm esc,\,2}^2\right) + 
                     {\mu} V_{\rm rel}^2.
\label{eq.Etot}
\end{equation}

As mentioned in Sec.~(\ref{sec.rates}), since $R_{\rm min} \ll R_{\rm max}$,
the collisional rate will be dominated at smaller radii. For
$M_{\bullet}=10^6\,M_{\odot}$, and for the adopted value of $\gamma=7/4$,
Eq.~(\ref{eq.Rmin}) yields $R_{\rm min}\sim 10^{-5}\textrm{pc}$, so that
$V_{\rm rel} \cong 20000 \textrm{km\,s}^{-1}$. One order of magnitude farther
away from the centre in radius, at $10^{-4}\textrm{pc}$, $V_{\rm rel} \cong
6500 \textrm{km\,s}^{-1}$.  For $M_{\bullet}=10^7\,M_{\odot}$, the minimum
radius is also $R_{\rm min}\sim 10^{-5}\textrm{pc}$ but $V_{\rm rel} \cong
65000 \textrm{km\,s}^{-1}$. At a distance from the MBH of $\sim
10^{-4}\textrm{pc}$ $V_{\rm rel} \cong 20000\textrm{km\,s}^{-1}$. 

At such high relative velocities, we can ignore the contribution of the binding
energy of the stars in Eq.~(\ref{eq.Etot}). To consider two limiting cases, a
$M_{\bullet}=10^7\,M_{\odot}$ at $R_{\rm min}\sim 10^{-5}\textrm{pc}$, yields
$E_{\rm tot} \approx 42\,\textrm{foe}$ ($4.2 \times 10^{52}\textrm{ergs}$),
while a $M_{\bullet}=10^6\,M_{\odot}$ at a distance of $10^{-4}\textrm{pc}$
yields $E_{\rm tot} \approx 0.42 \, \textrm{foe}$ ($4.2 \times
10^{50}\textrm{ergs}$).  A ``typical'' case would range between these two
limits; i.e. $E_{\rm tot} \approx 1\, \textrm{foe}$, which is the usual energy
release of a supernova (considering $V_{\rm rel} \cong
10000\textrm{km\,s}^{-1}$ at $10^{-4}\textrm{pc}$).

\subsection{Time evolution of the released energy and power}
\label{sec.energypower}

We define the loss of energy in the cloud as

\begin{equation}
\frac{dE}{dt} = - \frac{E}{T_{\rm diff}(t)}, 
\label{eq.dEdt}
\end{equation}

\noindent
with $T_{\rm diff}(t)$ as given by Eq.~(\ref{eq.Tdiff}). The physical meaning of the last
equation is that we are identifying $T_{\rm diff}(t)$ as the time for the photons to escape
the cloud as the main sink of energy of it and, hence, the right hand side is negative.
Therefore,

\begin{equation}
\frac{dE}{E} = - \frac{1}{\xi}\,t\,dt,
\label{eq.dEE}
\end{equation}

\noindent
with $\xi^{-1}:= cV_{\rm exp}/(\kappa M)$. The solution to Eq.~(\ref{eq.dEE}) is

\begin{equation}
E(t) = E(0)\left(\frac{\eta}{1}\right) \exp{\left[-\frac{1}{2\,\xi}t^2\right]}. 
\label{eq.Et}
\end{equation}

\noindent
Here $\eta$ is a parameter quantifying the amount of initial kinetic energy
$E(0)$ that goes into radiation.  The value of $\eta$ depends on the details of
the collision and in particular on the slowing down of the shock downstreams;
i.e. how the shock evolves during the collision will alter the relative
velocity of the parts of the stars which have still not collided and translate
into a total efficiency conversion of the kinetic into radiation. See for
instance the work of \cite{CalderonEtAl2020}, in particular their Figs. 4-10.
In this work they focus on relatively low velocities and stellar winds but it
illustrates the non-linearity of our problem. The derivation of this parameter
requires detailed numerical simulations. 

\noindent
We now introduce

\begin{equation}
T_{\rm E} \equiv \sqrt{\frac{\kappa\,M}{c\,V_{\rm exp}}}=\sqrt{T_{\rm diff}(0)\frac{R(0)}{V_{\rm exp}}}, 
\label{eq.TE}
\end{equation}

\noindent
as we can see from Eq.~(\ref{eq.Tdiff}). This
corresponds to $t$ in the approximation we did before, to obtain Eq.~(\ref{eq.Tdiffmonths}). Indeed,
for the values we adopted to derive Eq.~(\ref{eq.Tdiffmonths}), we have that $T_{\rm E} = 0.16\,\textrm{yr}$.
We can now rewrite Eq.~(\ref{eq.dEE}) as

\begin{equation}
E(t) = E(0)\left(\frac{\eta}{1}\right) \exp{\left[-\frac{1}{2}\,\left(\frac{t}{T_{\rm E}}\right)^2\right]}. 
\label{eq.Et}
\end{equation}

Normalizing to standard values, we have

\begin{equation}
E(t) = 10^{51}\,\textrm{ergs} \left(\frac{E(0)}{10^{51}\,\textrm{ergs}}\right)\left(\frac{\eta}{1}\right) 
                \exp\left[-\frac{1}{8}\left(\frac{t}{1\,\textrm{month}}\right)^2\right].
\label{eq.EtNor}
\end{equation}

\begin{figure}
\resizebox{\hsize}{!}
          {\includegraphics[scale=1,clip]{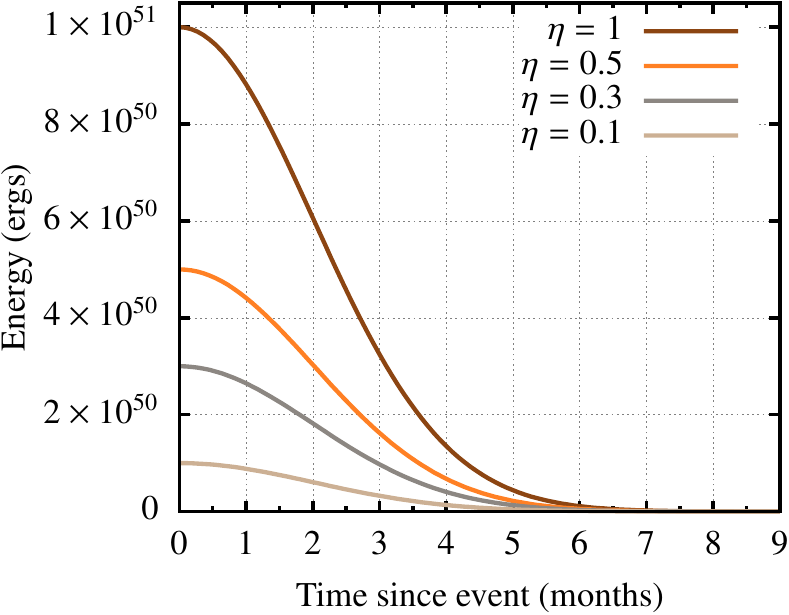}}
\caption
   {
Time evolution of the released energy for four different values of $\eta$, ranging from $1$ (uppermost curve)
to $0.1$ (lowest curve).
   }
\label{fig.E}
\end{figure}

In Fig.(\ref{fig.E}) we depict this time evolution for an initial energy of $E(0)=10^{51}\,\textrm{ergs}$.

With Eq.~(\ref{eq.Et}) we can obtain the emitted power by deriving this last equation,

\begin{equation}
P(t) = -\frac{dE}{dt} = \frac{E(0)}{T_{\rm E}^2}\,\left(\frac{\eta}{1}\right)\,t\,\exp\left[-\frac{1}{2}\,\left(\frac{t}{T_{\rm E}}\right)^2 \right]. 
\end{equation}

\begin{figure}
\resizebox{\hsize}{!}
          {\includegraphics[scale=1,clip]{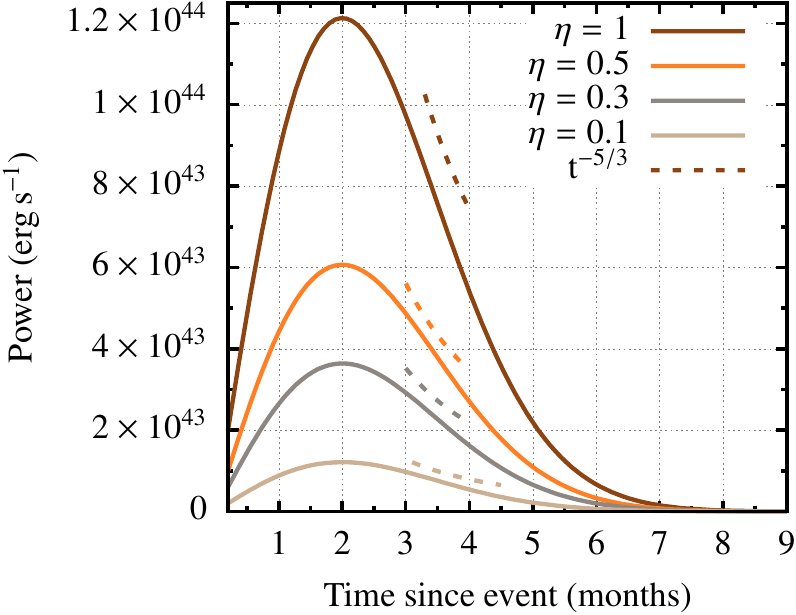}}
\caption
   {
Evolution of the power by a stellar disruption of masses $1\,M_{\odot}$ and $V_{\rm rel}=10^4\,\textrm{km\,s}^{-1}$,
corresponding to the default values of Eq.~(\ref{eq.Power}) for different efficiency parameters $\eta$. The uppermost
curve corresponds to the maximum value of $\eta$ and the lowermost to the minimum value. We add a power-law
curve proportional to $t^{-5/3}$, which is the typical value one expects from a stellar tidal disruption.
   }
\label{fig.P}
\end{figure}

We can normalize the equations by defining $\tau:=t/T_{\rm E}$ and $P_{\rm
norm}\equiv E(0)/T_{\rm E}$, so that

\begin{align}
E(\tau) & = E(0)\, \left(\frac{\eta}{1}\right)\,\exp\left[-\frac{\tau^2}{2}\right]\\
P(\tau) & = P_{\rm norm} \, \tau \, \left(\frac{\eta}{1}\right)\,\exp\left[-\frac{\tau^2}{2}\right].
\end{align}

\noindent
We note here that $\tau$ contains the information relative to the scattering
length of the environment, in $\kappa$, since the mean free path
$l = 1/(\rho_{\rm g}\,\kappa)$, as we can see in Eq.~(\ref{eq.l}), so that
encoded in $T_{\rm E}$ in Eq.~(\ref{eq.Et}) we have
the information about the location of the peak of the distribution, which is,
as we derived, after $2$ months.

Adopting typical values, we can express $P_{\rm norm}$ as follows,

\begin{align}
P_{\rm norm} \cong \times 10^{44}\,\textrm{erg\,s}^{-1} & \left( \frac{E(0)}{10^{51}\,\textrm{ergs}}         \right) 
                                                            \left( \frac{\kappa}{0.04\,\textrm{m}^2\textrm{kg}^{-1}}                         \right)^{-1/2} \nonumber \\
                                                          & \left( \frac{M}{1\,M_{\odot}}                      \right)^{-1/2}
                                                            \left( \frac{V_{\rm exp}}{10^4\textrm{km\,s}^{-1}} \right)^{1/2}.
\label{eq.Pnorm}
\end{align}

Therefore, the final equation for the evolution of power with time is

\begin{align}
P(t) \cong 10^{44}\textrm{erg\,s}^{-1} &\left(\frac{\eta}{1}\right) \left(\frac{t}{1\,\textrm{month}}\right) \exp\left[-\frac{1}{8}\left(\frac{t}{1\,\textrm{month}}\right)^2 \right] \nonumber \\
                                                          &  \left( \frac{E(0)}{10^{51}\,\textrm{ergs}} \right) \left( \frac{\kappa}{0.04\,\textrm{m}^2\textrm{kg}^{-1}}                         \right)^{-1/2} \nonumber \\
                                                          & \left( \frac{M}{1\,M_{\odot}}                      \right)^{-1/2}
                                                            \left( \frac{V_{\rm exp}}{10^4\textrm{km\,s}^{-1}} \right)^{1/2}.
\label{eq.Power}
\end{align}

In Fig.~(\ref{fig.P}) we depict this power for various values of $\eta$. Decreasing $\eta$ 
shifts the peak of the power, lowers its maximum and broadens the distribution, 
as expected from Eq.~(\ref{eq.dEdt}). We have added a line which follows a power-law
of time, $t^{-5/3}$, which corresponds to a stellar tidal disruption \citep[see
e.g.][]{Rees1988}. If the observation of the event takes place between the 3rd and 4th
month after the collision, it could easily be misinterpreted as a tidal disruption.
At later times the curves diverge, so that depending on the observational errors one
could discern the two, or not.

\section{Temperature and spectral power}
\label{sec.temperature}

\subsection{Effective temperature}

From the previous section, we can now estimate the evolution of the effective
temperature of the cloud which expands at a constant velocity $V_{\rm exp}$.
We use the approximation of Stefan–Boltzmann of black body radiation,
$P(t)=\sigma\,T_{\rm eff}^4\,4\pi R(t)^2$, with $\sigma$ the Stefan–Boltzmann
constant and $T_{\rm eff}$ the effective temperature of the body, and assume
that the radius of the cloud coincides with the photosphere.  The physical
interpretation of the definition of this temperature corresponds to the
observed temperature, i.e. what a telescope would measure from the moment of
the impact onwards.

\begin{figure}
\resizebox{\hsize}{!}
          {\includegraphics[scale=1,clip]{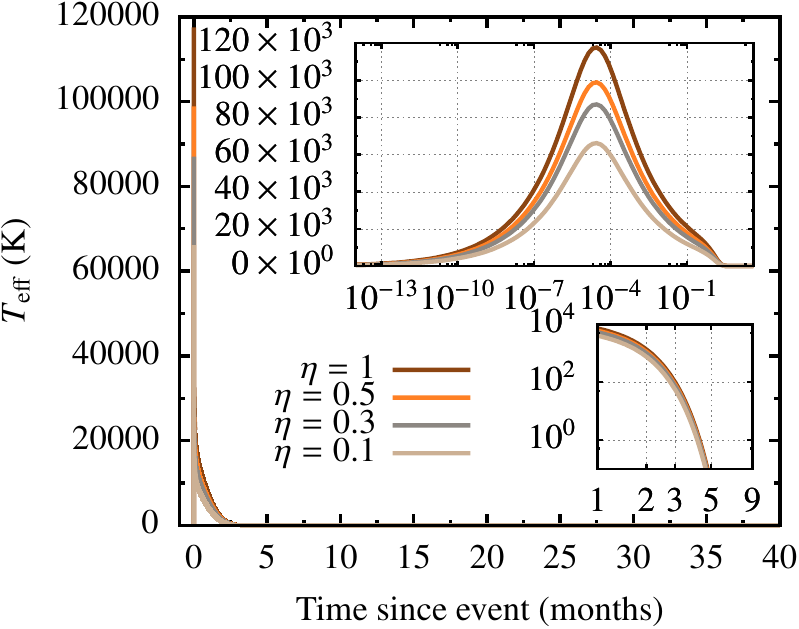}}
\caption
   {
Time evolution of $T_{\rm eff}$ for various values of $\eta$, following the
same order as in Fig.~(\ref{fig.P}). We include two
zooms; the top embedded zoom shows in logarithmic scale in the x-axis the whole range of
time, from $10^{-13}$ to $9$ months, and the bottom one in linear scale the 
last few months, from $1$ to $9$, in logarithmic scale in the y-axis.
   }
\label{fig.Teff}
\end{figure}

\noindent
From Eq.~(\ref{eq.Power}) we obtain

\begin{align}
T_{\rm eff} & \cong 2.32\times 10^{6}\,\textrm{K} \left(\frac{\eta}{1}\right)^{1/4} \left(\frac{t}{1\,\textrm{month}}\right)^{1/4} 
                                                    \exp\left[-\frac{1}{2}\left(\frac{t}{1\,\textrm{month}}\right)^2 \right]  \nonumber \\
                                                &  \left( \frac{E(0)}{10^{51}\,\textrm{ergs}} \right)^{1/4} 
                                                   \left( \frac{\kappa}{0.04\,\textrm{m}^2\textrm{kg}^{-1}}\right)^{-1/8} 
                                                   \left( \frac{M}{1\,M_{\odot}} \right)^{-1/8}                                \nonumber \\
                                                &  \left( \frac{V_{\rm exp}}{10^4\textrm{km\,s}^{-1}} \right)^{1/8} 
                                                   \left[1 + 37775 \left(\frac{V_{\rm exp}}{10^4\textrm{km\,s}^{-1}}\right) 
                                                   \left(\frac{t}{1\,\textrm{month}}\right) \right]^{-1/2},
\label{eq.Tefftime}
\end{align}

\noindent
where we have not neglected the $1$ in $R(t)=R(0)+V_{\rm exp}\,t$ in the last
bracket because this would lead to an artificial value of $T_{\rm eff}$ at $t=0$.
In Fig.~(\ref{fig.Teff}) we display the evolution of the effective temperature as a
function of time for the values of $\eta$ of Fig.~(\ref{fig.P}).

\begin{figure*}
\resizebox{\hsize}{!}
          {\includegraphics[scale=1,clip]{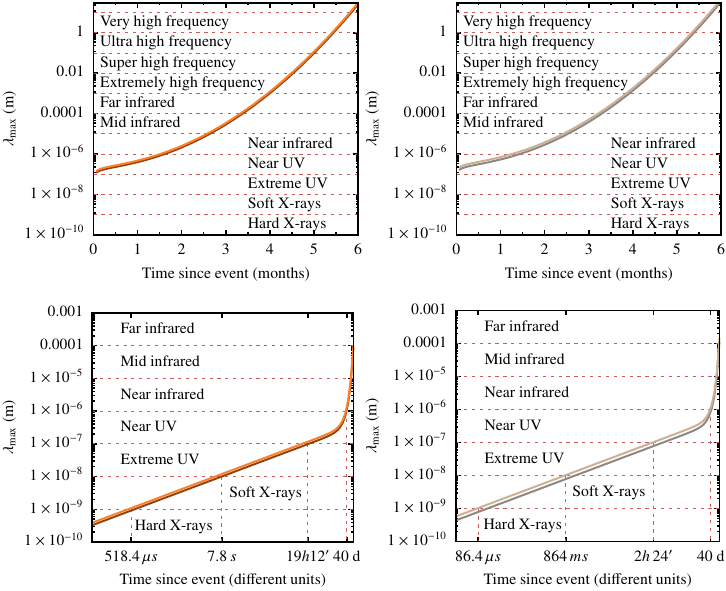}}
\caption
   {
\textit{Top, left panel:} Evolution of the peak wavelength $\lambda_{\rm peak}$ of the spectral radiance
for the cloud. We display the approximate ranges of the spectrum which it will
cover in time. The color scheme follows that of Fig.(\ref{fig.Teff}), meaning that
$\eta=1$ is the lower curve and the upper one corresponds to $\eta=0.5$.
\textit{Top, right panel:} Same as the left panel for $\eta=0.3$ (upper curve) and $\eta=0.1$ (lower curve).
\textit{Bottom, left panel:} Same as the top, left one, but for different time intervals. We add
vertical lines to delimit the different ranges in $\lambda_{\rm peak}$ in time.
   }
\label{fig.peak}
\end{figure*}

Since we are dealing with short wavelengths, we can calculate the peak
wavelength $\lambda_{\rm peak}$ of the spectral radiance of the cloud as a
function of time using an approximation. This is Wien's displacement law, which
relates the absolute temperature $T$ in $K$ and the peak wavelength as $T =
b/\lambda_{\rm peak}$, with $b\sim 2.89\,\times 10^{-3}\,\textrm{m\,K}$ Wien's
displacement constant. In Fig.~(\ref{fig.peak}) we show the evolution of
$\lambda_{\rm peak}$ in the different regimes of frequencies as a function of
time. 

\subsection{Kinetic temperature}

A different definition of temperature is the conversion of kinetic energy into
heat as a result of the impact of the stars. This definition will be useful for
the derivation of the sound velocity at the innermost region of the outcome of
the collision, which will be derived later.

Assuming an ideal gas, the energy and kinetic temperature of the environment are
linked via the usual equation

\begin{equation}
E = \frac{3}{2}\,N\,k \,T_{\rm kin},
\end{equation}

\noindent
with and $k$ the Boltzmann constant, $N=M_{\rm tot}/\mu$, $M_{\rm
tot}=2\,M_{\odot}$ is the total mass, $\mu = 0.6\,m_{\rm p}=5.05 \times
10^{-58}\,M_{\odot}$ the mean molecular mass for fully ionised matter, and
$m_{\rm p}$ the mass of the proton.  We adopt this value because it corresponds
to the radiative zone of a star with a mass similar to the Sun, where hydrogen
and helium constitute most of all elements. In the surface, where the
temperature drops significantly, this assumption would be wrong.

It follows from Eq.~(\ref{eq.EtNor}) that

\begin{equation}
T_{\rm kin}= 1.22 \times 10^{9}\,\textrm{K} \left(\frac{E(0)}{10^{51}\,\textrm{ergs}}\right)\left(\frac{\eta}{1}\right)
                   \exp\left[-\frac{1}{8}\left(\frac{t}{1\,\textrm{month}}\right)^2\right]. 
\label{eq.Tkin}
\end{equation}

\begin{figure}
\resizebox{\hsize}{!}
          {\includegraphics[scale=1,clip]{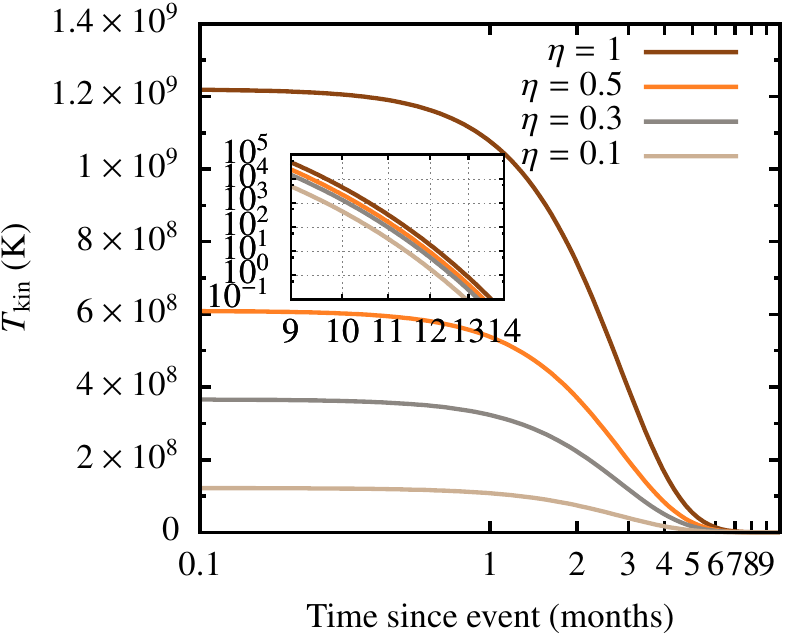}}
\caption
   {
Evolution of the kinetic temperature as the outcome of the collision with time, given by
Eq.~(\ref{eq.Tkin}). We include an embedded zoom of the last few months of evolution and
note that in it both axes are in log scale.
   }
\label{fig.Tkin}
\end{figure}

\noindent
We show the evolution of this last equation in Fig.~(\ref{fig.Tkin}).

\subsection{Spectral power}

In the previous sections we have estimated the total amount of energy released,
as well as the power and the peak wavelength, which can be used as an
approximation to understanding the distribution of energy over different
bandwidths. In this section we will derive how the power distributes over
different ranges of energy. For that, we first have to obtain the distribution
of energy in function of time $t$ and frequency $\nu$. Hence, we have to evaluate
the following quantity, which we will call the spectral power

\begin{equation}
\frac{dE}{dt\,d\nu} = P(t) \, b\left(T_{\rm eff}(t),\,\nu\right).
\label{eq.dEdtdnu}
\end{equation}

\noindent
In this equation the function $b\left(T_{\rm eff}(t),\,\nu\right)$ is the black body
spectrum normalised to $1$ for $T_{\rm eff}(t)$ (i.e. the ``observable temperature'') and $\nu$. In terms of
integration, $T_{\rm eff}(t)$ can be envisaged as a constant, because we have
to integrate in frequencies. I.e. the function corresponds to the
spectral radiance of the cloud for frequency $\nu$ at absolute temperature, Planck's law,
but normalised to one,

\begin{equation}
b\left(T_{\rm eff}(t),\,\nu\right) = \frac{B\left(T_{\rm eff}(t),\,\nu\right)}{C\left(T_{\rm eff}(t)\right)}. 
\label{eq.b}
\end{equation}

\noindent
Here $B\left(T_{\rm eff}(t),\,\nu\right)$ is

\begin{equation}
B\left(T_{\rm eff}(t),\,\nu\right) = \frac{2h\nu^3}{c^2}\frac{1}{e^{\,h\nu/(kT)}-1}, 
\label{eq.B}
\end{equation}

\noindent
with $h$ the Planck constant, $c$ the speed of light, 
and we are identifying $T \equiv T_{\rm eff}(t)$ for clarity.
The integral of this equation over the whole range of $\nu$ does not yield $1$, which
is why we need to obtain the normalization factor,

\begin{equation}
C\left(T_{\rm eff}(t)\right) = \int_{\nu=0}^{\nu=\infty} B\left(T_{\rm eff}(t),\,\nu\right) d\nu. 
\end{equation}

\noindent
If we change the variable $\alpha=h\nu/(kT)$ so that $d\alpha=hd\nu/(kT)$, we obtain

\begin{equation}
C\left(T_{\rm eff}(t)\right) = 2\frac{(kT)^4}{c^2h^3}\int_{0}^{\infty} \frac{\alpha^3}{e^{\,\alpha}-1}d{\alpha}. 
\label{eq.Cint}
\end{equation}

\noindent
The integral of Eq.~(\ref{eq.Cint}) is a special function and a particular case of a Bose–Einstein integral, the 
Riemann zeta function $\zeta (s)$, a function of a complex variable $s$. The integral is analytical and has the
solution

\begin{equation}
\int_{0}^{\infty} \frac{\alpha^3}{e^{\,\alpha}-1}d{\alpha} = \zeta(4)\,\Gamma(4),
\end{equation}

\noindent
with $\Gamma(n)$ the Gamma function, $\Gamma(n)=(n-1)!$ if $n$ is a positive integer. Hence,
$\zeta(4)\,\Gamma(4)=6\zeta (4) = {\pi^4}/15$ and so, Eq.~(\ref{eq.Cint}) becomes

\begin{equation}
C\left(T_{\rm eff}(t)\right) = \frac{2}{15} \frac{\left(T k \pi\right)^4}{c^2 h^3}. 
\end{equation}

\noindent
Plugging this result into Eq.~(\ref{eq.b}) and using Eq.~(\ref{eq.B}), we derive that

\begin{equation}
b\,(T,\,\nu) = 15 \left( \frac{h}{\pi kT} \right)^4 \frac{\nu^3}{e^{\,h\nu/(kT)}-1}. 
\end{equation}

\begin{figure*}
\resizebox{\hsize}{!}
          {\includegraphics[scale=1,clip]{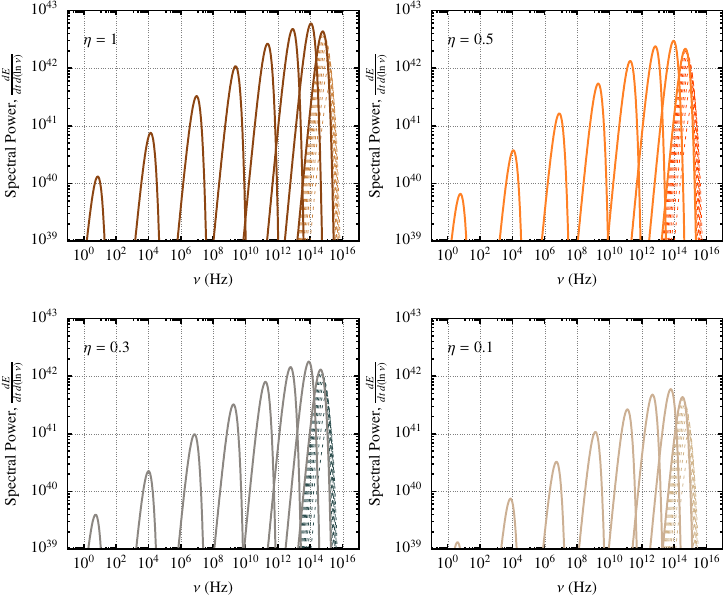}}
\caption
   {
The spectral power as a function of the frequency $\nu$ for the four different
values of the nonlinear parameter $\eta$ taken into consideration. In each of
the panels the different curves correspond to different moments in the
evolution of the expanding cloud after the stellar collision. From the right
(higher values of $\nu$) to the left, we show in dashed lines the first nine
tenths of the first month in the evolution, i.e. towards lower frequencies in
the first dashed curves there is a time increment of $1/10$ of a month. The first
rightmost solid curve corresponds to the spectral power range one month after
the event, the second rightmost one, achieving as expected the maximum value,
to the second month, etc. We display eight months in the evolution to show the
decrease in spectral power, although we note that $10^6\,\textrm{MHz}$
corresponds to the lowest frequency of present instruments.
   }
\label{fig.SpPower}
\end{figure*}

\noindent
Therefore, the spectral power of the cloud is

\begin{equation}
\nu \frac{dE}{dt\,d\nu} = \frac{dE}{dt\,d(\ln \nu)}  = \frac{15}{\pi^4}\,P(t) \frac{\left[h\nu/(kT)\right]^4}{e^{\,[h\nu/(kT)]}-1},
\label{eq.SpecPow}
\end{equation}

\noindent
where we have multiplied Eq.~(\ref{eq.dEdtdnu}) by $\nu$ to obtain the spectral power in $\ln \nu$,
and $P(t)$ is given by Eq.~(\ref{eq.Power}). In Fig.~(\ref{fig.SpPower}) we depict the
spectral power as a function of $\nu$ for the different values of $\eta$ taken into
consideration. With decreasing $\eta$ values, the spectral power is obviously lowered
but in the range of observable frequencies, i.e. from $10^6\,\textrm{MHz}$, the values
achieve relatively high values.

\subsection{Photometric colours and AB magnitude}

We now display the same information but in a different way. If we define a set
a set of passbands (or optical filters), with a known sensitivity to incident
radiation, we are in the position of comparing with real data taken from
surveys. For that, we first adopt Eq.~(\ref{eq.SpecPow}) and remove the factor
$\nu$ on the left-hand-side of the equation, so that we are left with this
integral to solve

\begin{equation}
\mathcal{C}(t) = \frac{15}{\pi^4}\,P(t) \left[h/(kT(t))\right]^4 
                                     \int_{\nu_\text{min}}^{\nu_\text{max}}
                                     \frac{\nu^3}
                                     {e^{\,[h\nu/(kT(t))]}-1}\,d\nu,
\label{eq.SpecPowColor}
\end{equation}

\noindent 
where we have identified $\mathcal{C}\equiv dE/dt$ as the ``colour''. Depending
on the range of frequencies of interest, we will be looking at different bands.
In particular, we define the following ranges for the bands of interest ($\nu$
is given in Hz): U-Band: $\nu_\text{min}=7.54\times 10^{14}$, $\nu_\text{max}=
9.04 \times 10^{14}$, B-Band: $\nu_\text{min}= 6.10\times 10^{14}$,
$\nu_\text{max}= 7.54 \times 10^{14}$, G-Band: $\nu_\text{min}= 5.68\times
10^{14}$, $\nu_\text{max}= 7.50\times 10^{14}$, V-Band: $\nu_\text{min}=
5.04\times 10^{14}$, $\nu_\text{max}= 5.92\times 10^{14}$, R-Band:
$\nu_\text{min}= 4.13\times 10^{14}$, $\nu_\text{max}= 5.09\times 10^{14}$.

\noindent 
The integral in Eq.~(\ref{eq.SpecPowColor}) is a non-trivial one. However,
since the ranges of frequencies that are of our interest are very narrow, 
what we can do is to approximate the integral by the value of the rectangle
delimited by those values. I.e. we simply calculate 

\begin{equation}
\frac{dE}{dt} = \frac{15}{\pi^4}\,P(t) \frac{\left[h\nu_\text{avrg}/(kT(t))\right]^4}{e^{\,[h\nu_\text{avrg}/(kT(t))]}-1} 
                \ln{\left( \frac{\nu_\text{max}}{\nu_{\text{min}}} \right)}.
\label{eq.colour}
\end{equation}

\begin{figure*}
\resizebox{\hsize}{!} 
          {\includegraphics[scale=1,clip]{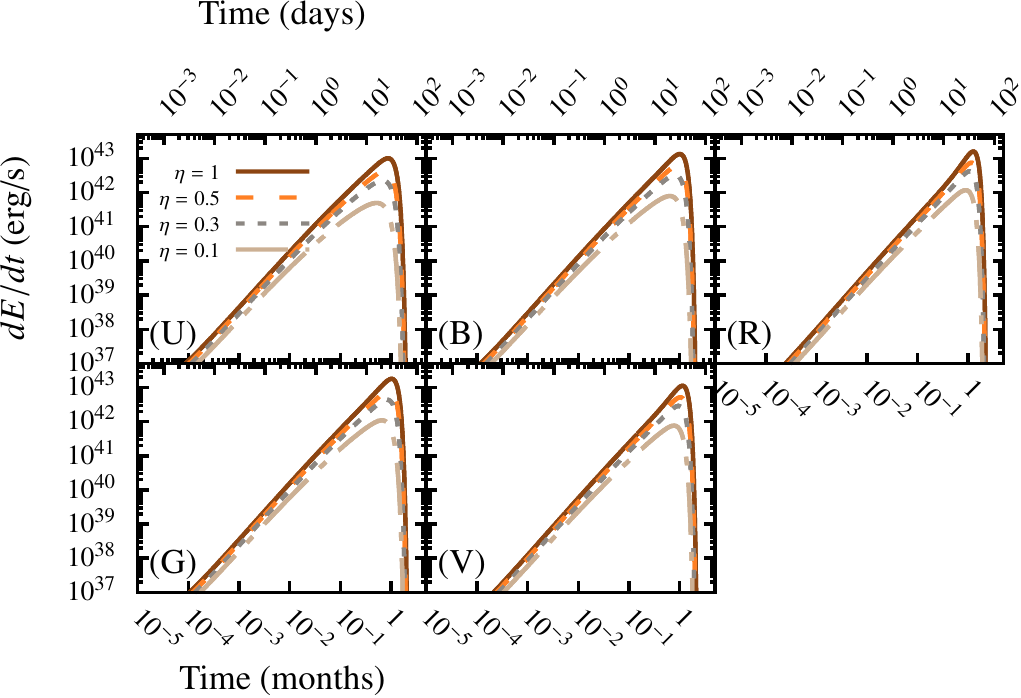}} 
\caption
   {
Photometric indeces U, B, G, V and R as a function of time (in months, lower x-axis and
days, upper x-axis) for the different values of the parameter $\eta$.
   }
\label{fig.PhotometryMultiplot}
\end{figure*}

\noindent 
In this expression, $\nu_\text{max}$ and $\nu_\text{min}$ are determined by the
colour of interest and $\nu_\text{avrg}$ is the characteristic frequency
associated with that particular band. We can obtain its value by knowing that
the length in nm for the various bands is in the U band $l=365$ nm, so that
$\nu_\text{avrg}=8.21\times 10^{14}$ Hz, in the B band $l=445$ nm, and hence
$\nu_\text{avrg}=6.74\times 10^{14}$ Hz, in the G band $l=464$,
$\nu_\text{avrg}=6.46\times 10^{14}$ Hz, in the V band $l=551$ nm,
$\nu_\text{avrg}=5.44\times 10^{14}$ Hz and in the R band $l=658$ nm,
$\nu_\text{avrg}=5.56\times 10^{14}$ Hz.  The conversion is straightforward,
since $\nu_\text{avrg}(l)=c/l=3\times 10^{8}/(l\times 10^{-9})$ to obtain Hz. 
This approximation has an error of about $10\%$ as compared to a numerical integration.
In Fig.~(\ref{fig.PhotometryMultiplot}) we show the different evolutions of the
photometric indeces as a function of time.

In order to derive the absolute magnitude (AB magnitude), we remind the reader that it is usually
defined as the logarithm of the spectral flux density which defines a zero point value at 3631 Jy. 
By defining the spectral flux density as $\mathcal{F}$, the AB magnitude can be calculated in
cgs units as

\begin{equation}
m_{\text{AB}}=-2.5\log_{10}\mathcal{F}-48.60.
\end{equation}

\noindent 
The bandpass AB magnitude spanning across a continuous range of wavelengths is
usually defined in such a way that the zero point corresponds to
$\mathcal{F}\sim 3631\,\text{Jy}$. Hence,

\begin{equation}
m_{\text{AB}}\approx -2.5\log_{10}\left[{\frac {\int \mathcal{F}{(h\nu)}^{-1}e(\nu)\,
\mathrm{d}\nu }{3631\int {(h\nu)}^{-1}e(\nu)\,\mathrm{d}\nu }}\right].
\end{equation}

\noindent 
In this expression, $e(\nu)$ is the filter response function and the term 
${(h\nu)}^{-1}$ accounts for the photon-counting device.

\begin{figure*}
\resizebox{\hsize}{!}
          {\includegraphics[scale=1,clip]{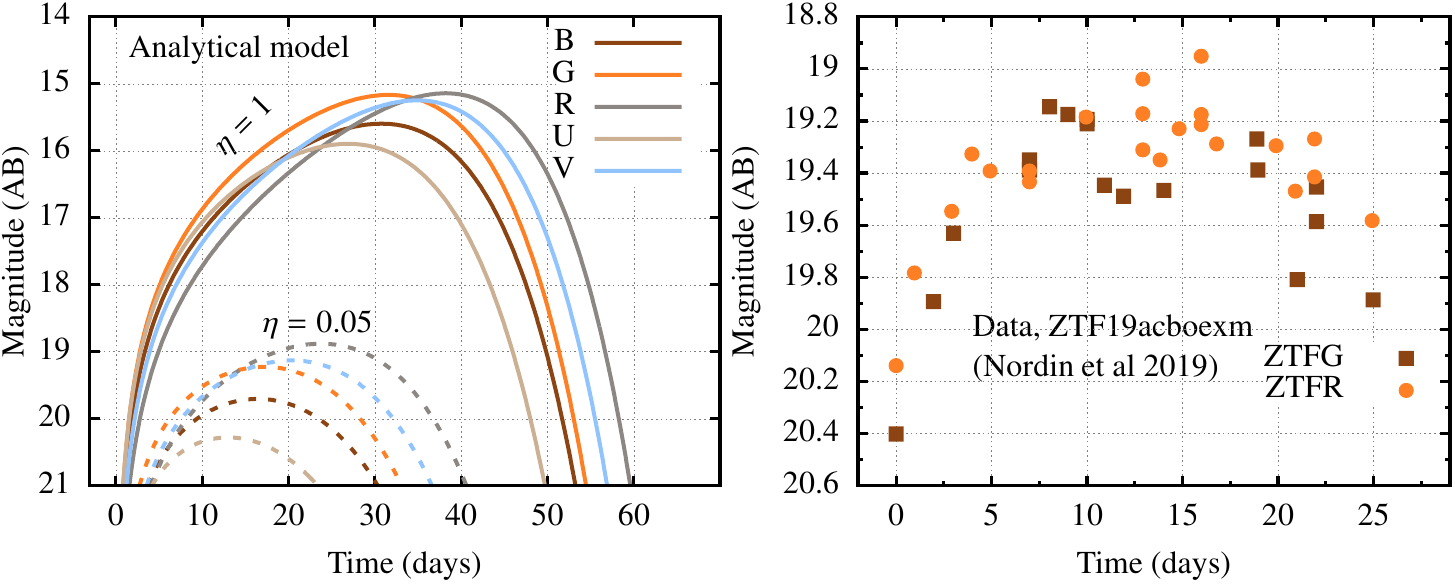}} 
\caption
   {
\textit{Left panel: } AB magnitude as calculated from the theoretical 
model at a distance of 194.4 Mpc. We give the extreme values that we
have adopted in this work for the free parameter $\eta$, i.e. 1 and also 0.05. 
\textit{Right panel: } Zwicky Transient Facility (ZTF) report for 2019-10-07
corresponding to the object ZTF19acboexm by \cite{NordinEtAl2019}.
The data taken with ZTFG are marked with squares and the data taken with
ZTFR with circles. 
If the transient was the result of a stellar collision,
it would seem to correspond to a value of $\eta \lesssim 0.05$.
   }
\label{fig.ZTF19acboexm_theory_data}
\end{figure*}

In Fig.~(\ref{fig.ZTF19acboexm_theory_data}) we display the AB magnitude for a
typical collision located at a distance of 194.4 Mpc to be able to compare it
to the object ZTF19acboexm from the ZTF transient discovery report of
\cite{NordinEtAl2019}. If this transient had inded its origin in a stellar
collision, then the free parameter responsible for the efficiency of the energy
conversion should be of about $\eta 0.05$.

\section{Gravitational waves and multimessenger searches}
\label{sec.GW}

If we calculate the binding energy of the cores of the stars which initially
are on a hyperbolic orbit and compare it to the total kinetic energy of the
system as derived in~Sec.(\ref{sec.tot_energy}), we obtain that the binding
energy is of about one order of magnitude below the total kinetic energy. This
is a natural consequence of our choice of the problem, since in this work we
are focusing on totally disruptive collisional events, which are the most
energetic ones. 

However, for lower relative velocities, of about $V_{\rm rel} \leq
2500\,\textrm{km\,s}^{-1}$, a fraction of the stellar collisions are such that
the inner cores survive the impact and form a temporary binary embedded in a
gaseous medium. In this section we will consider a fixed relative velocity of
$V_{\rm rel}=1000\,\textrm{km\,s}^{-1}$.

With this new value, when evaluating Eq.~(\ref{eq.Etot}), we find that the
total kinetic energy involved is of $T_\text{K}\sim 9.94\times 10^{48}$ ergs,
while the binding energy of the two stars is of $E_\text{bind}\sim1.57\times
10^{49}$ ergs (i.e.  $\sim 7.6\times 10^{48}$ ergs per star).  Therefore, after
the collision, one has a gaseous cloud which is expanding very quickly plus two
surviving pieces of the stars.

If we assume that $T_\text{K}$ is distributed equally among the two colliding
stars, then each receives an input of $T_\text{K}/2=4.97\times10^{48}$ ergs. This
means that after the collision, there would be a leftover of binding energy per
star of approximately 40\% the initial binding energy of one star.

Since the core is the densest part of the star, it stands to reason that this
40\% represents the core which is surviving. The core of the Sun has a mass of
$\sim 0.34\,M_{\odot}$. So all we have after the collision is two cores in a
gaseous cloud which is expanding.

The luminosity of a naked core of a Sun-like star radiates at
$\sim4\times10^{33}$ ergs but the total initial kinetic energy radiated right
after the collision is of $\sim 10^{49}$ ergs. We could think that the gaseous
cloud will radiate away this energy in such a short timescale that we are left
with the two cores which will continue radiating. However, as we will see, the
cores will merge before this happens. Therefore we will neglect this extra
luminosity of the cores when evaluating the properties of such a ``flare'' in
the following sections.

This kind of collisions is a subfraction of the subset of almost head-on
collisions, i.e. for small impact parameters \citep[private communication of
Marc Freitag, as published in his PhD thesis, but see][as well]{FB05}. In this
section we will adopt a representative value of $V_{\rm rel} =
10^3\,\textrm{km\,s}^{-1}$, i.e. one order of magnitude smaller than before,
which is of the order of the velocity dispersion in these environments. We note
that the derivation of the absolute rates, however, as derived previously,
remain the same, since the assumptions we used still hold for our current
choice of $V_{\rm exp}$, even if it is one order of magnitude smaller, as
explained in section~(\ref{sec.rates}).  Nonetheless, Eq.~(\ref{eq.GammaColl})
should be multiplied by a fraction number $f_{\rm bin}$ of those simulations
which lead to the temporary formation of a core binary. This is the second free
parameter of this article (the first is $\eta$, responsible for the
non-linearity), which would require dedicated numerical simulations since this
information is not contained in \cite{FB05} or elsewhere to the best of our
knowledge.

In this section we consider a low-velocity disruptive collision which firstly
leads to a source of electromagnetic radiation.  We rederive the quantities and
figures of the previous sections for this smaller value of $V_{\rm rel}$.
Later, we derive the properties of the binary to then address the evolution of
the source of gravitational waves and the prospects for its detection because,
as we will see, it could mimic a binary of two supermassive black holes in vacuum,
although it should be straightforward to tell them apart. 

\subsection{Electromagnetic signature of low-velocity collisions}

Because we are interested in the electromagnetic precursor of the gravitational
wave, we reproduce the previous figures for the effective temperature, energy
release, power output and spectral power for the new value of $V_{\rm rel} =
10^3~\textrm{km\,s}^{-1}$, because they change and could be of interest in a
search in observational data.

To derive the time evolution of the released energy and power, we must note that
Eq.~(\ref{eq.TE}) now is $T_{\rm E} \sim 0.52 ~\textrm{yr} \sim 6.2~\textrm{months}$
and that $E(0)\sim 10^{49}~\textrm{ergs}$. Hence,

\begin{align}
E(t) & \cong 10^{49}\,\textrm{ergs} \left(\frac{E(0)}{10^{49}\,\textrm{ergs}}\right)\left(\frac{\eta}{1}\right) \nonumber \\
     &           \exp\left[-\frac{13}{2000}\left(\frac{t}{1\,\textrm{month}}\right)^2\right].
\label{eq.EtNor_low_vel}
\end{align}

\begin{figure}
\resizebox{\hsize}{!}
          {\includegraphics[scale=1,clip]{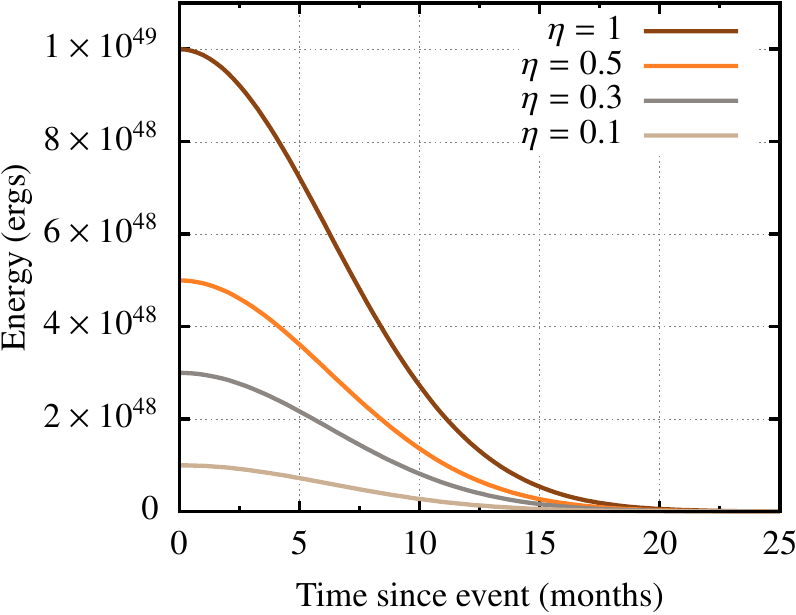}}
\caption
   {
Same as Fig.~(\ref{fig.E}) but for $V_{\rm exp} = 10^3~\textrm{km\,s}^{-1}$.
   }
\label{fig.E_low_vel}
\end{figure}

\begin{figure}
\resizebox{\hsize}{!}
          {\includegraphics[scale=1,clip]{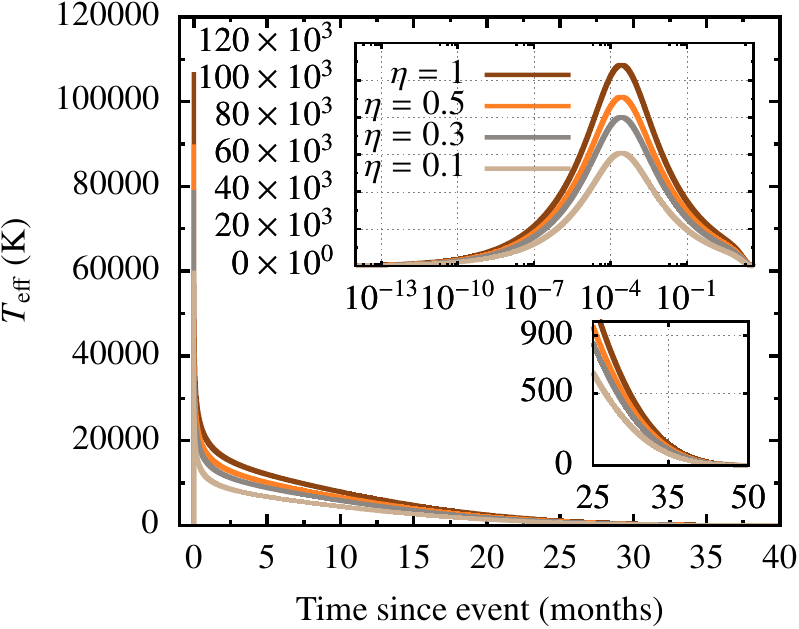}}
\caption
   {
Same as Fig.~(\ref{fig.Teff}) but for $V_{\rm exp} = 10^3~\textrm{km\,s}^{-1}$.
   }
\label{fig.T_low_vel}
\end{figure}

\noindent
We can see this graphically in Fig.~(\ref{fig.E_low_vel}). The initial values are significantly
lower but the time in which the source is radiating is extended to almost two years in the decay.
In Fig.~(\ref{fig.T_low_vel}) we depict the same as in Fig.~(\ref{fig.Teff}) but for the new velocity.

\begin{figure}
\resizebox{\hsize}{!}
          {\includegraphics[scale=1,clip]{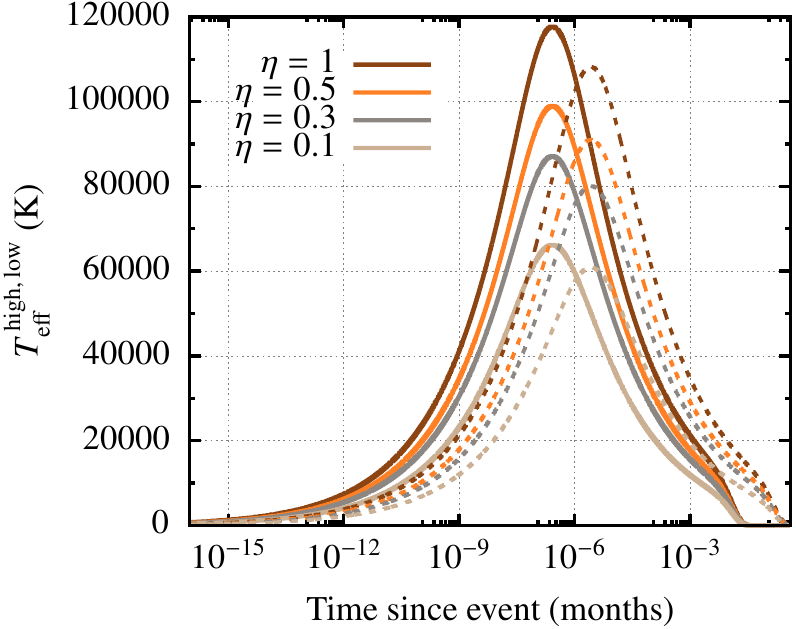}}
\caption
   {
Comparison of $T_{\rm eff}(\tau)$ for the $V_{\rm exp} =
10^4~\textrm{km\,s}^{-1}$ case (solid lines, $T_{\rm eff}^{\rm high}(\tau)$)
and $V_{\rm exp} = 10^3~\textrm{km\,s}^{-1}$ ($T_{\rm eff}^{\rm low}(\tau)$,
dashed curves).
   }
\label{fig.Teff_Comparison}
\end{figure}

In Fig.~(\ref{fig.Teff_Comparison}) we display a comparison between the two different
cases we are treating, the high-velocity one and the low one. As expected, the temperature
peak decreases in the case of low velocity, and lasts longer, so that it is shifted towards
later times.

As for the emitted power, Eq.~(\ref{eq.Pnorm}) becomes

\begin{align}
P_{\rm norm}  \sim 6.32 \times 10^{41}\,\textrm{erg\,s}^{-1} & \left( \frac{E(0)}{10^{49}\,\textrm{ergs}}         \right)
                                                            \left( \frac{\kappa}{0.04\,\textrm{m}^2\textrm{kg}^{-1}}                         \right)^{-1/2} \nonumber \\
                                                          & \left( \frac{M}{1\,M_{\odot}}                      \right)^{-1/2}
                                                            \left( \frac{V_{\rm exp}}{10^3\textrm{km\,s}^{-1}} \right)^{1/2},
\label{eq.Pnorm_low_vel}
\end{align}

\noindent
and so, the emitted power in the collision of two stars at low velocity is

\begin{align}
P(t) & \sim 6.32\times 10^{41}\textrm{erg\,s}^{-1} \left(\frac{\eta}{1}\right) \left(\frac{t}{1\,\textrm{month}}\right) \nonumber \\
                                                          & \exp\left[-\frac{13}{2000}\left(\frac{t}{1\,\textrm{month}}\right)^2 \right] \nonumber \\
                                                          &  \left( \frac{E(0)}{10^{49}\,\textrm{ergs}} \right) \left( \frac{\kappa}{0.04\,\textrm{m}^2\textrm{kg}^{-1}}                         \right)^{-1/2} \nonumber \\
                                                          & \left( \frac{M}{1\,M_{\odot}}                      \right)^{-1/2}
                                                            \left( \frac{V_{\rm exp}}{10^3\textrm{km\,s}^{-1}} \right)^{1/2}.
\label{eq.Power_low_vel}
\end{align}

\begin{figure}
\resizebox{\hsize}{!}
          {\includegraphics[scale=1,clip]{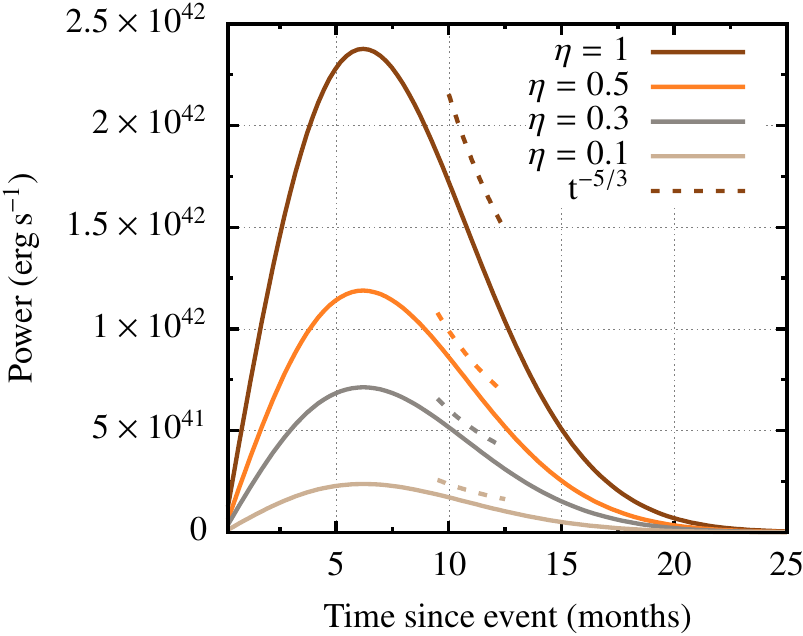}}
\caption
   {
Same as Fig.~(\ref{fig.P}) but for $V_{\rm exp} = 10^3~\textrm{km\,s}^{-1}$.
   }
\label{fig.P_low_vel}
\end{figure}

We can see this in Fig.~(\ref{fig.P_low_vel}). Thanks to this last expression, as explained in the
previous section, we can now derive $T_{\rm eff}$ for the low-velocity collision,

\begin{align}
T_{\rm eff} & \cong 1.2\times 10^{6}\,\textrm{K} \left(\frac{\eta}{1}\right)^{1/4} \left(\frac{t}{1\,\textrm{month}}\right)^{1/4} \nonumber \\
                                                &   \exp\left[-3.25\times 10^{-3}\left(\frac{t}{1\,\textrm{month}}\right)^2 \right]
                                                  \left( \frac{E(0)}{10^{49}\,\textrm{ergs}} \right)^{1/4} \nonumber \\
                                                &   \left( \frac{\kappa}{0.04\,\textrm{m}^2\textrm{kg}^{-1}}\right)^{-1/8} 
                                                   \left( \frac{M}{1\,M_{\odot}} \right)^{-1/8}  
                                                    \left( \frac{V_{\rm exp}}{10^3\textrm{km\,s}^{-1}} \right)^{1/8} \nonumber \\
                                                &   \left[1 + 3777 \left(\frac{V_{\rm exp}}{10^3\textrm{km\,s}^{-1}}\right) 
                                                   \left(\frac{t}{1\,\textrm{month}}\right) \right]^{-1/2},
\label{eq.Tefftime_low_vel}
\end{align}

\begin{figure*}
\resizebox{\hsize}{!}
          {\includegraphics[scale=1,clip]{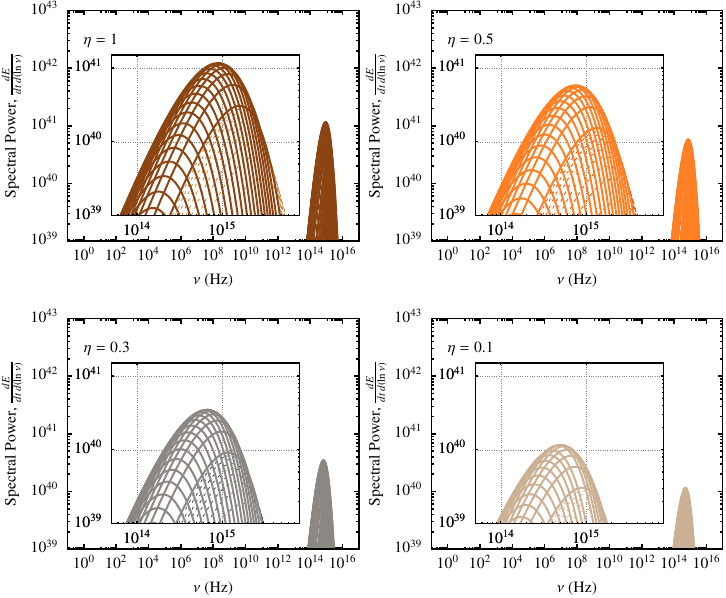}}
\caption
   {
Same as Fig.~(\ref{fig.SpPower}) but for $V_{\rm exp} = 10^3~\textrm{km\,s}^{-1}$.
The solid lines now range from the first month to the 24th and the dashed lines
represent the same fraction of time as in Fig.~(\ref{fig.SpPower}). We add a zoom
in of each $\eta$.
   }
\label{fig.SpPower_low_vel}
\end{figure*}

Finally, in Fig.~(\ref{fig.SpPower_low_vel}) we show the corresponding of
Fig.~(\ref{fig.SpPower}) but for the lower value of $V_{\rm exp}$. The
Eq.~(\ref{eq.SpecPow}) needs no modification, but we need to take the correct
values for $T_{\rm eff}$ and $P(t)$ into account, i.e.
Eq.~(\ref{eq.Tefftime_low_vel}) and Eq.~(\ref{eq.Power_low_vel}) respectively.
We note how the spectral power is now concentrated over a much shorter span
of frequencies.

The kinetic temperature can be estimated as in Eq.~(\ref{eq.Tkin}), but this time the
values are accordingly lower,

\begin{align}
T_{\rm kin} \cong& 1.22 \times 10^{7}\,\textrm{K} \left(\frac{E(0)}{10^{51}\,\textrm{ergs}}\right)\left(\frac{\eta}{1}\right) \nonumber \\
                 &  \exp\left[-\frac{13}{2000}\left(\frac{t}{1\,\textrm{month}}\right)^2\right]. 
\label{eq.Tkin_low_vel}
\end{align}

\begin{figure}
\resizebox{\hsize}{!}
          {\includegraphics[scale=1,clip]{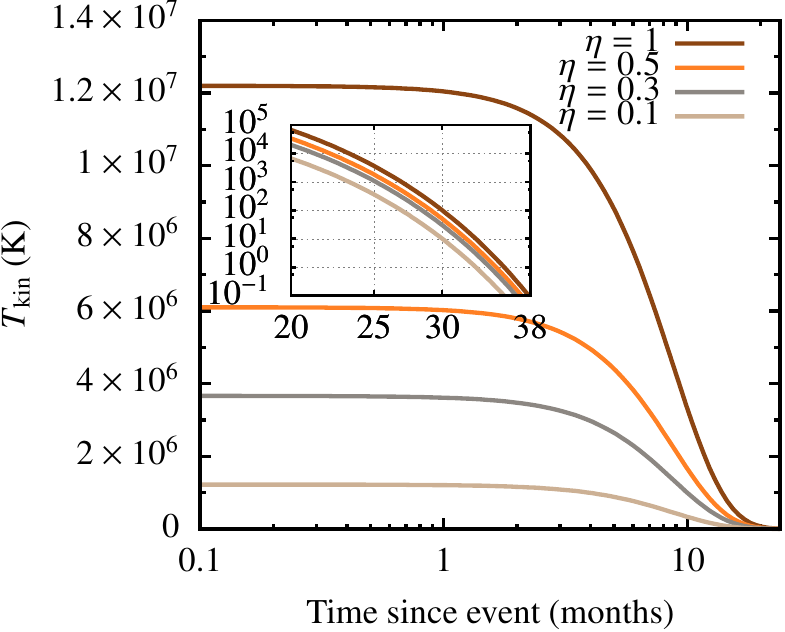}}
\caption
   {
Same as in Fig.~(\ref{fig.Tkin}) but for $V_{\rm exp} = 10^3~\textrm{km\,s}^{-1}$.
   }
\label{fig.Tkin_low_vel}
\end{figure}

\noindent
In Fig.~(\ref{fig.Tkin_low_vel}) we depict this evolution. We can see that the values remain higher
at later times.

\subsection{Point particles in vacuum}

Let us now consider the evolution of two point particles in perfect vacuum with
the masses of the cores starting at a given semi-major axis and evolving only
due to the emission of gravitational radiation.  Thanks to the the
approximation of non-precessing but shrinking Keplerian ellipses of
\cite{Peters64} we can derive an estimate for the associated timescale for a
binary of (point) masses $m_1 = m_2 = m_{\rm core}$ and semi-major axis $a$ to
shrink only via emission of gravitational radiation, 

\begin{equation}
T_{\rm GW} \equiv \frac{a}{|\dot{a}_{\rm GW}|} = \frac{5}{128}\frac{c^5\,a^4}{G^3m_{\rm core}^3}F(e)^{-1}.
\label{eq.TGW}
\end{equation}

Normalising to the values we are using,

\begin{equation}
T_{\rm GW} \cong 5\times 10^8\textrm{yrs} \left(\frac{m_{\rm core}}{0.34 M_{\odot}}\right)^{-3} \left( \frac{a}{R_{\odot}/2} \right)^4 F(e)^{-1}, 
\label{eq.Tgw}
\end{equation}

\noindent
where we have chosen the semi-major axis of the cores to be roughly $a \sim
d_{\rm min} = R_{\odot}/2$, from Eq.~(\ref{eq.fcoll}). We will however see that
this initial choice has little to no impact on the merging time when gas is taken
into account. 

We have chosen $m_{\rm core}=0.34 M_{\odot}$ by assuming that the core radius
of the Sun is located at about a distance of $r_{\textrm{core}}\sim
0.2\,R_{\odot}$, following the data in table 3 of \cite{AbrahamIben1971}, and
we note that the correction factor $Q$ to multiply this timescale introduced by
\cite{ZwickEtAl2019} can be neglected, because $Q \sim 1$ in our case. However,
as we will see later, the final results are to some extent independent of the
choice of initial and final semi-major axes.  In the equation we have
introduced

\begin{equation}
F(e):=(1-e^2)^{-7/2}\left(1+\frac{73}{24}e^2+\frac{37}{96}e^4\right). 
\end{equation}

\noindent
For a very eccentric orbit, $e=0.9$, $F(e)^{-1}\sim 2\times 10^{-3}$. I.e.  we
shorten the timescale by two orders of magnitude. However, even if the
eccentricity at binary formation is very large, it circularizes in a very few
orbits (see the SPH simulations of \citealt{FB05}). We will hence assume
$F(e)^{-1}=1$.

Nevertheless, Eq.~(\ref{eq.Tgw}) is nothing but an \textit{instantaneous} estimation of the
(order of magnitude) time for merger due solely to the emission of gravitational radiation.
This means that, for a given, fixed, semi-major axis, we obtain a timescale. Nonetheless,
the axis shrinks as a function of time, so that $T_{\rm GW}$ will become shorter as well,
because it is a function of time. From Eq.~(\ref{eq.Tgw}), and taking into account the original
negative sign of \cite{Peters64}, we can derive that

\begin{equation}
\int a^3\,da = - \frac{128}{5}\frac{G^3\,m_{\rm core}^3}{c^5}\int dt. 
\end{equation}

\noindent
Hence,

\begin{equation}
\frac{a^4}{4} = -\frac{128}{5}\frac{G^3\,m_{\rm core}^3}{c^5}\,t + \textrm{constant}. 
\end{equation}

\noindent
We can obtain the value of the constant by setting $t=0$, which leads to $\textrm{constant}=a(0)^4/4$.
Since we have chosen $a(0)\equiv a_0 = R_{\odot}/2$, we derive that the evolution of the semi-major axis
of the binary due only to the emission of gravitational waves is

\begin{align}
a(t) \cong & \Big[ \frac{1}{16} \left( \frac{a_0}{R_{\odot}/2} \right)^4 \nonumber \\
           &  - 4.4\times 10^{-11} \left(\frac{m_{\rm core}}{0.34\,M_{\odot}}\right)^{3} \left(\frac{t}{1\,\textrm{month}}\right)
\Big]^{1/4}\,R_{\odot}. 
\label{eq.atGW}
\end{align}

\begin{figure}
\resizebox{\hsize}{!}
          {\includegraphics[scale=1,clip]{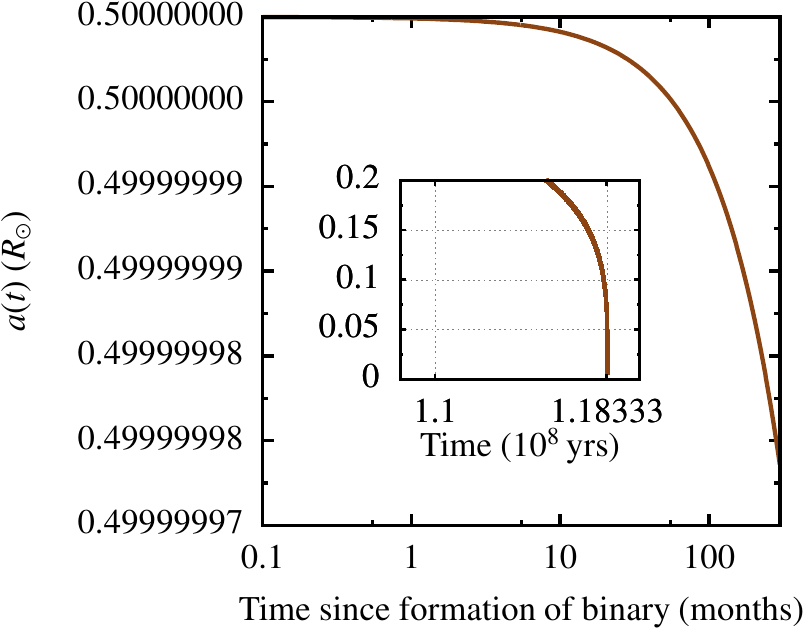}}
\caption
   {
Evolution of the semi-major axis of the binary since formation, as described by
Eq.~(\ref{eq.atGW}). The embedded panel allows us to see the evolution of the last
$0.2\,R_{\odot}$ in units of $10^8\,\textrm{yrs}$. The semi-major axis reaches
0 (assuming point-particles) at $t=1.42\times 10^9 \,\textrm{months}$, as can
be derived by setting Eq.(\ref{eq.atGW}) to zero.
   }
\label{fig.semimajor}
\end{figure}

In Fig.~(\ref{fig.semimajor}) we show the evolution of Eq.~(\ref{eq.atGW}). 

\begin{figure}
\resizebox{\hsize}{!}
          {\includegraphics[scale=1,clip]{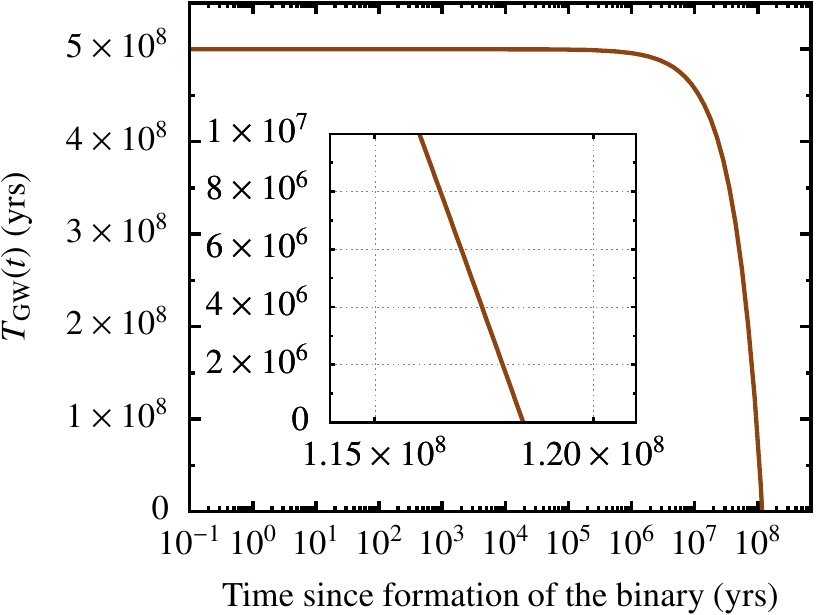}}
\caption
   {
Evolution of the characteristic timescale $T_{\rm GW}$ as a function of time.
The inset allows us to see when it reaches zero. Note that the x-axis in it is in
linear scale.
   }
\label{fig.TGW}
\end{figure}

Replacing Eq.~(\ref{eq.atGW}) in Eq.~(\ref{eq.TGW}) leads to

\begin{align}
T_{\rm GW}(t)  \cong & \,5 \times 10^8\, \textrm{yrs} \left(\frac{m_{\rm core}}{0.34\,M_{\odot}}\right)^{-3} \times \Big[ \left( \frac{a_0}{R_{\odot}/2} \right)^4 \nonumber \\
                    & - 7.04 \times 10^{-10} \left(\frac{m_{\rm core}}{0.34\,M_{\odot}}\right)^{3} \left(\frac{t}{1\,\textrm{month}}\right) \Big]
\label{eq.TGW_t}
\end{align}

\noindent
whose evolution we can see in Fig.~(\ref{fig.TGW}).

\subsection{Cores embedded in a gaseous medium}

For a stellar object of mass $m_{\rm obj}$ moving through a homogeneous
isothermal gaseous medium of constant density $\rho$ along a straight line with
a velocity $V_{\rm obj}$, \cite{Ostriker1999} derives that for a supersonic
motion, the drag force provided by dynamical friction as derived by
\cite{Chandrasekhar1943} must be modified and is 

\begin{equation}
F_{\rm drag} \sim 4\pi \rho \left(\frac{Gm_{\rm obj}}{V_{\rm obj}}\right)^2.
\end{equation}

\noindent
Her results have been confirmed numerically by the work of
\cite{SanchezSalcedoBrandenburg1999}.
Hence, for the velocity of one of the
two cores to be decreased by one e-folding in the gaseous cloud, the associated
timescale is

\begin{equation}
T_{\rm gas} \equiv \frac{V_{\rm core}}{dV_{\rm core}/dt} = \frac{dt}{d\ln{V_{\rm core}}} \cong
            \frac{m_{\rm core}V_{\rm core}}{F_{\rm drag}},
\label{eq.Tefolding}
\end{equation}

\noindent
where $V_{\rm core}$ is the velocity of the core. The last term in the equation
is momentum divided by force, which gives an estimate of order of magnitude for
the characteristic timescale, the timescale to change $\ln{V_{\rm core}}$ by
one dex.  We normalise it to the relevant values for this work as

\begin{align}
T_{\rm gas} & \cong 8.4\times 10^{-6}~\textrm{yrs}~\left( \frac{n}{10^{24}~\textrm{cm}^{-3}} \right)^{-1} \nonumber \\
            &  \left(\frac{m_{\rm core}}{0.34M_{\odot}} \right)^{1/2} 
\left(\frac{a}{R_{\odot}/2} \right)^{-3/2}. 
\label{eq.Tgas}
\end{align}

\noindent
This timescale agrees with the results found by \cite{AntoniEtAl2019}, in particular their Eq.~(37).
This is about two orders of magnitude shorter than the orbital period of the
binary with the default values in Eq.~(\ref{eq.Tgas}), 
$P_{\rm orb}= 2\pi \sqrt{a^3/(2\times G\,m_{\rm core})}$,  

\begin{equation}
P_{\rm orb} \sim 1.4\times 10^{-4}\,\textrm{yrs} \left(\frac{a}{R_{\odot}/2}\right)^{3/2}\left(\frac{m_{\rm core}}{0.34\,M_{\odot}}\right)^{-1/2},
\label{eq.Porb}
\end{equation}

\noindent
which means that the binary would not be able to do one orbit before the cores
sink and merge due to the gas. To derive Eq.~(\ref{eq.Tgas}), we have
taken as average density that of the Sun, $\rho_{\odot} \sim
1\,\textrm{gr\,cm}^{-3}$, which translates into a numerical density of
$10^{24}\textrm{cm}^{-3}$ for the mass of the proton.  The amount of gas
contained within the orbit can be easily calculated; this is important because,
should it be larger than the mass of the cores, then one should use this mass
to calculate the orbital velocity. However, for the kind of semi-major axis
that we are considering, the mass in gas contained in the orbit of the cores is
$M_{\rm gas,\,orb}=\bar{\rho}_{\odot}\times V_{\rm gas,\,orb}\sim 5\times
10^{-3}M_{\odot} < 2\times m_{\rm core}$, with $V_{\rm gas,\,orb}$ the volume inside
of the orbit and $\bar{\rho}_{\odot}$ the average solar density in the radiative zone, assumed to be $\bar{\rho}_{\odot}=10\,\textrm{g\,cm}^{-3}$. This means
that the velocity to take into account to derive Eq.~(\ref{eq.Tgas}) is $V_{\rm
core}$, as we have done. 

Nonetheless, this derivation of $T_{\rm gas}$ does \textit{not} take into
account the fact that the cores are not moving into a straight line, but they
form a binary and hence the density wake around them modifies the drag force
\citep[see e.g.][]{SanchezSalcedoBrandenburg2001,EscalaEtAl04,KimKim2009}.  If
the semi-major axis is smaller than the Bondi accretion radius,

\begin{equation}
R_{\rm Bondi} = \frac{2\,Gm_{\rm core}}{C_{\rm s}}, 
\label{eq.Rbondi}
\end{equation}

\noindent
with $C_{\rm s}$ the sound speed of the cloud, one needs to correct the gas
density around the cores by multiplying $n$ in Eq.~(\ref{eq.Tgas}) by $(R_{\rm
Bondi}/a)^{3/2}$ \citep[as realised by][]{AntoniEtAl2019}. Since we are
assuming almost head-on collisions, we have chosen the semi-major axis for the
cores to be of about $R_{\odot}/2$, also motivated by the outcome of the SPH
simulations of \cite{FB05}.  

Assuming an ideal gas, we can estimate $C_{\rm s}=\sqrt{\gamma_\text{ad}\,P/\rho_{\rm
g}}$, with $\gamma_\text{ad}$ the adiabatic index of the gas, which we assume to be a
fully ionized plasma, so that $\gamma_\text{ad}=5/3$, and $P$ the pressure, and so
$C_{\rm s} = \sqrt{\gamma_\text{ad} T(t) k/m}$, with $m = 0.6\,m_{\rm p}=
1.004\times10^{-27}\,\textrm{kg}$ and $T(t)$ the temperature of the environment.
This temperature is \textit{not} the effective temperature $T_{\rm eff}(t)$,
but the kinetic temperature $T_{\rm kin}(t)$, i.e. the temperature around the
cores in the environment in which they are embedded, whose properties we
approximate to be those of the radiative zone in the Sun, in terms of fully
ionised matter but also of density, as we will see later. Since we are
interested in low-velocity collisions, from Eq.~(\ref{fig.Tkin_low_vel}), we
derive that Eq.~(\ref{eq.Tgas}) is

\begin{align}
T_{\rm gas} \cong 6.4\times 10^{-4}~\textrm{yrs} & \left(\frac{n}{10^{24}\,\textrm{cm}^{-3}}\right)^{-1}
                                                   \left(\frac{m_{\rm core}}{0.34\,M_{\odot}}\right)^{-1} \nonumber \\
                                                 & \left(\frac{C_{\rm s}}{20\,\textrm{kms}^{-1}}\right)^{3},
\label{eq.Tgas2}
\end{align}

\noindent
where we have used the value of $C_{\rm s}$ at $T_{\rm kin} = 5 \times 10^2\,K$ as an illustrative example.
However, $T_{\rm kin}$ is a function of time, and hence $C_{\rm s}$ as well, so that plugging in
Eq.~(\ref{eq.Tkin_low_vel}),

\begin{align}
C_{\rm s}(t) \cong& 5.29 \times 10^{2}\,\textrm{kms}^{-1} \left(\frac{E(0)}{10^{49}\,\textrm{ergs}}\right)^{1/2}\left(\frac{\eta}{1}\right)^{1/2} \nonumber \\
                 &  \exp\left[-\frac{13}{2000}\left(\frac{t}{1\,\textrm{month}}\right)^2\right].
\label{eq.Cstime}
\end{align}

\noindent
In Fig.~(\ref{fig.Cs_low_vel}) we depict its evolution with time. It follows
the trend of Fig.~(\ref{fig.T_low_vel}); i.e.  because of the temperature
quickly drops, so does $C_{\rm s}(t)$ too.  We note that this value is in
agreement with the results of \cite{Vorontsov1989} (Fig. 2) for the Sun at a
radius of about $\sim 1\,R_{\odot}$, with the proviso that the radius is
roughly that of the Sun, i.e. at values of $t \sim 0$, which is our departure
assumption. 

Before we derive the final expression for $T_{\rm gas}(t)$, we note that the
density around the cores is not constant; it will decrease with time, since the
gaseous cloud is expanding at $V_{\rm exp}$. The SPH simulations of \cite{FB05}
show that when the cores form a binary, the gaseous density around them is of
about one order of magnitude lower than the density in the cores. 

To derive the initial value of the density around the cores, i.e. at $t=0$, we
take the Sun as a reference point. Most of its mass is enclosed in the
radiative zone, because the convective zone only represents about
$0.3\,R_{\sun}$ and the density in that region is negligible.  Following the
work of \cite{AbrahamIben1971}, we note that for the mass we have adopted for
the cores, $M_{\rm core} = 0.34\,M_{\odot}$, the corresponding radius is of
$R_{\rm core}\sim 0.2\,R_{\odot}$ and, according to their table 3, the
corresponding density in that region is of $\rho \sim 150\,
\textrm{g\,cm}^{-3}$.  Therefore, we will assume that the density around the
cores (corresponding to that of the radiative zone) should be of $\rho_{\rm
rad} \sim 15 \textrm{g\,cm}^{-3}$ (and hence use the tag ``rad''), which
corresponds to a numerical density of $n_{\rm rad}\sim
10^{25}\,\textrm{cm}^{-3}$. {We therefore only consider a radius of
$0.7\,R_{\odot}$, because we are assuming that all mass is in the radiative
zone}. Taking these considerations into account, {plus assuming that the
convective zone is fully ionised hydrogen, with the of the proton $m_p \sim 1.7\times 10^{-24}$ g,} the time evolution of the
numerical density around the cores follows the expression

\begin{align}
n_{\rm rad}(t) \sim & 10^{25}\,\textrm{cm}^{-3} \left(\frac{M}{1\,M_{\odot}}\right) \nonumber \\
                    & \left[{\bf \frac{7}{10}} + \frac{19}{5}{\bf \times 10^3}\,\left(\frac{V_{\rm exp}}{10^{3}\textrm{km\,s}^{-1}}\right) \left(\frac{t}{1\,\textrm{month}}\right) \right]^{-3}.
\label{eq.nrad}
\end{align}

\begin{figure}
\resizebox{\hsize}{!}
          {\includegraphics[scale=1,clip]{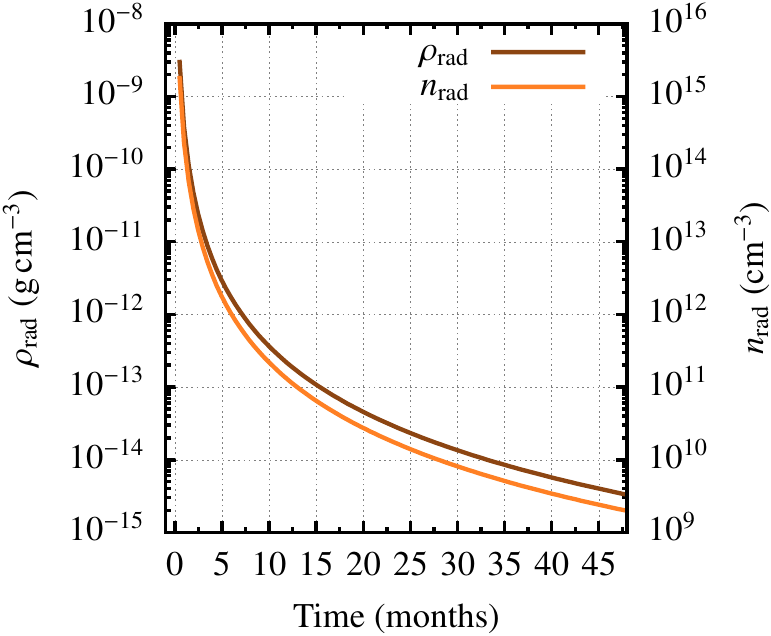}}
\caption
   {
Evolution of the physical (lower curve) and numerical density (upper curve) of
the radiative zone with time, assuming a total mass of $1\,M_{\odot}$ and
$V_{\rm exp}=10^{3}\,\textrm{km\,s}^{-1}$.  The values corresponding to the
physical density, in $\textrm{g\,cm}^{-3}$, are to be read on the left y-axis,
and those to the numerical density on the right axis.
   }
\label{fig.rho_time}
\end{figure}

\begin{figure}
\resizebox{\hsize}{!}
          {\includegraphics[scale=1,clip]{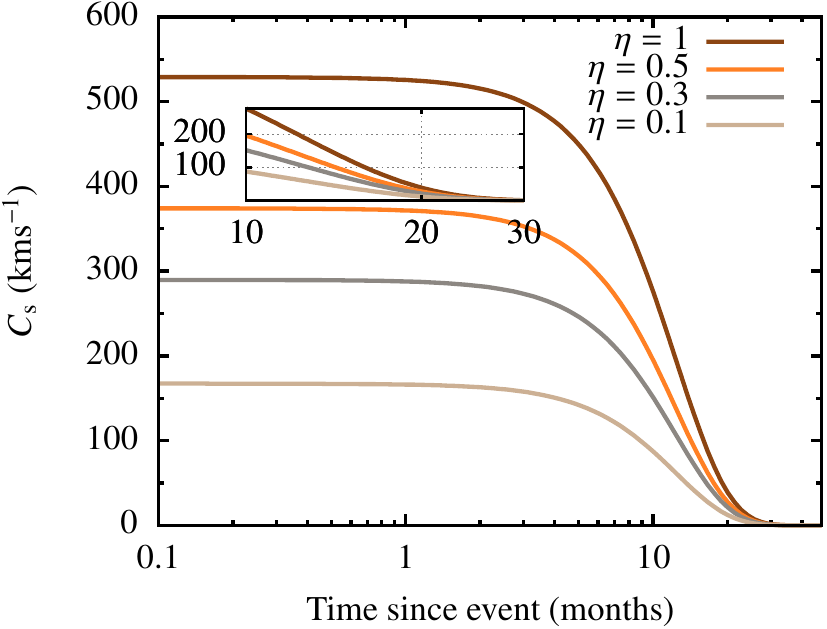}}
\caption
   {
Evolution of the sound speed in the cloud as a function of time. We include
a zoom in between months 10 and 30, when it drops to zero. The uppermost curve
corresponds to $\eta=1$ and the lowermost to $\eta=0.1$. We add an inset to 
show the convergence of the models when the sound speed is zero.
   }
\label{fig.Cs_low_vel}
\end{figure}

\noindent
We can see this evolution, as well as the evolution of the physical density, in
Fig.~(\ref{fig.rho_time}). In a few months the density decreases significantly, so
that assuming a constant value would be wrong.

We are now in the position of deriving the time dependency of $T_{\rm gas}(t)$ by
replacing Eq.~(\ref{eq.Cstime}) and Eq.~(\ref{eq.nrad}) in Eq.~(\ref{eq.Tgas2}),

\begin{align}
T_{\rm gas}(t) & \cong 473.7\,\textrm{yrs} \left( \frac{M}{1\,M_{\odot}} \right)^{-1} \left(\frac{m_{\rm core}}{0.34\,M_{\odot}}\right)^{-1} 
                 \left(\frac{\eta}{1}\right)^{3/2}  \nonumber \\
               & \left( \frac{E(0)}{10^{49}\,\textrm{ergs}} \right)^{3} 
                 \exp\left[-\frac{39}{2000} \left(\frac{t}{1\,\textrm{month}}\right)^2 \right] \nonumber \\
               & \left[1 + \frac{19}{5} \left(\frac{V_{\rm exp}}{10^3\textrm{km\,s}^{-1}}\right) \left(\frac{t}{1\,\textrm{month}}\right) \right]^{3} \nonumber \\ 
\label{eq.Tgastime}
\end{align}

\begin{figure}
\resizebox{\hsize}{!}
          {\includegraphics[scale=1,clip]{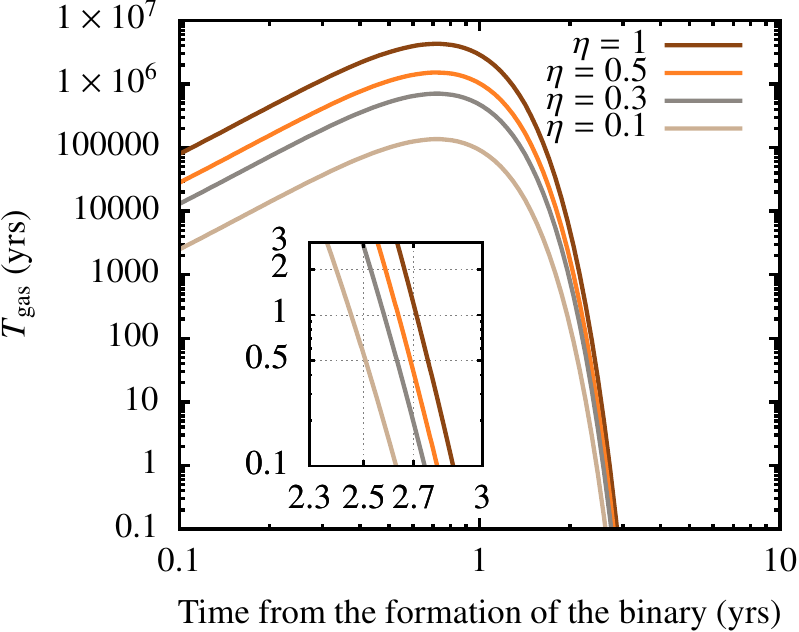}}
\caption
   {
Evolution of $T_{\rm gas}$ as a function of time, see Eq.~(\ref{eq.Tgastime}). The embedded zoom
has linear scale in the x-axis ranging from 2.3 years fter the formation of the binary to 3 years.
We can see that all of the four models follow a similar behaviour, but the difference between them 
is not linearly proportional to $\eta$.
   }
\label{fig.Tgas_low_vel}
\end{figure}

In this result, the power of 2 in the exponential for the time stems from the
cooling of the cloud via the sound speed, Eq.~(\ref{eq.Tgas2}).  This quickly
decays, as we can see in Fig.~(\ref{fig.Cs_low_vel}), and is in power law of 3.
The power of 3 in the last term reflects the fact that in our model we assume
that the cloud has a volume expanding at a constant rate over time. These are
competitive effects responsible for the behaviour of the curve, which we can
see see in Fig.~(\ref{fig.Tgas_low_vel}), where we display
Eq.(\ref{eq.Tgastime}). The function initially increases until about
$1.5~\textrm{yrs}$ from the formation of the binary to then decay. The shape of
the curve allows us to estimate when the binary will merge. Since $T_{\rm
GW}(t) \gg T_{\rm gas}(t)$, we can ignore the effects of gravitational
radiation in the shrinkage of the binary. By evaluating
Fig.~(\ref{fig.Tgas_low_vel}) we can obtain a rough approximation for the
binary to merge via gas friction when the elapsed time (i.e. the abscissa, time
since the formation of the binary) is larger than $T_{\rm gas}$ and $T_{\rm
gas}$ is not increasing in time.  We see in the inset of the figure that this requirement is met
approximately when $t \sim 2.7~\textrm{yrs}$ (for $\eta = 1$), which
corresponds to $T_{\rm gas} = 1~\textrm{yr}$. From that
point, i.e. $(x,\,y)=(2.7,\,1)\,\text{yrs}$, (i) $t > T_{\rm gas}$ and (ii) $T_{\rm gas}$ is only decreasing in time.
Hence, if after $t \sim 2.7~\textrm{yrs}$ the binary has not yet merged, it
should do so in about $T_{\rm mrg} \sim 1$ year, as an upper limit, as for all
other values of $\eta$.

To derive a more accurate value for the merger time $T_{\rm mrg}$, we
need to derive the evolution of the semi-major axis of the binary due to the
drag force of the gas. The differential equation can be derived by taking into
account that, for a circular orbit, $V^2 \propto 1/a$, which means that
$\dot{a}/a = -2 \dot{V}/V$. Since we have identified in
Eq.~(\ref{eq.Tefolding}) $V/\dot{V}=T_{\rm gas}$, we have that
$\dot{a}/a=-2/T_{\rm gas}$, and hence

\begin{equation}
\int_{R_{\odot}/2}^{a_{\rm mrg}} a^{-1}\,da = -2 \int_{0}^{T_{\rm mrg}} T_{\rm gas}^{-1}(t)\,dt,
\label{eq.horribleint}
\end{equation}

\noindent
since we are integrating from the initial semi-major axis
$a_0=R_{\odot}/2$ to $a_{\rm mrg}$. This final value of
the semi-major axis, $a_{\rm mrg}$ is reached when the separation between
the cores reaches $R_{\rm core}$. I.e. $a_{\rm mrg}=  0.2\,R_{\rm core}$

\begin{equation}
\ln\left( \frac{a_{\rm mrg}=0.2\,R_{\odot}}{a_0=R_{\odot}/2} \right) = -2 \int_{0}^{T_{\rm mrg}} T_{\rm gas}^{-1}(t)\,dt,
\label{eq.MonsterWithSemi}
\end{equation}

\noindent
we need to evaluate the right-hand side of the last equation to find the time $t$ for which
$a_{\rm mrg}= R_{\rm core}$, although, a priori, from Fig.~(\ref{fig.Tgas_low_vel}), we already predict
that this time is of about 1 yr. Nonetheless, as we already mentioned before, the solution is relatively
independent of the initial and final semi-major axis.

In the integral, $T_{\rm gas}$ is given by Eq.~(\ref{eq.Tgastime}) and $T_{\rm
mrg,\,m}:= T_{\rm mrg}/(\textrm{month})$, and we introduce $\tau := t/(\textrm{month})$,
so that $d\tau = dt/(\textrm{month})$. Hence,

\begin{equation}
\int_{0}^{T_{\rm mrg}} T_{\rm gas}^{-1}(t)\,dt = \frac{1}{\alpha\,\eta} \int_{0}^{T_{\rm mrg,\,m}} e^{\,c\,\tau^2}
                                                  \left(1 + b\,\tau\right)^{-3}\,d\tau.
\label{eq.RHSMonsterIntegral}
\end{equation}

\noindent
We have introduced $\alpha \equiv 5684.4~\textrm{months}$ (see Eq.~(\ref{eq.Tgastime})), $b\equiv {19}/{5}$, and $c \equiv {39}/{2000}$.

\noindent
The integral given by Eq.~(\ref{eq.RHSMonsterIntegral}) can be solved
analytically, as we show in Appendix 1. The result is

\begin{align}
    I(x)    & = \frac{1}{2b}\left[1-\frac{1}{(1+bx)^2}\right] + \left(\frac{c}{b^3} + \frac{2c^2}{b^5}\right)e^{\,c/b^2}\ln(1+bx) \nonumber \\
            & - \frac{1}{2bx}\frac{e^{\,cx^2}-1}{(1+bx)^2} -\frac{cx}{b^2}\frac{e^{\,cx^2}}{1+bx} \nonumber\\ 
            & + \sum_{n=1}^\infty \frac{n(2n-1)c^{\,n}}{n!} F_n(x).
\label{eq.ResultMonsterIntegral}
\end{align}

\noindent
With

    \begin{equation}
        F_n(x) = 
        \begin{cases} 
            0 & n=1 \\
            \sum_{k=1}^{2n-2} {2n-2 \choose k} \frac{(-1)^{\,k}}{k}\left[(1+bx)^{\,k} - 1\right] & n > 1 ,
        \end{cases}
    \end{equation}

\noindent
where we have defined $x \equiv T_{\rm mrg}$ for legibility.  
The solution agrees with standard numerical Gauss-Kronrod quadrature methods 
to evaluate the value of the integral at different values of $\tau$.

\begin{figure}
\resizebox{\hsize}{!}
          {\includegraphics[scale=1,clip]{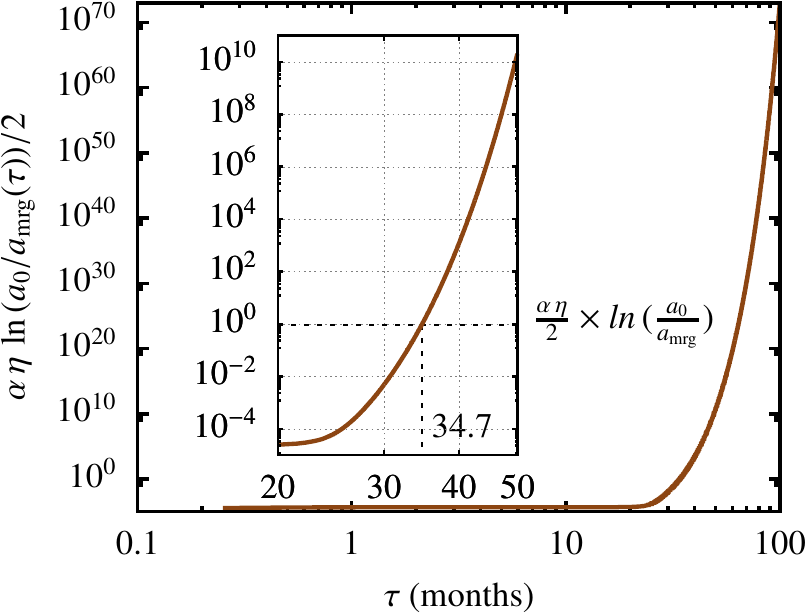}}
\caption
   {
Evolution of $\alpha \, \eta \, \ln\,(a_{\rm mrg}(\tau)/a_0)/2$ as a function of
$\tau$ (i.e. in months).  In the zoom, with a dashed line, we show the values
corresponding to $\alpha \eta \times \ln(a_0/a_{\rm mrg})/2$. This corresponds
to $T_{\rm mrg,\,m}=34.7~\textrm{months}$, i.e. $2.892$ yrs, which is off by a
value of $0.192$ yr from the value predicted by analysing
Fig.~(\ref{fig.Tgas_low_vel}). We can see that $\tau_{\rm mrg}$ varies very
little as a function of $a_0$ and $a_{\rm mrg}$, because a big change in
distance in the y-axis turns into a small change in the x-axis. This means that
the result does not depend (much) on the choice of the initial semi-major axis,
which was chosen here to be $R_{\odot}/2$. We display only the value $\eta=1$
because the other values are virtually identical.
   }
\label{fig.Integral}
\end{figure}

\noindent
Since 

\begin{equation}
\ln\left(\frac{a_{\rm mrg}=0.2}{a_0=0.5}\right) = -0.916291 = - \frac{2}{\alpha\,\eta}\int_{0}^{T_{\rm mrg,\,m}} I(\tau)\,d\tau,
\end{equation}

\noindent
with $I(\tau)$ the integrand of Eq.~(\ref{eq.RHSMonsterIntegral}), and
$\ln(a_{\rm mrg}/a_0)=-\ln(a_0/a_{\rm mrg})$, we plot $log(a_{\rm mrg}/a_0)$
as a function of $\tau$ and look for the value at which

\begin{equation}
0.916291 = \frac{2}{\alpha\,\eta}\int_{0}^{T_{\rm mrg,\,m}} I(\tau)\,d\tau, 
\end{equation}

\noindent
to find $T_{\rm mrg,\,m}$.
In Fig.~(\ref{fig.Integral}) we show the evolution of the right-hand side of
Eq.~(\ref{eq.RHSMonsterIntegral}).  We can see that from the month 20th the
exponential behaviour dominates the evolution of the function and the integral
reaches values as high as $10^{70}$.  Although mathematically correct, this is
a result of the infinite summation of Eq.~(\ref{eq.ResultMonsterIntegral}),
which is physically only realistic up to the moment at which we consider that
the binary forms, i.e. at the value of $\tau$ for which $a_0 = 0.5\,R_{\odot}$,
which is $\tau = 34.3\,\textrm{months}$. From that moment upwards, the result
of the integral is physically meaningless for our purposes. As a consequence of
the exponential behaviour, we note that the result is relatively independent of
the initial semi-major axis. More precisely, this means that, if we e.g.
mutiply by a factor 3 the initial semi-major axis, the result in the x-axis
will be larger by a small factor $\epsilon$,

\begin{equation}
\ln\left(\frac{3 \times 0.5\,R_{\odot}}{a_{\rm mrg}}\right) = 2\times I(T_{\rm mrg} + \epsilon)/(\alpha\,\eta).
\end{equation}

\subsection{Supermassive black hole mimickers}
\label{sec.BBHs}

The drag force acting on to the cores has a direct impact on the observation of
the mass of the source in gravitational waves, as shown by \cite{ChenShen2019},
more precisely on the chirp mass, as introduced by \cite{CutlerFlanagan1994}

\begin{equation}
M_{\rm chirp} := \frac{\left(m_1 m_2\right)^{3/5}}{\left( m_1 + m_2\right)^{1/5}}, 
\label{eq.Mchirp}
\end{equation}

\noindent
which reduces in our case to the following trivial expression, since $m_1 = m_2 = m_{\rm core}$,

\begin{equation}
M_{\rm chirp} = \frac{1}{2^{1/5}}m_{\rm core}= 0.29\,M_{\odot}. 
\label{eq.Mchirp2}
\end{equation}

On the detector, however, the evolution of the gravitational wave frequency is
affected by the timescale in which the gas shrinks the binary in such a way
that the observed chirp mass is not given by Eq.~(\ref{eq.Mchirp}) but for 

\begin{equation}
M_{\rm chirp,\,obs}(t) = \Big[1+ \Lambda(t) \Big]^{3/5}M_{\rm chirp},  
\label{eq.MchirpObs}
\end{equation}

\noindent
with $\Lambda(t):=T_{\rm GW}(t)/T_{\rm gas}(t)$. This can be seen from
Eq.~(3) of \cite{ChenEtAl2020,ChenShen2019}, and is due to the fact that the frequency $f$ and
its time derivative $\dot{f}$ now do not evolve solely because of the
gravitational radiation \citep[and see also][]{CaputoEtAl2020}. In our case, however
$T_{\rm gas}$ is a function of time, given by Eq.~(\ref{eq.Tgastime}), and
$T_{\rm GW}(t)$ is given by Eq.~(\ref{eq.TGW_t}). The full expression for $\Lambda(t)$
is

\begin{align}
\Lambda(t) & \cong 10^6 \left( \frac{M}{1\,M_{\odot}} \right) \left(\frac{m_{\rm core}}{0.34\,M_{\odot}}\right)^{-2} \left(\frac{\eta}{1}\right)^{-3/2}  \nonumber \\
           & \left( \frac{E(0)}{10^{49}\,\textrm{ergs}} \right)^{-3} \exp\left[\frac{39}{2000} \left(\frac{t}{1\,\textrm{month}}\right)^2 \right] \nonumber \\
           & \left[1 + \frac{19}{5} \left(\frac{V_{\rm exp}}{10^3\textrm{km\,s}^{-1}}\right) \left(\frac{t}{1\,\textrm{month}}\right) \right]^{-3} \nonumber \\
           & \Big[ \left( \frac{a_0}{R_{\odot}/2} \right)^4 - 7.04 \times 10^{-10} \left(\frac{m_{\rm core}}{0.34\,M_{\odot}}\right)^{3} \left(\frac{t}{1\,\textrm{month}}\right) \Big]
\label{eq.Lambda}
\end{align}

\begin{figure}
\resizebox{\hsize}{!}
          {\includegraphics[scale=1,clip]{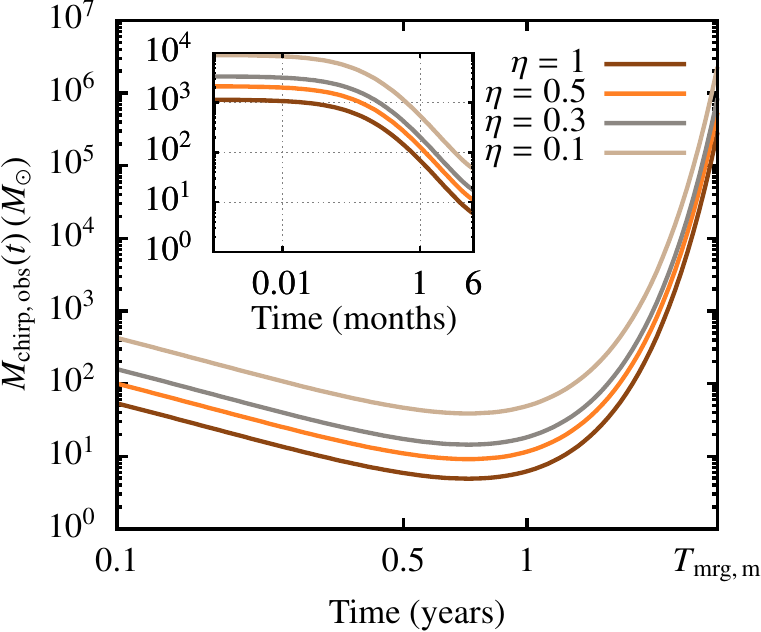}}
\caption
   {
The observed chirp mass for a binary of two cores of masses $0.34\,M_{\odot}$
each in function of time for the usual four values of $\eta$ (with the highest
value in the lowermost curve), as given by Eq.~(\ref{eq.MchirpObs}). We stop
the plot at $T_{\rm mrg,\,m}=2.917\,\textrm{yrs}$, which corresponds to the coalescence time,
as derived previously, and include an embedded zoom corresponding to the range
$10^{-3}$ months ($1.8$ minutes) to $6$ months.
   }
\label{fig.MchirpObs}
\end{figure}

\noindent
From this and (\ref{eq.MchirpObs}), we observe in Fig.~(\ref{fig.MchirpObs})
the increase of the chirp mass as observed by a gravitational-wave detector
such as LIGO/Virgo, the Einstein Telescope or LISA (depending on the observed
chirp mass). 

The fact that the chirp mass reaches a minimum to then again increase again to
higher values is due to the fact that we are taking into account the Bondi
radius, Eq.~(\ref{eq.Rbondi}), since the cores will be surrounded by a region
of overdensity, a ``wake'' around them.  Since the sound speed decreases over
time, as we can see in Fig.~(\ref{fig.Cs_low_vel}), $R_{\rm Bondi}$ increases.
Moreover, the semi-major axis decreases with time, and since we are multiplying
Eq.~(\ref{eq.Tgas}) by $(R_{\rm Bondi}/a)^{3/2}$, this translates into an
increase over time of the chirp mass.

An advantage of gravitational wave data analysis is that, since the time
evolution of the frequency will be very different as compared to the vacuum
case, as we show in this article, so that it will become clear that these
sources correspond to stellar collisions. This will be the first evidence. The
second one is that the merger will be very different to that of a binary of two
black holes because there is no event horizon.  Last, and also due to the fact
that these objects due have a surface, there will be an afterglow.

From the work of \cite{ChenShen2019,ChenEtAl2020}, the observed distance in
gravitational waves due to the same effect has the correction

\begin{equation}
D_{\rm obs}(t) = \big[ 1+ \Lambda(t) \big] D,
\label{eq.Dt}
\end{equation}

\noindent
with $D$ the real distance to the source, as derived in \cite{ChenEtAl2020}.
Assuming a vacuum binary of masses $m_1=m_2=0.34\,M_{\odot}$, semi-major axis
$R_{\odot}/2$ and a particular value of the eccentricity, $e=0$, the horizon
distance can be estimated to be $D \sim 108\,\textrm{Mpc}$ using the
approximant waveform model IMRPhenomPv2 \citep{KhanEtAl2019}, a
phenomenological model for black-hole binaries with precessing spins, at a flow
frequency of $10\,\textrm{Hz}$ with PyCBC \citep{PyCBC2020}, an open-source
software package designed for use in gravitational-wave astronomy and
gravitational-wave data analysis.

\begin{figure}
\resizebox{\hsize}{!}
          {\includegraphics[scale=1,clip]{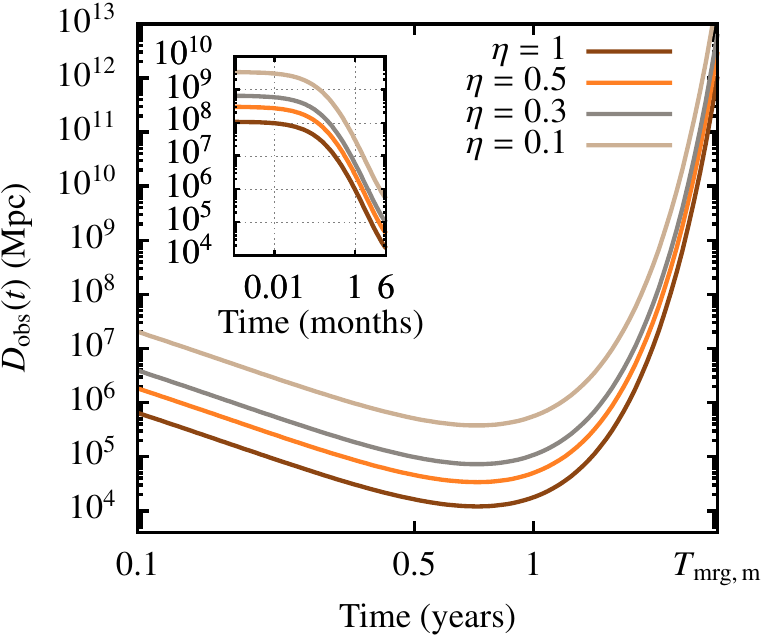}}
\caption
   {
Evolution of the observed distance to the source, $D_{\rm obs}(t)$ in Mpc
as a function of time, following the same nomenclature as in Fig.~(\ref{fig.MchirpObs}).
   }
\label{fig.DistanceObs}
\end{figure}

In Fig.~(\ref{fig.DistanceObs}) we can see the evolution of $D_{\rm obs}$ as
given by the Eq.~(\ref{eq.Dt}) with $D \sim 108\,\textrm{Mpc}$. Again, this is
a consequence of $\dot{f}$ being different from what you expect in vacuum. As
with the chirp mass, the distance will diverge from what is expected in vacuum
very quickly. The big missmatch in the chirp mass and the too large distance to
the source, but in particular the frequency evolution represent the identifiers
of the actual physical origin of the source; namely two colliding stars instead
of a binary of two black holes.

\subsection{Polarizations in vacuum and in gas}

We can relate the polarizations of the waveform amplitude to the chirp mass and
the distance to the source in an approximate, Newtonian way as given by the
Eqs.(4.30, 4.31, 4.32) of \cite{Maggiore2008}, which we reproduce here for
convenience. 

\begin{align}
h_{+}(\tau) & = \frac{1}{r} \left( \frac{G\,M_c}{c^2} \right)^{5/4} \left( \frac{5}{c\,\varsigma} \right)^{1/4}
         \left( \frac{1+\cos^2(\iota)}{2} \right)\, \cos\left[ \Phi \left( \varsigma \right)\right] \nonumber \\
h_{\times}(\tau) & = \frac{1}{r} \left( \frac{G\,M_c}{c^2} \right)^{5/4} \left( \frac{5}{c\,\varsigma} \right)^{1/4}
         \cos(\iota)\,\sin\left[ \Phi \left( \varsigma \right) \right].
\label{eq.Amplitudes}
\end{align}

\noindent
In this equations $\tau$ is our usual definition of $\tau = t/\textrm{month}$,
$M_c$ is the chirp mass, $\varsigma:= (T_{\rm mrg} - \tau)$, $r$ the distance
to the source and $\iota$ is the inclination to the source. Finally,
the phase of the gravitational wave $\Phi(\varsigma)$ is the following function,

\begin{equation}
\Phi(\varsigma) = -2 \left( \frac{5GM_c}{c^3} \right)^{-5/8} \varsigma^{5/8} + \Phi_0, 
\label{eq.Phi0}
\end{equation}

\noindent
with $\Phi_0$ the value of $\Phi(\varsigma=0)$, and $r$ is the
distance to the source, $D$.
The value of the constant of Eq.~(\ref{eq.Phi0}) can be derived by setting
$\tau=T_{\rm mrg}$. With this we find that in vacuum, the value of $\Phi_0$ is

\begin{figure}
\resizebox{\hsize}{!}
          {\includegraphics[scale=1,clip]{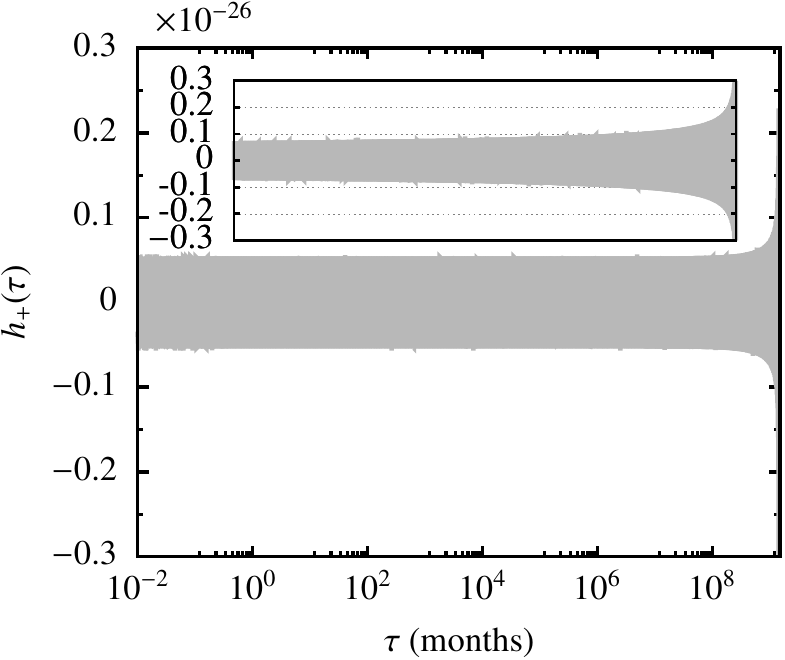}}
\caption
   {
Plus polarization of the gravitational wave produced by the cores, assuming an inclination of $\iota={45}^\circ$.  The
grey, background curve corresponds to the vacuum waveform. We add a zoom figure showing 
the interval $10^9~\textrm{months}$ to $\tau=T_{\rm mrg,\,m}=1.42 \times 10^9~\textrm{months}$.
We note that both y-axis need to be multiplies by $10^{-26}$, as displayed in
the left, uppermost corner. The small spikes in the waveform are an artifact
of the sampling of the plotting program.
   } 
\label{fig.hplus} 
\end{figure}

\begin{equation}
\Phi_0 \cong -1.44\times 10^{8}.
\label{eq.bothPhi0}
\end{equation}

\noindent
Hence, replacing $T_{\rm mrg,\,m}$, we have in vacuum

\begin{equation}
\Phi(\tau) \cong -1.56\times 10^7 \big(35-\tau \big)^{5/8} + \Phi_0,
\label{eq.bothPhi}
\end{equation}

\noindent
with $\Lambda(\tau)$ given by Eq.~(\ref{eq.Lambda}) we employ our usual definition of $\tau \equiv t/(\textrm{1~month})$.

Taking into account that we have chosen $D=108\,\textrm{Mpc}$ and Eq.~(\ref{eq.Mchirp}), and setting
$T_{\rm mrg,\,m}=1.42\times 10^9~\textrm{months}$,
Eqs.~(\ref{eq.Amplitudes}) become

\begin{align}
h_{+}(\tau)      & \cong 1.65\times 10^{-25} \Big[ \big(1.42\times 10^9 - \tau \big) \Big]^{-1/4}                               \nonumber\\
                           & \times \left( \frac{1+\cos^2(\iota)}{2} \right)\, \cos\left[ \Phi \left( \tau \right)\right],      \nonumber\\
h_{\times}(\tau) & \cong 1.65\times 10^{-25} \Big[ \big(1.42\times 10^9 - \tau \big) \Big]^{-1/4}                               \nonumber\\
                           & \times \cos(\iota) \, \sin\left[ \Phi \left( \tau \right)\right],
\label{eq.haches}
\end{align}

\noindent
with $\Phi(\tau)$ given in Eq.~(\ref{eq.bothPhi}) and $\Phi_0$ in Eq.~(\ref{eq.bothPhi0}).

In Fig.~(\ref{fig.hplus}) we display as an example the plus polarization of
the Eqs.~(\ref{eq.haches}) in vacuum. 

In order to derive an expression for the evolution of the polarizations in the
case in which we consider the influence of the gas, what we have to do is to
analyse the evolution of the semi-major axis of the binary under the influence
of the gas, which is given by Eq.~(\ref{eq.horribleint}). In this case,
however, we do not integrate up to the merger, i.e. $a=a_\text{mrg}$,
$t=T_\text{mrg}$, but up to some semi-major axis $\hat{a}$ in $R_{\odot}$ and
some time $\hat{\tau}$ in units of months. Therefore, we have

\begin{equation}
\hat{a} = \left(\frac{R_{\odot}}{2}\right) \exp\left[- \frac{2}{\alpha\,\eta}\int_{0}^{\hat{\tau}} I(\tau)\,d\tau\right],
\end{equation}

\noindent 
with $I(\tau)$ given by Eq.~(\ref{eq.ResultMonsterIntegral}). Therefore, for
each value of $\hat{\tau}$, we can derive $\hat{a}$ and, with it and Eq.~(\ref{eq.TGW}), we can obtain
what is the time $\varsigma$ that we need to use in the set of
Eqs.~(\ref{eq.Amplitudes}), 

\begin{equation}
\varsigma = \frac{5}{128}\frac{c^5\,\hat{a}^4}{G^3m_{\rm core}^3}F(e)^{-1}.
\end{equation}

\begin{figure}
\resizebox{\hsize}{!}
          {\includegraphics[scale=1,clip]{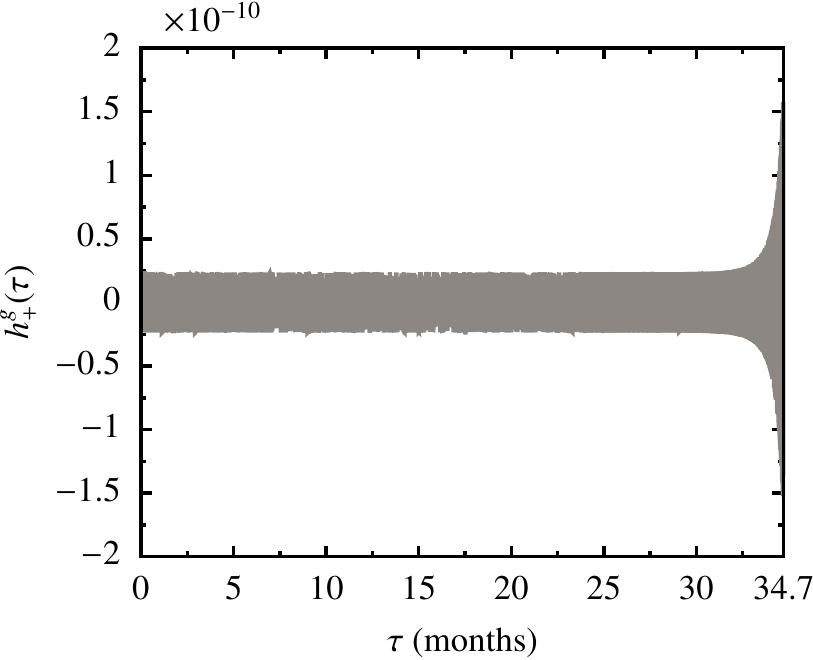}} 
\caption
   {
Plus polarization for the binary embedded in gas. We note that, contrary to Fig.~(\ref{fig.hplus}),
the X-axis is in linear scale.
   }
\label{fig.h_plus_polarization_in_gas}
\end{figure}

\noindent 
I.e. we are deriving the characteristic timescale for an evolution due to
gravitational radiation in a case in which the semi-major axis is shrinking at a rate given by the friction
with the gas. In Fig.~(\ref{fig.h_plus_polarization_in_gas}) we show the result, which is the
counterpart of Fig.~(\ref{fig.hplus}). We can see that the time has significantly reduced, as well as the
width of the oscillations.

\subsection{Characteristic strain in vacuum and in gas}

So as to compare the vacuum case with the one in which the cores are embedded in the gaseous cloud, we
will derive the characteristic strain as approximated by Eq.~(10.146) of \cite{Maggiore022018},

\begin{equation}
h_c(f) = \frac{1}{D} \sqrt{\frac{2}{\pi^2}\frac{G}{c^3}\frac{dE}{df}}, 
\end{equation}

\noindent
with $dE/df$ the energy spectrum in the inspiraling phase in the Newtonian approximation, see e.g. Eq.~(4.41)
of \cite{Maggiore2008},

\begin{equation}
\frac{dE}{df} = \frac{\pi^{2/3}}{3G}\frac{\left( G\,M_c \right)^{5/3}}{1+z}\,f^{-1/3}. 
\end{equation}

\noindent
We have then

\begin{equation}
h_c(f) = \frac{\sqrt{2/3}}{\pi^{2/3}\,c^{3/2}}\frac{\left( G\,M_c \right)^{5/6}}{D \sqrt{1+z}}\,f^{-1/6}. 
\label{eq.approx_hc}
\end{equation}

Now, the characteristic strain can be expressed in terms of the amplitude in frequency $A(f)$, the frequency itself $f$
and its time derivative $\dot{f}$ as follows \citep[see Eq.~16.21 of][]{Maggiore022018},

\begin{equation}
h_c(f) = A(f) \frac{f}{\dot{f}^{1/2}}, 
\end{equation}

\noindent
the only thing we need to do is to take the ratio of the characteristic strain affected by the gas, $h_c^g(f)$, and that in
vacuum, $h_c^v(f)$. Since the amplitudes and the frequencies are the same, we are left with

\begin{align}
h_c^{\,g}[f(t)] & = h_c[f(t)] \big[ \Lambda(t) \big]^{-1/2} \nonumber \\
         & = \frac{\sqrt{2/3}}{\pi^{2/3}\,c^{3/2}}\frac{\left( G\,M_c \right)^{5/6}}{D \sqrt{1+z}} \big[ \Lambda(t) \big]^{-1/2}\,f(t)^{-1/6},
\label{eq.hcg}
\end{align}

\noindent
with $\Lambda(t)$ given, as usual, by Eq.~(\ref{eq.Lambda}), and $f(t)$ the
associated frequency of the source, which is a function of time as well and
accordingly needs to be evaluated at the same time as $\Lambda(t)$. This expression,
Eq.~(\ref{eq.hcg}) gives us the instantaneous value of $h_c^g[f(t)]$ at a given
moment $t$.

To derive $f(t)$ we need to take into account two things. First, the driving
mechanism in the evolution of the binary, as we have seen previously, is the
friction of the binary with the gas, rather than the loss of energy via
gravitational radiation, so that in Eq.~(\ref{eq.hcg}) time derivatives must be done in the context of
gas friction. Second, in our derivation of Eq.~(\ref{eq.horribleint}) we used
the fact that  $\dot{a}/a = -2 \dot{V}/V$ and $\dot{a}/a=-2/T_{\rm gas}$.
Hence, since the frequency associated to any GW source can be expressed in the
Newtonian limit as

\begin{equation}
f = \frac{1}{\pi} \sqrt{ \frac{G\,M_{\rm tot}}{a^3}}, 
\end{equation}

\noindent
where $M_{\rm tot} = 2\,m_{\rm core}$ and we are omitting the time dependence. 
The time derivative can be calculated to be

\begin{equation}
\dot{f}_{\rm gas} = - \frac{3}{2\pi} \sqrt{\frac{G\,M_{\rm tot}}{a^3}} \frac{\dot{a}_{\rm gas}}{a}. 
\end{equation}

\noindent
To derive this expression we have used the chain rule and the fact that what
induces a change in the semi-major axis is the gas, so that $da/dt\equiv
\dot{a}_{\rm gas}$. I.e. the physical process that induces time changes is the
friction with the gas, so that we need to derivate respect to the time the
quantities related to it. Hence,

\begin{equation}
\dot{f}_{\rm gas} = 3\,\frac{f}{T_{\rm gas}}.
\end{equation}

\noindent
I.e. we need to solve

\begin{equation}
\int f^{-1}\,df = \ln[f(t)] = 3 \int T_{\rm gas}^{-1}(t')\,dt'. 
\end{equation}

As before, in Eq.~(\ref{eq.RHSMonsterIntegral}), $T_{\rm gas}$ is given by
Eq.~(\ref{eq.Tgastime}), $\tau := t/(\textrm{month})$, so that $d\tau =
dt/(\textrm{month})$ and so

\begin{equation}
\ln[f/f_0] = 3 \int T_{\rm gas}^{-1}(t)'\,dt' = \frac{3}{\alpha\,\eta} \int e^{\,c\,\tau^2}
                                                  \left(1 + b\,\tau\right)^{-3}\,d\tau.
\label{eq.RHSMonsterIntegral2}
\end{equation}

\noindent
With the same values of $\alpha$, $b$ and $c$. In this equation, $f_0$ is the initial
frequency from which we start to measure the source, and the ratio is $f/f_0$ because
it is a positive integral.

The result of the previous integral is $3\times I(\tau)$, with $I(\tau)$ given
by Eq.~(\ref{eq.ResultMonsterIntegral}) and $\tau$ is a moment of time in
months before the merger, i.e. at merger $\tau=T_{\rm mrg,\,m}$. Therefore, we
have that the integrated characteristic strain from the moment of formation of
the binary at a frequency $f_0$ and an ulterior given time in months $\tau$ is

\begin{align}
h_c^{\,g} = &   \frac{\sqrt{6}}{\pi^{2/3}\,c^{3/2}}\frac{\left( G\,M_c \right)^{5/6}}{D \sqrt{1+z}} 
                     \big[ \Lambda(\tau) \big]^{-1/2}\nonumber \\
                 &   \times f_0^{\,-1/6}\exp{\left[-\frac{I(\tau)}{2\,\alpha\,\eta}\right]},
\label{eq.hcg_result}
\end{align}

\noindent 
with $I(\tau)$ given by Eq.~(\ref{eq.ResultMonsterIntegral}) and $\Lambda(\tau)$ by
Eq.~(\ref{eq.Lambda}), as usual. As for $f_0$, we can derive it from the initial
semi-major axis of the binary and the masses of the cores. Since the gravitational-wave
frequency is twice the orbital frequency, we have that $f_0 \cong 4.7\times 10^{-4}\,\textrm{Hz}$.

\begin{figure}
\resizebox{\hsize}{!}
          {\includegraphics[scale=1,clip]{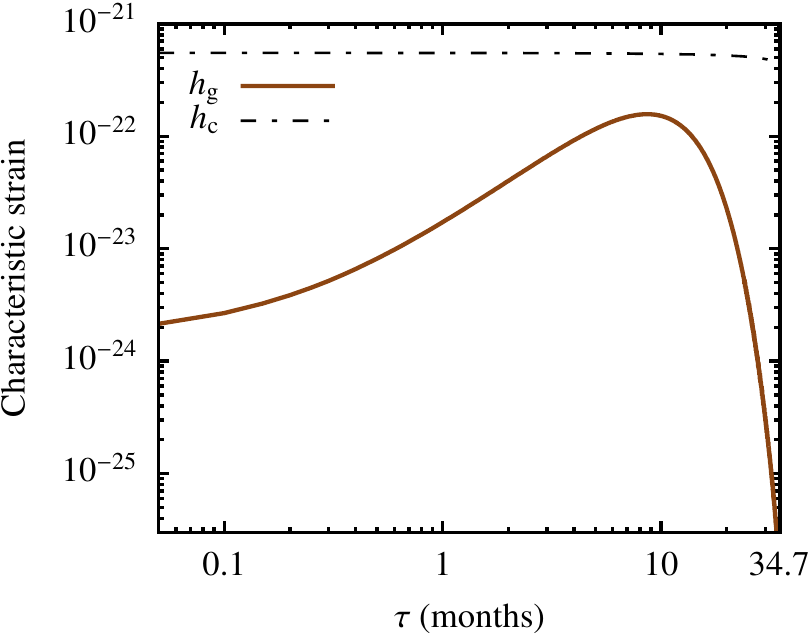}}
\caption
   {
Evolution of the characteristic strain in vacuum, $h_{\rm c}$, and after the
collision, i.e. in a gaseous environment, $h_{\rm c}^{\rm g}$. The two curves
correspond to the latter case for two different initial frequencies $f_0$,
while the former is depicted with a dashed, straight line which does not depend
on the initial frequency.
   }
\label{fig.Integral_hgas_Time}
\end{figure}

\begin{figure}
\resizebox{\hsize}{!}
          {\includegraphics[scale=1,clip]{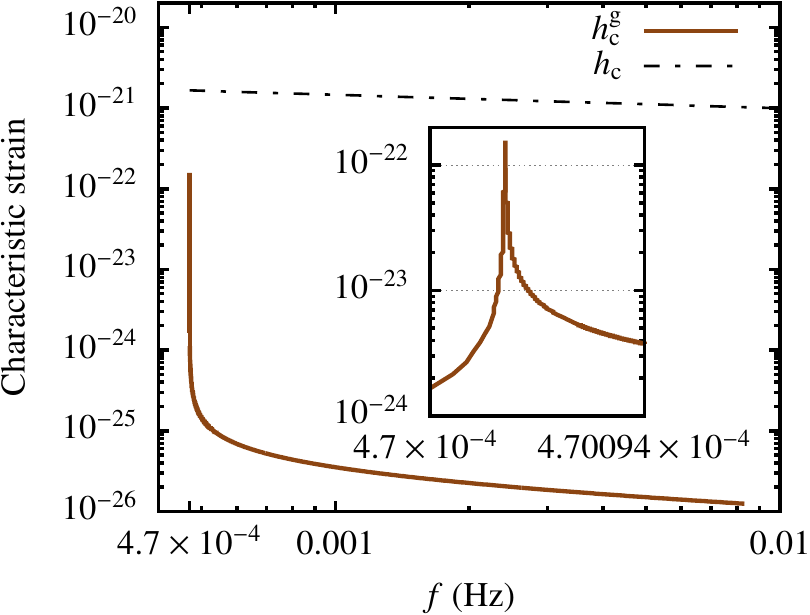}}
\caption
   {
Same as Fig.~(\ref{fig.Integral_hgas_Time}) but in frequency domain. We include
a zoom in for the characteristic strain of the two cores in the gaseous 
environment between the range of frequencies $f \in [4.7,\,4.70094]\,\times10^{-4}$
Hz.
   }
\label{fig.Integral_hgas_Frequency}
\end{figure}

We can express the gravitational wave frequency in vacuum of a binary with the
same chirp mass as a function of time in months by assuming a Keplerian,
circular orbit which shrinks over time via gravitational loss. In the
quadrupole approximation and for circular orbits the source orbital frequency
$\nu_s$ (given via Kepler's laws) and the gravitational-wave frequency
$\nu_{\rm GW}$ are related via $\nu_{\rm GW}=2\,\nu_s$. We hence can find from
the orbital energy and the fact that $2\pi\,f_{\rm GW} = \nu_{\rm GW}$ that

\begin{equation}
f(t) = \frac{1}{\pi} \left(\frac{G\,M_c}{c^3}\right)^{-5/8} \left(\frac{5}{256}\frac{1}{\varsigma} \right)^{3/8}, 
\end{equation}

\noindent
with $\varsigma:= (T_{\rm mrg} - \tau)$. See e.g. Sec. 4.1 of
\cite{Maggiore2008} for an explicit derivation of this result.  We now
substitute this result in Eq.~(\ref{eq.approx_hc}) and obtain that

\begin{equation}
h_c    = \frac{\sqrt{{2}/{3}}\left({5}/{256}\right)^{-3/48}}{\pi^{1/2}c^{29/16}} \frac{\left(GM_c\right)^{15/16}}{D\sqrt{1+z}}
         \left( \frac{1}{\varsigma} \right)^{-3/48}. 
\end{equation}

\noindent 
If we adopt $D=108\,\textrm{Mpc}$, $z=0$, $M_c=0.29\,M_{\odot}$ and introduce
$T^{m}_{\rm mrg}:=T_{\rm mrg}/\textrm{month}$ and $\tau$, $h_c(t)\cong
4.42\times 10^{-22} \left(T^{m}_{\rm mrg} - \tau \right)^{3/48}$.

In Fig.~(\ref{fig.Integral_hgas_Time}) we can see the differences in the time
evolution of the different characteristic strains. The vacuum case corresponds
to a straight line as one would expect, since we are working in the inspiral
approximation of the quadrupole for circular orbits (as is the case). The cores
embedded in the stellar debris, however, evolve in a very different fashion
even for the very short timescales related to the problem (of months). At the
initial time we see that the strains differ in about three orders of magnitude,
as Eq.~(\ref{eq.hcg}) suggests for the default values given in
Eq.~(\ref{eq.Lambda}).  In a similar way, in
Fig.~(\ref{fig.Integral_hgas_Frequency}) we depict the frequency evolution of
the two strains. Again, in the very short interval of frequencies, the strain
in vacuum does not change significantly, while the one corresponding to the
gaseous case has a completely different behaviour.

It is interesting to see the propagation in
Figs.~(\ref{fig.Integral_hgas_Time},\,\ref{fig.Integral_hgas_Frequency}) of
the combined effect of the evolution of the speed of sound and the fact that
the cloud is expanding over time, as we mentioned in the paragraph following
Eq.~(\ref{eq.Tgastime}).

We present a sketch of a possible strategy to calculate the mismatch between the vacuum- and the
gas sources in Appendix 2.

\section{Red giants}
\label{sec.RedGiants}

So far we have focused on main sequence stars and looked at the high-energy
emission and the potential production of an associated gravitational wave
source.  A particularly interesting kind of star for which the previous
analysis can be applied, however, are red giants. This is so because their
masses are also of the order of $1\,M_{\odot}$, even if they have much larger
radii. When the red giants collide, they will also be a powerful source of
high-energy. The presence of a degenerate core at the centre of the star, makes
it more appealing from the point of view of gravitational radiation, and when
the two degenerate cores collide, this will again turn into a strong source of
electromagnetic radiation, which has been envisaged as a possible explanation
for Type Ia supernova, such as SN 2006gy (\citealt{SmithEtAl2007} and see
\citealt{Gal-Yam2012}). We hence would have a precursor electromagnetic signal
announcing the gravitational-wave event followed by another posterior, very
violent electromagnetic emission.

Contrary to supernovae, red giants come with a different spectrum of masses and
radii, and the total mass of the resulting degenerate object would not be
constrained by the Chandrasekhar limit. As a consequence, one cannot use them
as standard candles. If what is interpreted as Type Ia supernova is mostly the
outcome of two colliding red giants, this would have important implications, as
we will see.

\subsection{Event rate of collisions between red giants}

The process of giganterythrotropism, as coined by Peter Eggleton, means that
the kind of main sequence stars we have been dealing with in this article will
tend to get large and red as they evolve. The main sequence stars we are
considering here, of light mass, spend a percentage of their lives in the form of
a red giant. 

To derive the amount of time spent in the different phases, we refer to the work
of \cite{VassiliadisWood1993}, in which they estimate that the amount of time
spent in the first giant branch (FGB) is of $\sim 3.62 \times 10^9\textrm{yrs}$,
i.e. $24\%$ of the total life of their one-solar mass star of metallicity $Z=0.016$
in their table 1.

Later, the star will reach the asymptotic giant branch (AGB), and during this
stage the star's radius can reach as much as $\sim 215\,R_{\odot}$
\citep{VassiliadisWood1993}. The amount of time spent in the AGB, for a
solar-like star is $\tau_{\rm AGB} \sim 2.5\times 10^7\textrm{yrs}$ according
to \cite{VassiliadisWood1993}, which in their model represents $0.17\%$ of the
total life of the star 

In order to be conservative on the derivation of the rates, this means that the 
event rates, as derived in the Eq.~(\ref{eq.GammaColl}) must be multiplied by a factor of $10^{-2}$ to take this into account, since we
need two stars. We pick up a $1\,M_{\odot}$ main sequence star, which in its
red-giant phase and a few numerical timesteps before the triple-alpha process
has a mass of $M_{\rm RG}=0.953\,M_{\odot}$ and an associated radius of $R_{\rm
RG} = 25\,R_{\odot}$\footnote{P. Eggleton, private communication.}. We choose these
as representative values of our default red giant in the red-giant branch, where
the $1\,M_{\odot}$ main-sequence star will stably fuse hydrogen in a shell for
about 10\% of its entire life.

\begin{figure}
\resizebox{\hsize}{!}
          {\includegraphics[scale=1,clip]{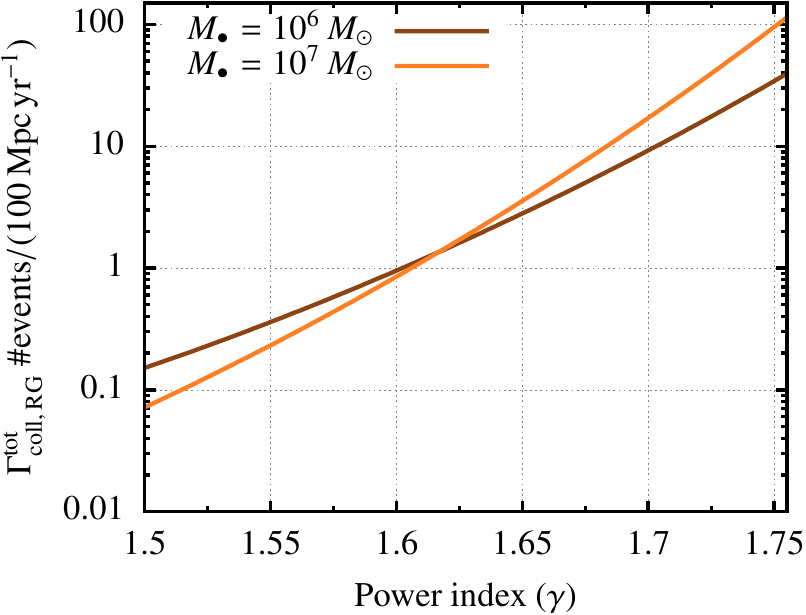}}
\caption
   {
Same as in Fig.~(\ref{fig.G}) but for red giants of masses $M_{\rm RG}\sim
0.953\,M_{\odot}$ and radii $R_{\rm RG} = 25\,R_{\odot}$, taking into account
that we have adopted the occupation fraction in phase space for the two giants
to be of $f_{\rm RG} = 10^{-2}$. This stems from the fact that we are only
considering giants in the asymptotic giant branch, where they spend about
$0.17\%$ of their life. We do not consider the first giant branch, where they
spend up to $24\%$ of their lifetime in order to derive lower-limit quantities.
   }
\label{fig.G_RedGiants}
\end{figure}

This has a significant impact on the geometrical cross-section. As we can see
in Eq.~(\ref{eq.GammaColl}), this leads to an enhancement factor of $\sim 600$
without taking into account the first term enclosed in the square brackets
which is, however, basically negligible as compared to the second term in the
square brackets as we discussed in that section.  We do lose a small factor in
terms of mass but, in total, the rates are significantly enhanced. In
Fig.~(\ref{fig.G_RedGiants}) we show the equivalent of Fig.~(\ref{fig.G}) but
for the collision of two red giants with the above-mentioned properties.

It is interesting to note that, even though red giants are fully convective,
the treatment we have derived in the previous sections regarding the
electromagnetic signature still applies to them because the thermodynamics of
the gas will not be different from that of the main stars after the collision
as soon as the red giants collide, i.e.  as soon as they are not in
thermodynamical equilibrium.

The compact binary forming in the collision will be surrounded by gas in any
case. Even if the impact parameter  was exactly zero, there will be gas because
the merging time of the compact cores due to the gas drag is much shorter than
the timescale in which the gas dissipates.

However, it is interesting to address the formation of the binary which forms
because, as we will see in the next section, it is a particular one.

\subsection{Structure of the red giants}

The nature of the red giant plays a role however in the evolution of the cores
in the resulting gaseous cloud that emerges as a result of the collision.  This
is important for us because we want to understand what source of gravitational
radiation the collision will produce after the collision between the two red
giants has taken place, with the proviso that the relative speed does not
exceed $V_{\rm rel} \leq 2500\,\textrm{km\,s}^{-1}$, as noted in
Sec.~(\ref{sec.GW}).  For this we need to know (i) the average density of the
medium in which the cores will be embedded after the collision, (ii) the
density of the H-fusing shell around the cores (see ahead in the text), (iii)
the masses of the cores and (iv) an estimate of the initial semi-major axes.
We will set the mass of the red giant to $M_{\rm RG}=0.953\,M_{\odot}$, which
comes from the numerical simulation of a $1\,M_{\odot}$ main-sequence star
before reaching the helium flash, where it spends most of its life, and $R_{\rm
RG} = 25\,R_{\odot}$ (see previous footnote).

In general, a red giant can be envisaged as a self-gravitating, degenerate core
embedded in an extended envelope.  This is a consequence of the decrease of
hydrogen in the inner regions of the star, so that if a main sequence star
consumes it, the convective core gives place to an isothermal one. The
helium-filled core collapses after reaching a certain maximum
\citep{SchoenbergChandrasekhar1942} which releases energy that expands the
outer layers of the star.  However, as proven analytically in the work of
\cite{EggletonCannon1991}, it is not possible to simply add an envelope fusing
H at its base on to a wholly degenerate white dwarf core.  One needs to have an
(almost) isothermal non-degenerate shell below the fusing shell and above the
degenerate core\footnote{We note here that, although the article has in its
title ``A conjecture'' it is in reality a proper theorem, as demonstrated in
the appendix of the work.} The work of \cite{EggletonCannon1991} proves that the
fact that a red giant's envelope expands, after shell burning is established,
is not related to the nature of the envelope, and even of the burning shell,
but to the ostensibly small isothermal non-degenerate shell between the
degenerate core and the fusing shell.

Since we are interested in the collision and characteristics of the cores when
they form a binary and eventually merge via emission of gravitational waves,
we need to evaluate the properties of this shell.

We hence consider a red giant as a star with a He-degenerate core, a H-fusing
shell around it as the only energy source, transiting through a thin radiative
zone to the fully convective, extended envelope.  Assuming an ideal gas in the
H-fusing shell, the equation of state is 

\begin{equation}
P = P_{\rm gas} + P_{\rm rad} = \frac{\Re}{\mu}\rho_{\rm sh} T_{\rm sh} + \frac{a}{3}T_{\rm sh}^4, 
\end{equation}

\noindent
with $\rho_{\rm sh}$ the density in the shell and $T_{\rm sh}$ its
temperature, the radiation density constant $a=7.56\times
10^{-15}~\textrm{erg}/(\textrm{cm}^3\,\textrm{K}^4)$ and the universal gas
constant $\Re=8.31\times 10^7~\textrm{erg}/\textrm{(K\,g)}$.  Usually one
introduces $\beta:=P_{\rm gas}/P$, the constant ratio of gas pressure $P_{\rm
gas}$ to total pressure $P$, so that $1-\beta = P_{\rm rad}/P$. We can now
solve for $\rho_{\rm sh}$,

\begin{equation}
\rho_{\rm sh} = \frac{a \mu}{3\Re} T_{\rm sh}^3 \frac{\beta}{1-\beta}. 
\label{eq.rhoHf}
\end{equation}

\noindent
We hence have to derive an estimate for the temperature to obtain the density.
For this we follow the derivation of the gradient of temperature with radius as
in e.g. \cite[][, their section 5.1.2]{KippenhahnWeigertStellarStructure}. We
consider the flux of radiative energy $F$ in spherical symmetry in the shell
and make an analogy with heat conduction, so that (see Eq.~5.11 of
\citealt{KippenhahnWeigertStellarStructure})

\begin{equation}
\frac{dT_{\rm sh}}{dr} = -\frac{\kappa \rho_{\rm sh}\,L}{4 \pi a c r^2 T_{\rm sh}^3},
\end{equation}

\noindent
where we have absorbed the flux into the luminosity, $L=4\pi r^2 F$ and
$\kappa$ is considered again to be constant, but in this case
$\kappa=0.2(1+X)$ for electron scattering. Since $P_{\rm rad}=aT^4_{\rm sh}/3$,

\begin{equation}
\frac{dP_{\rm rad}}{dr} = -\frac{1}{4\pi c} \frac{\kappa \rho_{\rm sh}\,L}{r^2}. 
\label{eq.dPrad}
\end{equation}

The equation of hydrostatic equilibrium is

\begin{equation}
\frac{dP}{dr} = - \frac{Gm(r)}{r^2}\rho_{\rm sh}, 
\end{equation}

\noindent
and we approximate $m(r)\sim M_{\rm core}$. Hence

\begin{equation}
dP = \mathcal{C} \, dP_{\rm rad}, 
\label{eq.dPdPrad}
\end{equation}

\noindent
with $\mathcal{C}$ constant.  We integrate this last equation and take into
account that we can neglect the integration constant deep inside the radiative
zone, as noted by Paczy{\'n}ski\footnote{This approximation is explained in the
unpublished work of Bohdan Paczy{\'n}ski. See the small note in Appendix 2.}, so that
$P/P_{\rm rad}=\mathcal{C}\equiv 4 \pi c GM_{\rm
core}/(\kappa\,L)=1/(1-\beta)=L_{\rm Edd}/L$. where $L_{\rm Edd}\equiv4 \pi c G
M_{\rm core}/\kappa$ is the Eddington luminosity, the maximum luminosity that
the source can achieve in hydrodynamical equilibrium
\citep{RybickiLightman1979}. If this luminosity was to be exceeded, then
radiation pressure would drive the outflow. From Eq.~(\ref{eq.dPrad}) and
Eq.~(\ref{eq.dPdPrad}), we obtain

\begin{equation}
\frac{dT_{\rm sh}}{dr} = - \frac{\kappa L\mu}{16 \pi c \Re}\left( \frac{\beta}{1-\beta} \right)\frac{1}{r^2}. 
\end{equation}

\noindent
Since we have the expression for $(1-\beta)$,

\begin{equation}
\frac{dT_{\rm sh}}{dr} = - \frac{\mu \beta G M_{\rm core}}{4 \Re r^2}. 
\label{eq.dTdr}
\end{equation}

\noindent
We integrate this equation and neglect the constant of integration for the same reasons as we
did previously to find

\begin{equation}
T_{\rm sh} = \frac{\mu \beta G M_{\rm core}}{4 \Re R_{\rm core}}. 
\end{equation}

Finally, we obtain that the density can be expressed as

\begin{equation}
\rho_{\rm sh} \cong 6 \times 10^{-3}~\textrm{g\,cm}^{-3}\frac{\left(\beta\,\mu\right)^4}{1-\beta}
                       \left(\frac{M_{\rm core}}{M_{\odot}}\right)^3\left(\frac{R_{\rm core}}{R_{\odot}}\right)^{-3},
\label{eq.rhoHf1}
\end{equation}

\noindent
and we note that we have used the radius of the core $R_{\rm core}$ to normalize
the last term, although we are referring to the density in the shell. However,
the thickness of the H-fusing shell, $R_{\rm sh}$ extends only a bit farther than
the radius of a white dwarf from the center (we use here the letter $R$ for the thickness instead of 
$T$ because it could be misinterpreted with temperature). This is so because the shell is not (yet) degenerate,
but we will also derive the value of $R_{\rm sh}$ later.

We can rewrite Eq.~(\ref{eq.rhoHf1}) because $\beta$ is constant in the shell,
as we have seen previously, so that it can be approximated with a
polytrope of index $n=3$, thanks to Eddington's quartic equation (Eq. 22 of
\citealt{Eddington1924}), which can be written as

\begin{equation}
\frac{1-\beta}{\mu^4 \beta^4} = \frac{a}{3\Re^4}\frac{\left(\pi G \right)^3c_1^2}{z_3^6}M^2,
\end{equation}

\noindent
with $M$ the total mass of the stellar object, in our case $M=M_{\rm core}$,
and $z:=A\,r$ ($A$ a constant) the usual dimensionless variable for the radius
introduced to derive the Lane-Emden equation. The value of $z_3$ (polytrope of
index $n=3$) has to be derived numerically, and is $z_3\sim6.897$
\citep{ChandrasekharStellarStructure}. Finally, the constant $c_1$ can be
obtained thanks to the relation between central density and average density
which one obtains from the Lane-Emden equation, e.g. Eq.~(19.20) of
\cite{KippenhahnWeigertStellarStructure}, $c_1=12.93$. Therefore,

\begin{equation}
\frac{1-\beta}{\mu^4 \beta^4} \cong 3 \times 10^{-3} \left( \frac{M_{\rm core}}{M_{\odot}} \right)^2,
\label{eq.quartic}
\end{equation}

\noindent
and so, Eq.~(\ref{eq.rhoHf1}) becomes

\begin{equation}
\rho_{\rm sh} \cong 2 \times 10^4~\textrm{g\,cm}^{-3}\left( \frac{M_{\rm core}}{0.3\,M_{\odot}} \right) 
                          \left( \frac{R_{\rm core}}{3\times 10^{-2}\,R_{\odot}} \right)^{-3}. 
\label{eq.rhoHf}
\end{equation}

\noindent
This result is not unexpected, since the density of a white dwarf ranges
between $10^4$ and $10^7~\textrm{g\,cm}^{-3}$, and the H-fusing shell supports
pressures very close to that of the degenerate core itself.

We can obtain the mass enclosed between the radius of the white dwarf ($R_{\rm WD}$) and that of the core ($R_{\rm core}$)
by integrating Eq.~(\ref{eq.rhoHf1}),

\begin{equation}
M_{\rm sh} \cong 2\times 10^{-3}\,M_{\odot} \left(\frac{\beta^4}{1-\beta}\right)
                                       \left(\frac{M_{\rm core}}{M_{\odot}}\right)^3\ln\left(\frac{R_{\rm core}}{R_{\rm WD}}\right). 
\label{eq.M12}
\end{equation}

\noindent
From Eq.~(\ref{eq.quartic}) and $\mu=0.5$ for pure hydrogen, we have that

\begin{equation}
\frac{1-\beta}{\beta^4} \sim 1.7\times 10^{-5}\left( \frac{M_{\rm core}}{0.3 M_{\odot}} \right)^2, 
\end{equation}

\noindent
so that Eq.~(\ref{eq.M12}) can be rewritten as

\begin{equation}
M_{\rm sh} \cong 3.2 \,M_{\odot}\left(\frac{M_{\rm core}}{0.3 M_{\odot}}\right)^3\ln\left(\frac{R_{\rm core}}{R_{\rm WD}}\right).  
\end{equation}

\noindent
The natural logarithm between the two radii and the total mass means that
$R_{\rm sh}$ is a minor amount that extends beyond the radius of the
degenerate core, approached by a white dwarf in our work.

\begin{figure}
\resizebox{\hsize}{!}
          {\includegraphics[scale=1,clip]{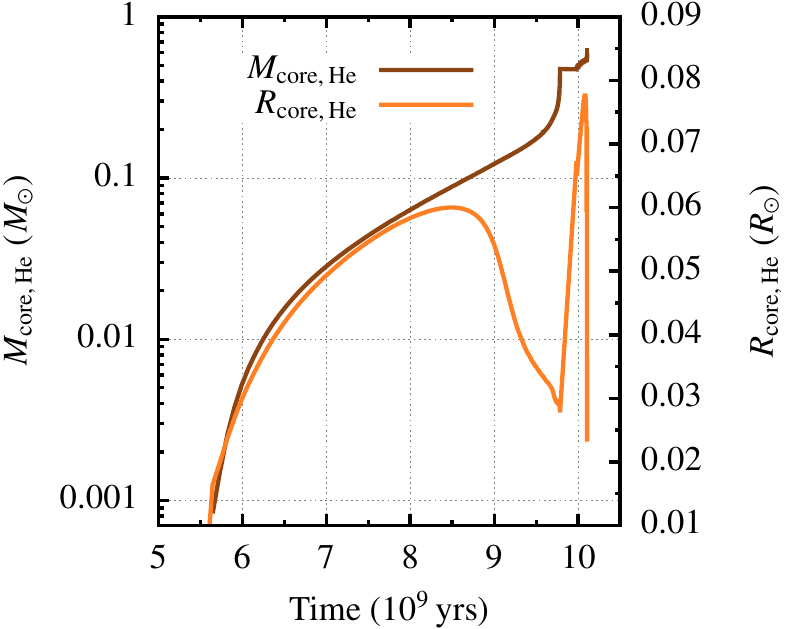}}
\caption
   {
Evolution of the mass and radius of the He core of a red giant which initially
had a $1\,M_{odot}$. The left Y-axis shows the mass of the core in $M_{\odot}$
and the right one the radius in $R_{\odot}$. We can see that, in its evolution,
the mass of the core can span three orders of magnitude.
   }
\label{fig.MassRadiusHeCore}
\end{figure}

\begin{figure}
\resizebox{\hsize}{!}
          {\includegraphics[scale=1,clip]{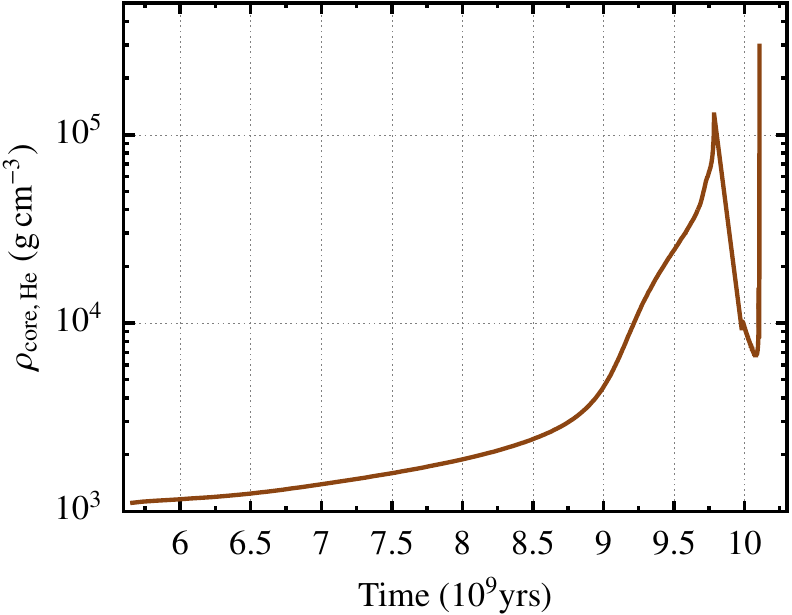}}
\caption
   {
Same as Fig.~(\ref{fig.MassRadiusHeCore}) but for the density of 
the core. In its evolution, the different densities can span over
two orders of magnitude.
   }
\label{fig.DensityHeCore}
\end{figure}

Therefore, and to first order, we can consider that the properties of the two
degenerate objects taking place in the collisions are those of the He core.
The numerical code of \cite{Eggleton1971} allows us to obtain the properties of
our fiducial model, which is a $1\,M_{\odot}$ red giant. In
Fig.~(\ref{fig.MassRadiusHeCore}) we show the evolution of the mass and radius
of the He core, while in Fig.~(\ref{fig.DensityHeCore}) we depict the evolution
of its density.

We can see that in particular the mass (and hence the density) significantly
vary in the lifetime of the star, while the radius can change by almost one
order of magnitude. This means that, when the two degenerate cores form a
binary and merge, the properties of the electromagnetic radiation will
considerably change depending on which stage of the evolution the red
giants are.

In principle we could choose a given mass and radius for the red giants
participating in the collision and repeat the whole electromagnetic analysis we
have done in the first sections, when we were addressing main sequence stars.
This is so because, even if from the point of view of the Eddington standard
model of stellar structure a main-sequence star and a red giant are vew
different (treated as radiative objects and fully convective, respectively),
the gaseous debris after the collision will be similar.

However, because the masses and radii change so much, we decide not to do this
exercise just now because we are not aiming at comparing with observational
data in this work. It is likely that later we will follow this idea elsewhere.

\section{Stellar collisions in globular clusters}
\label{sec.globular}

We have focused so far on galactic nuclei. Covering globular clusters is
interesting because the rates are potentially larger due to the smaller
relative velocities between the stars participating in the collision which is
of the order of the velocity dispersion, as mentioned in the introduction.
Indeed, the Table 2 of \cite{BaumgardtHilker2018} contains a catalogue of
velocity dispersion profiles of 112 Milky Way globular clusters.  The average
yields $6.57\,\text{km\,s}^{-1}$, so that we will fix the relative velocity of
the stars participating in the collision to the average velocity dispersion of
$\sigma = 7\,\text{km\,s}^{-1}$.

\subsection{Rates}
\label{sec.rates_cl}

While it would be straightforward to repeat the calculations we have presented
in Sec.~(\ref{sec.rates}) by assuming the presence of an intermediate-mass
black hole with a given mass at the centre of the globular cluster, we prefer
not to do it. The uncertainty regarding the mass, position (we cannot longer
assume it to be fixed at the centre of the system, so that the calculations
become more complex) and even existence of such objects would make the rate
determination exercise too unconvincing. 

However, to motivate this section, the following is a brief summary of the most
relevant work that has been done in this context. The problem on the origin of
blue stragglers \citep[][]{Maeder1987,Bailyn1995,Leonard1989} is a good choice to
try to infer the amount of stellar collisions in globular clusters, since these
are very likely the outcome of such collisions.

\cite{Leonard1989} derives a collisional rate of $10^{-8}\,\text{yr}^{-1}$
assuming that a small fraction of main-sequence stars are in primordial
binaries. If we take the Milky Way as a reference point, then a galaxy should
have of the order of 100 globular clusters, so that the rate is of
$10^{-6}\,\text{yr}^{-1}$ per galaxy. This number might be larger, because
collisions of binaries are more important \cite{LeonardFahlman1991}.
It is important to note here that the average number of globular clusters
correlates with the mass of the central massive black hole
\citep{BurkertTremaine2010} in early-type galaxies. In their Fig.~(1) we can
see that this number can go up by many orders of magnitude depending on the
mass of the supermassive black hole.  For instance, NGC 4594 has about $2\times
10^3$ globular clusters. 

A few years later, \cite{SigurdssonPhinney95} carried out a detailed
theoretical and numerical study of stellar collisions, and their results
suggest a rate that ranges between $10^{-6}$ and $10^{-4}$ main-sequence
stellar collisions per year and galaxy (assuming 100 globular clusters).  For
the arbitrary reference distance that we have adopted of the order of 100 Mpc,
we have many clusters of galaxies such as the Virgo Cluster, with about $10^3$
galaxies, the Coma Cluster (Abell 1656), also with over $10^3$ identified
galaxies, and superclusters such as the Laniakea Supercluster
\citep{TullyEtAl2014} with about $10^5$ galaxies and the CfA2 Great Wall
\citep{GellerHuchra1989}, one of the largest known superstructures, at a mere
distance of $\sim 92\,\text{Mpc}$.  Regardless of what the rates are, if we
took an average of 1000 clusters and the larger rate of $10^{-4}$ of
\cite{SigurdssonPhinney95}, the number of collisions would be a thousand times
larger as compared to 100 clusters per galaxy and the rate of $10^{-6}$.
Although the authors did not address red giant collisions, the much larger
cross section and the smaller relative velocities in globular clusters are an
evidence that their rates must be, as in the case of galactic nuclei, much
larger.

\subsection{Low relative velocities and impact parameters}

Until now we have had the advantage of dealing with collisions that
kinematically are very powerful, so that after the collision we have no
surviving parts of the star (section~\ref{sec.energy}) or just the core
(section~\ref{sec.GW}). However, at a typical relative velocity of
$7\,\text{km\,s}^{-1}$, the collision will have a much lower impact on the
structure of the stars. We are looking at a different scenario.

On Sec.~(\ref{sec.rates}) we mentioned that we neglect gravitational focusing
in the case of galactic nuclei. For globular clusters we cannot do this anymore
because of the low relative velocity. 

\begin{figure}
\resizebox{\hsize}{!} 
          {\includegraphics[scale=1,clip]{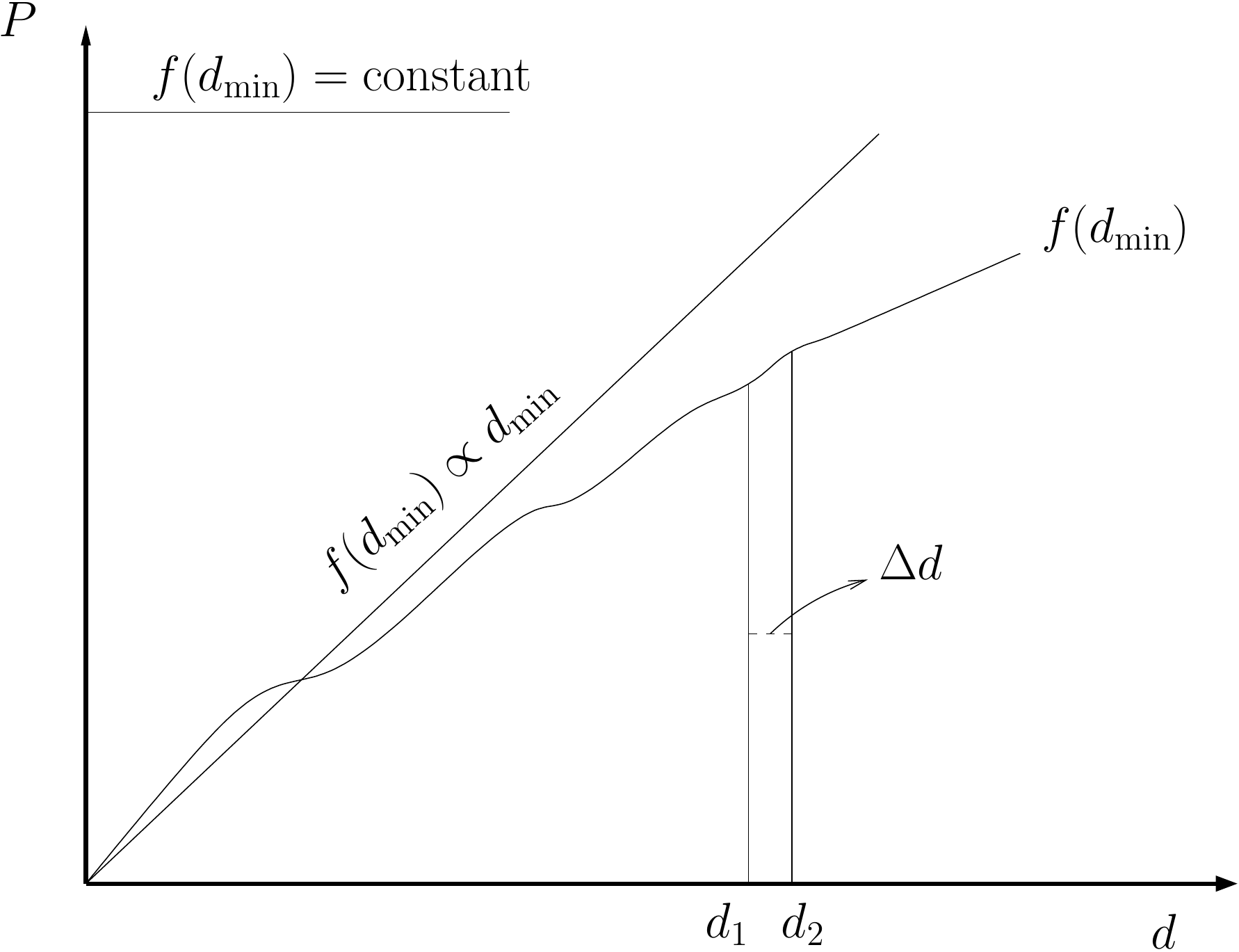}} 
\caption
   {
Probability and probability density as a function of the impact parameter.
We depict a generic curve and the two limiting cases we are addressing in
this study, namely the case in which $V_\text{rel} \ll V_\text{esc}$
and $V_\text{rel} \gg V_\text{esc}$.
   }
\label{fig.prob_dens_xfig}
\end{figure}

The probability of having a collision for a parameter $d_\text{min}$, as
introduced in Eq.~(\ref{eq.dmin}) with values ranging between $d_1$ and $d_2$
is

\begin{equation}
P_{d_1 \to d_2} = \int_{d_1}^{d_2} \frac{dP}{d\left(d_\text{min}\right)} 
                  d\left(d_\text{min}\right),  
\end{equation}

\noindent 
where $f(d_\text{min})={dP}/{d\left(d_\text{min}\right)}$ is the probability
density. If we consider a range of $\Delta d := d_2 - d_1 \ll d_\text{min}$,
then we can approximate the integral by $P_{d_1 \to d_2} \cong f(d_\text{min})
\Delta d$, as we can see in Fig.~(\ref{fig.prob_dens_xfig}).

When we consider the limit in which $V_\text{rel} \gg V_\text{esc}$, which
corresponds to a galactic nucleus, then $f(d_\text{min}) \propto d_\text{min}$,
which is shown in Fig.~(\ref{fig.prob_dens_xfig}). We can see that in this
case, then, the probability of having a collision with $d_\text{min}<d$ is
proportional to $d^2$ (i.e. it is proportional to the ``surface''). On the
contrary, in the case of a globular cluster, $V_\text{rel} \ll V_\text{esc}$,
so that all impact parameters have the same probability.

What this means is that in a galactic nucleus grazing collisions are more
probable than head-on ones, while in a globular cluster a grazing collision
and a pure head-on impact have exactly the same probability.

The parameters we used in the previous two sections remain the same but for the
relative velocity, which allows us to infer that the kinetic energy deposited
on to one star (again, assuming that it is distributed equally) is of
$T_\text{K}/2 \sim 2.43 \times 10^{44}\,\text{ergs}$. Hence, after the
collision, the star receives an amount of energy equivalent to $3\times10^{-3}\%$
its initial binding energy.  This amount of energy is small enough so that we
can investigate the evolution of one of the stars perturbatively.

We will start exploring this situation in its simplest possible form. For that,
we consider one collision at a
$d_\text{min}=\epsilon\left(R_{\text{half},\,1}+R_{\text{half},\,2}\right)$
such that $\epsilon \gtrsim 1$, which leads to contact between the stars after
the first close encounter (when they are not bound). The fact that even if
$d_\text{min} > R_{\text{half},\,1}+R_{\text{half},\,2}$ leads to a potential
collision due to the formation of a binary is because of tidal resonances
\cite{FabianEtAl1975}, because the cross-section is then as large as 1--2 times
that of collisions. 

The stars we are considering are main-sequence, Sun-like ones. If we consider
them (1) to be in hydrostatic equilibrium, (2) to be described by an equation
of state of an ideal gas and (3) to be spherical symmetric, then our
dynamically stable star reacts on a time given by the hydrostatic timescale

\begin{equation}
\tau_{\mathrm{hydr}} \approx\left(\frac{R_{\odot}^3}{GM_{\odot}}\right)^{1/2} 
                      \approx \frac{1}{2}(G {\varrho}_{\odot})^{-1/2},
\end{equation}

\noindent 
where ${\varrho}$ is the mean density of the star, which we assume to be like
our Sun, so that $\tau_{\mathrm{hydr}} \approx 30\,\text{min}$, orders of
magnitude shorter than the Kelvin-Helmholtz timescale, which in the case of the
Sun is $\tau_\text{KH}\sim 1.6 \times 10^{7}\,\text{years}$. This timescale is
interesting because it can be envisaged as an approximation to the
characteristic timescale of a thermal fluctuation, i.e. a thermal adjustment of the
star to a perturbation (in the simplistic picture which we are assuming now,
since we do not take into account the internal structure). If
we are talking about a red giant of mass $1\,M_{\odot}$ and a radius
$100\,R_{\odot}$, then $\tau_{\mathrm{hydr}} \approx 18\,\text{days}$.

\subsection{Dynamical stability in the adiabatic approach}
\label{sec.AdiabaticAppr}

Let us consider the collision to induce a small perturbation in the star. 
After the collision, we will assume for simplification that the
energy is equally distributed over all the surface of the star, which therefore
becomes denser because it is compressed. Since we are assuming this compression
to be adiabatic and homologous, the star will abandon its hydrostatic
equilibrium. The pressure in one layer of mass of the star can be obtained by
evaluating the integral $P=\int^M_m Gm\,dm/(4\,\pi r^4)$. Because of homology
and adiabaticity, by inspecting both sides of this equation we obtain that

\begin{equation}
\left(\frac{\varrho'}{\varrho} \right)^{\gamma_\text{ad}} = \left( \frac{R'}{R} \right)^{-3\gamma_\text{ad}},
\end{equation}

\noindent 
where primes represent the values after the collision; i.e. we are dealing with
Eq. 25.24 of \cite{KippenhahnWeigertStellarStructure}. This expression tells us
that after the collision the star will be dynamically stable in the adiabatic
regime if $\gamma_\text{ad}>4/3$ because the pressure's growth is more
important than the weight's increase. Since we are assuming that the stars
participating in the collision are Sun-like, we could draw the conclusion that
they are stable after the collision since one can approach
$\gamma_\text{ad}=5/3(>4/3)$. Indeed, in the case of the Sun the layer affected
would the convective one, located between $0.7\,R_{\odot}$ and the surface.
However, this is a very crude approach in the evaluation of the dynamical
stability which needs to be improved because the critical value depends on the
simplifications we have adopted in this section (but for the exception of
homology, since the threshold for $\gamma_\text{ad}$ is the same one for
non-homologous scenarios). Moreover, even if the stars are dynamically stable,
it is not discarded that they will be instable vibrationally or secularly. We
have addressed the dynamical stability because timescale associated is
the shortest one.

\subsection{Adiabatic pulsations after the collision and considerations about
binary formation}
\label{sec.pulsations}

Since 1638 we have observed that stars pulsate thanks to the observations of
Johannes Phocylides Holwarda of Mira. Arthur Ritter proposed in 1879 that these
variations are due to radial pulsations \citep{Gautschy1997}, and
\cite{Shapley1914} suggested that the temperature and brightness of Cepheid
variables originated in radial pulsations. Later, \cite{Eddington1917}, with
his piston analogy gave a working frame to describe them. In this valve
approximation, the radial pulsation period ${\Pi}_{r}$ can be estimated by
calculating the time that a sound wave will need to pass through the star, i.e.
${\Pi}_{r}=2R_{\odot}/C_\text{s}$.

We can determine $C_\text{s}$ from the pressure $P$ and (mean) density of the
star, $C_\text{s}^2=\gamma_\text{ad}\,P/\varrho$, where $\varrho$ is the
average density and $\gamma_\text{ad}$ is the adiabatic index, the heat
capacity ratio or Laplace's coefficient. It can be envisaged as a measure of
the stiffness of the configuration \citep[see e.g.  section 38.3
of][]{KippenhahnWeigertStellarStructure}.

\noindent 
Assuming that $\varrho$ is the actual value of the density througout the whole
star and requiring hydrostatic equilibrium, so that ${dP}/{dr}=-{G
M\varrho}/{r^2}=-G\left({4 \pi r^3}/{3} \varrho\right){\varrho}/{r^2}=-{4 G \pi r
\varrho^2}/{3}$, and requiring that $P=0$ at $r=0$, we derive that $P(r)={2}\pi
G p^2\left(R^2-r^2\right)/3$. Therefore we can obtain that

\begin{align}
{\Pi}_{r}&=2 \int_0^R \frac{dr}{\sqrt{{2}\gamma_\text{ad} \pi G \rho\left(R^2-r^2\right)/3}} 
\approx \sqrt{\frac{3 \pi}{2} \left(\gamma_\text{ad} G \varrho\right)^{-1}} \nonumber \\
         & \sim 44.5\,\text{min},
\end{align}

\noindent 
for $\varrho \sim 5.9\,\text{gr\,cm}^{-3}$ after the first collision.

In the adiabatic, spherical approximation, the pulsation is stable and has an
associated timescale of about 45 minutes. However, it would be interesting to
know if more pulsations can be produced to maintain the rhythm of oscillations
typical of the Cepheids. One possible way are further collisions.

\subsection{Maintened pulsations}

In this section we will quantitatively elucidate possible ways to produce
repeated pulsations in a main sequence star that is not in the instability
strip through dynamical phenomena, i.e. collisions. 

One first idea is that of recurrent collisions due to the formation of a binary
after the first impact.  The amount of energy loss per collision is $2\times
\delta\,E$, with $\delta\,E = T_\text{K} \sim 2.44 \times
10^{44}\,\text{ergs}$, as we have estimated before.  If we just look at the
energy, the question whether the two stars will form a binary seems too simple.
If the stars are initially on a parabolic orbit, the orbital energy of the
system, considered as two mass points (i.e. without taking into account the
binding energy of each star) is zero at the beginning (since the relative
velocity at infinity is zero, as is the gravitational energy). Any collision
-in fact even a close pass without any kind of physical contact which produces
tidal effects- will convert kinetic energy into thermal energy and thus leave
the stars with negative orbital energy, thus forming a binary. 

The real question is how this binary will evolve once it has formed. And this
is not a question which can be solved analytically in detail. It is worth to
note however that if there is a real contact at the first pericentre passage, a
collision, this will make the stars expand, so that further impacts will take
place, probably more violent at each successive orbit. The possibility that the
binary survives for a long time before the two stars merge is probably low.

These considerations are regarding main-sequence stars, whose envelopes are
rather dense.  In the case of red giants, it is likely that the the collisions
lead to the ejection of the envelope and we are left with a stable binary
consisting of the two cores which will then follow the previous scheme:
Evolution via gas drag, detection via gravitational radiation and an afterglow
when they eventually collide.

This reasoning is for main-sequence stars, whose envelope is quite dense. For
giants, perhaps the collisions lead to the ejection of the envelope and we are
left with a stable binary consisting of the two cores.

A possible first estime from an energy point of view would be to look at the
binding energy of the envelope of the giant; i.e. how much energy leads to an
ejection of the envelope and then compare that energy to the orbital energy
decrease from the parabolic trajectory to a circular binary formed by the two
cores. This would allow us to estimate the semi-major axis of the final binary
but this reasoning does not involve the impact parameter at all and is hence
simplistic.

The binding gravitational energy of stars in isolation is hence not a useful
quantity for studying the formation of binaries. The interesting point has
already been addressed: If the relative orbital energy of the two stars is
smaller than the sum of the binding energies, it is impossible to destroy both
stars completely.

In a globular cluster, it is unlikely for a completely destructive collision
to occur, because the relative velocities at infinity are very
low. And even if there is enough energy to destroy the stars, we need also a
very small impact parameter.

Therefore, for main-sequence stars in a globular cluster, most collisions lead
to the formation of a binary star that rapidly merges (in the classical
meaning, not the relativistic one). A smaller subset of collisions, those with
small impact parameters, produce a direct merger. Very little mass is ejected.
But there is a possibility of non-colliding binaries forming due to tidal
resonances \citep{FabianEtAl1975}.

Hence, it is difficult to assess analytically the duration and potential
periodicity of such pulsations originating from stellar collisions. If they are
vibrationally unstable, then we need to input a given amount of energy to
maintain the pulses, since the oscillations will damp. The input of energy
might (i) come from further collisions with the other star, if they build a
binary, (ii) from other stars in the cluster or (iii) internally from the
structure of the star, if we have amplitudes increasing in time because the
vibrational or thermal instability have excited the star. 

Addressing this problem is out of the scope of this paper but it is important
to note that pulsating stars are also used as another rung in the standard
candle ladder, as pointed out by Henrietta Swan Leavitt \citep{Fernie1969}. Since
the implications are potentially important, it would be interesting to
investigate the collisional pulsating nature of stars in globular clusters.

This would not be the first time that there is the need to revisit the cosmic
ladder argument due to anomalies found in globular clusters. Indeed, if we
consider two stars, one of population I (classical Cepheids) and another of
population II in the instability strip, they will pulsate due to the $\kappa$
mechanism \citep[see e.g.][]{KippenhahnWeigertStellarStructure}. Having
different masses but same radii because they are located at the same place in
the Hertzsprung-Russell diagram.  The lighter stars have lower $\varrho$ and,
hence, in principle, a longer period than classical Cepheids, even if they have
the same luminosity. This is not correct, and the derivation of the correct
periods led to Baade to realise that the cosmic distance scale was to be
multiplied by a factor of 2 \citep{Baade1944}.

\subsection{A scheme to study the injection of energy into the star}

Because in globular clusters the relative velocity at infinity is lower than
the stellar escape velocity, of the order $500-1000\,\text{km\,s}^{-1}$, the
relative velocity at contact is similar to the thermal velocity of stellar
matter. Hence, such collisions are only mildly supersonic and entropy is nearly
conserved. The entropic variable A defined as $A:= P/\rho^{\gamma_\text{ad}}$
(with $P$ the pressure) of a fluid element is subject to increase because of
the heat produced during the shock. Nonetheless, because the speed at contact
is similar to the speed of sound in the stars participating in the collision,
the shocks must have Mach numbers of about unity and hence a weak heating
production during the shock.  The important point here is that for these
reasons, the considered fluid element will have a constant entropic variable
during the collisional process, as demonstrated by \cite{LombardiEtAl2002}.
This allows us to treat the collision with a semi-analytical approach which is
derived by conservation laws of the process. This scheme yields very good
results when compared to three-dimensional computer simulations, including
shock heating, hydrodynamic mixing, mass ejection, and angular momentum
transfer
\citep{LombardiEtAl1996,LombardiEtAl2002,LombardiEtAl2002b,LombardiEtAl2003}.

\begin{figure}
\resizebox{\hsize}{!}
          {\includegraphics[scale=1,clip]{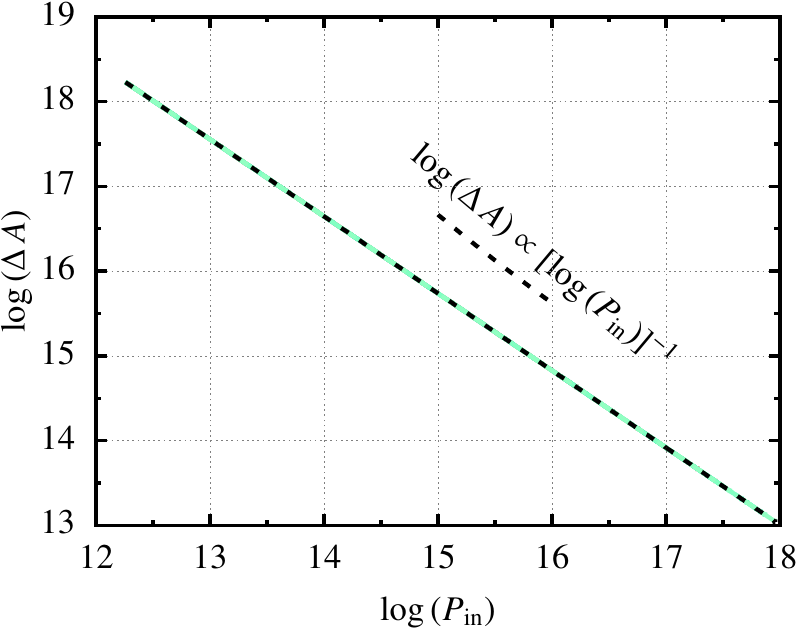}} 
\caption
   {
Difference of the entropic index as a function of the initial pressure of the
stars for the following values of the distance of closest approach: $d_{\rm
min}/(R_{1}+R_{2})=0,\,0.01 ,\,0.05 ,\,0.10 ,\,0.15 ,\,0.20 ,\,0.25 ,\,0.30
,\,0.35 ,\,0.40 ,\,0.45 ,\,0.50 ,\,0.55 ,\,0.60i \allowbreak ,\,0.65 ,\,0.70
,\,0.75 ,\,0.80 ,\,0.85 ,\,0.90 ,\,0.95 ,\,0.99 ,\,0.999$. We cannot see the
different curves because they all follow the same power-law relation, as given
with the black, dashed line. In all of the calculations we have assumed
$M_{1}=M_{2}=0.8\,M_{\odot}$, a relative velocity at infinity of $7
\text{km\,s}^{-1}$ and an initial separation normalized to the sum of the
parent star radii of 5.
   }
\label{fig.DeltaEntropicIndex_InitialPressure}
\end{figure}

In Fig.~(\ref{fig.DeltaEntropicIndex_InitialPressure}) we show the correlation
between the initial and final (i.e. after the collision) entropic index $\Delta
A:=A_{\rm fin} - A_{\rm in}$ as a function of the initial pressure of one of
the parent stars, $P_{\rm in}$. This finding was already presented in
\cite{LombardiEtAl2002b}, their Fig.~3, using smoothed-particle hydrodynamics.
It is interesting to see that the fluid sorting algorithm gives a result which
is very close to what three-dimensional computer simulations yield. We can see
that there is a proportion between both quantities such that $\log(\Delta A)
\propto \log(1/P_{\rm in})$. 

We can use this correlation to our benefit to understand how a collision will
add energy to one of the stars after they have gone through an interaction. In
particular

\begin{equation}
\log(A_{\rm fin}-A_{\rm in}) = b -\log(P_{\rm in}),
\end{equation}

\noindent
so that

\begin{equation}
A = A_{\rm in} + \frac{10^{b}}{\log{P_{\rm in}}} := A_{\rm in} + \frac{B}{\log{P_{\rm in}}}.
\end{equation}

\noindent 
In this equation, $b$ is a constant which contain information about the
properties of the collision. For instance, the larger $b$, the more energy will
be deposited on to the surface of one of the two stars, and we have defined
$B:=10^{\,b}$, which has units of pressure times $A$ (i.e. units of
$P^{\,2}/\rho^{\gamma_{\rm ad}}$).

If we consider a weak interaction, we assume that the entropy will be added
instantaneously on to the star, and that the density profile will not change.
The final pressure is hence

\begin{equation}
P_{\rm fin} = \rho^{\gamma_{\rm ad}}\,A_{\rm fin} = 
               P_{\rm in} + B\,\left(\frac{\rho^{\gamma_{\rm ad}}}{P_{\rm in}}\right),
\end{equation}

\noindent 
and therefore the specific internal energy profile is

\begin{equation}
u = \frac{3}{2} \frac{P_{\rm fin}}{\rho}= 
    u_{\rm in} + u_{\rm fin} :=  u_{\rm in} + \frac{3}{2} \left(\frac{B}{P_{\rm in}}\right) \rho^{2/3},
\end{equation}

\noindent 
because we have adopted $\gamma_{\rm ad} = 5/3$. Because the density profile is
unchanged, the gravitational potential energy is unchanged as well, which means
that only the thermal energy changes, since we are neglecting rotation as a
first approach. Therefore, the energy added over the star after the first ``hit'' 
is the following integral evaluated over the entire star

\begin{equation}
E_{\rm hit} = \int u_{\rm fin}(m) dm = 
              \int 6\pi \left(\frac{B}{P_{\rm in}(r)}\right) \rho(r)^{5/3} r^2 dr, 
\label{eq.Ehit}
\end{equation}

\noindent 
because $dm=\rho\,4\,\pi(r)\,r^2\,dr$ in spherical symmetry, which we are assuming.

Eq.~(\ref{eq.Ehit}) allows us to determine $B$ by evaluating the unperturbed
parent star. I.e. we solve the equation while setting $B=1$ and then we can choose
B to be the desired energy input divided by the result of the equation.

This scheme allows us to then evaluate the propagation of the energy through
the star and the induced pulsations. Unfortunately the analytical calculations
require solving the eigenvalue problems of the Sturm-Liouville type to calculate
the overtones if we want to consider non-adiabatic, non-radial oscillations,
although rotation might help with shearing deformation. 

Given that we have seen that any impact parameter has the same probability, we
consider that it is not worth extending this article any further than we are
already doing. We will therefore study this problem separately in a future
publication, either analytically or numerically with the energy injection
scheme we have outlined in this section.

\section{The cosmic ladder argument}
\label{sec.Ladder}

The event rate of colliding red giants and their observational nature is
telling us that we might be misinterpreting SNe Ia observations and and be
wrong by calling what we observe ``standard'' candles.  Also, their collisions
in globular clusters might trigger pulsating stars which are also used as
reference points when deriving cosmological scales, as we just pointed out in
the last section. 

There might be ways to tell them apart in the case of the SNe Ia observations,
though.  One unique observational signature for WD-WD collisions is the
double-peak profile of Cobalt and Iron lines in late-time spectra (also called
``nebular spectra'') of SNe Ia \cite{DongEtAl2015}. At late times, supernova
ejecta become optically thin, so that the line profiles reflect the underlying
velocity distributions.  Since both, Cobalt and Iron are decay product of Ni56
which is synthesized in the WD-WD merger, the profiles of these Co- and Fe
nebular lines show the velocity distribution of Ni56 in the ejecta. The authors
studied a sample of some 20 well-observed SNe Ia with nebular spectra, and
found in the sample 3 objects showing double peaks and an additional one with a
flat-top profile (i.e.  departing from a single-peak profile).

This bimodal velocity distribution is a feature of WD-WD mergers (see e.g. the
top panels of their Fig.~5). These results are supported by the work of
\cite{KushnirEtAl2013}, which shows from two-dimensional simulations of WD-WD
mergers that the full range of $\sim 0.1-1\, M_{\odot}$ Ni56 can be produced
from (exactly head-on) collisions of WDs with masses between $\sim 0.5-1
M_{\odot}$. However, other models, such as the one by \cite{vanRossumEtAl2016},
their Fig.~(13), do not predict such double peaks although, as noted by
\cite{DongEtAl2015}, the observed line profiles depend on the view angle, as
well as in other parameters\footnote{Dong Subo, personal communication.}, and their
data is not homogeneous, statistically speaking.

Another feature, as shown in \cite{SuboEtAl2018} is that for SNe Ia at the very
low end of luminosity function, Ni56 ejecta show significantly off-center
distribution at about $\sim 1000\,\textrm{km/s}$, which can be explained by
WD-WD mergers with significant mass ratios. We note that sub-Chandrasekhar
merger models, the delay detonation model, can also produce a large off-center
distribution, but not a bi-modal distribution.

It is interesting to note that \cite{WygodaEtAl2019a,WygodaEtAl2019b} also
explore the WD-WD merger scenario of \cite{KushnirEtAl2013} and they find that
the Ni56 column density distribution of the SNe Ia population can be explained
in terms of it. Also, \cite{LivnehKatz2020} find that the key signatures of SNe
Ia near the peak, i.e. the diverse distribution of Si II line width
distribution, which is usually referred to as the so-called ``branch plot'',
and widely used to classify SNe Ia population, can be explained by asymmetry in
ejecta from WD-WD mergers.

We note that in supernova searches, galactic nuclei are usually left out from
the survey because they are complex systems. However, (i) sometimes the whole
galaxy is too small in the data to be able to tell apart the nucleus and (ii)
as we have mentioned in the introduction, in this work we are focusing on
galactic nuclei to evaluate the lower-number case. In globular clusters
collisions should happen more frequently due to the lower velocity dispersion,
which approximately corresponds to the relative velocity of stars in the
system. The lower the relative velocity, the more likely that a gravitational
deflection ends up in a collision due to the larger exchange of energy and
angular momentum.

\section{Conclusions}
\label{sec.conclusions}

In this work we have made an analytical study of the electromagnetic- and
gravitational radiation implications of stellar collisins between stars in
dense stellar systems such as galactic nuclei and globular clusters, whether
main-sequence or red giants.  

In the case of galactic nuclei, we analyse the remaining gaseous cloud which
forms after the impact and its electromagnetic features, while taking into
account the ulterior dynamical evolution of the gas, which is expanding and
cooling down. In particular, we address the time evolution of the released
energy and find that it resembles that of a stellar tidal disruption. 

Since we are interested in the observational prospects of detecting this
phenomenon, we also describe the time evolution of the effective temperature,
the evolution of the peak wavelength of the spectral radiance, as well as the
evolution of the kinetic temperature as the outcome of the collision and the
spectral power as a function of the frequency. 

We find that the electromagnetic traces left by these violent and transient
processes strongly resemble over time periods tidal disruptions but also SNe Ia
supernovae. 

Our complete analysis depends only on two free parameters, one appears in the
electromagnetic study and the other one in the gravitational-waves one.  In the
part dedicated to the electrodynamics, the free parameter is responsible for
the non-linearity of the collision, i.e. the transmission of the shocks and
hence of the total efficiency conversion of kinetic energy into radiation.  The
second one, which is relevant for the total rates of gravitational-wave
sources, is the number fraction of main-sequence stars whose cores form a
binary. We parametrise the solution in terms of the non-linearity parameter and
explore four different values. In order to derive this parameter one would need
dedicated numerical simulations.

From among the colliding stars, a subgroup of them leads to the formation of a
binary consisting of their cores. This subgroup is interesting because it
leads to the formation of a binary of two objects that is sufficiently massive
and compact to detectably emit gravitational waves.

We find that the friction exerted by the gas accelerates the approach of the
surviving cores and brings them closer to eventually merge, with an
electromagnetic afterglow such as in the case of binaries of neutron stars
merging. Due to the time-varying properties of the gas (which our analytical
model takes into account in all calculations), the observed appearance of the
gravitational waves is very different from any known source. In particular, two
nuclei of very low masses, $0.34\,M_{\odot}$, will be perceived as two black
holes of initially slightly above stellar masses, which later increase to
become, apparently, two merging supermassive black holes. Something similar
happens to the luminance distance, which apparently decreases and then
increases very significantly.  

As noted in Sec.~(\ref{sec.BBHs}), the fact that the frequency evolution is
different from the vacuum one, will be the first evidence that these are not
black holes emitting gravitational radiation, but a stellar collision. Later,
the absence of event horizon will make it obvious and, finally, the
electromagnetic afterglow will confirm this. In this sense, the gravitational
waves are a perfect tool to identify the nature of the source.  We sketch in
the second appendix a possible strategy to address the gravitational wave data
analysis of the collisions.

We calculate analytical characteristic strains and polarisations of the nuclei
in vacuum, as a reference point, and then derive them in the gaseous case, also
analytically. The changes are evident and very pronounced, differing by
orders of magnitude, although the overall behaviour in the gas case captures,
or rather tries to mimic, the behaviour of gravitational radiation emission.

As the gravitational merger time is drastically reduced, electromagnetic and
gravitational wave detection go practically hand in hand.  This means that the
collisions of main-sequence stars and red giants represent two multi-messenger
probes that complement each other. This is particularly interesting in the
case of red giants, since the core is a degenerate object that will be a more
interesting source of gravitational radiation.

In the case of red giants, we calculate the importance of the H-burning shell
in the process, as this calculation was not found in the literature, to the
best of our knowledge. This is important because this layer around the cores
could strongly influence the further evolution of the binary of the two
degenerate objects. However, we derive that the role of this shell can be
disregarded in this study. 

According to our results, these degenerated cores, which can be envisaged as
white dwarves, embedded in the host red giants, have a collisional event rate
which can be of up to some hundreds of them a year within a volume of
$100\,\textrm{Mpc}$.  The properties of the collision will strongly vary in
function of the mass of the cores and the impact parameter, which depends on
the radii of the cores. The properties of these collisions are very similar to
SNe Ia.  In view of the event rate, this could pose a problem to the
interpretations of SNe Ia, which are referred to as ``standard candles''
following the idea of Henrietta Swan Leavitt \citep{Fernie1969} as a way to
derive cosmological distances following the ladder argument. This is because,
as we have just explained, stellar collisions are not standard at all.

Finally, collisions in globular clusters lead to different phenomena, in
particular they might lead to stellar pulsation like in the classic problem of
the Cepheids.  The periodicity of these pulsations is to be investigated
because the formation of a binary which is long-lived seems to be unlikely, but
collisions arising from other stars can be a way to sustain the pulsations, or
vibrational or thermal instabilities triggered in the interior of the star
after the first collision.  We have shown that their pulsations are stable in
the case of the adiabatic, spherical special case, but it is worth to
investigate (i) the non-adiabaticity of spherical pulsations and (ii)
non-radial oscillations, in both the $\kappa$ and $\epsilon$ mechanisms.  We
think this is an interesting question because these pulsations are considered
to be another rung in the cosmological ladder and, as noted in
Sec.~(\ref{sec.pulsations}), a misclassification of these has already had an
important impact in the past, also in globular clusters.  We have not addressed
this for the sake of the length of this article, but this is a part of current
work and will be presented elsewhere.

Finally, it is worth mentioning that our Galactic Centre is a known region of
heightened cosmic ray abundance. Naively, one would have thought that the
increase in cosmic ray abundance we observe there would be brought about by a
larger abundance of supernovae in this region. However, no such over abundance
of supernova is observed in this region. Furthermore, the quiescence level of
the supermassive black hole activity in this region casts doubt on an accretion
episode being responsible for the cosmic rays. Consequently, a heightened
cosmic ray abundance in galactic nuclei appears peculiar, demanding the
existence of a regular non-thermal energy source within this region, which
seems to be natural to be linked to stellar collisions.

\acknowledgments

We thank Marc Freitag for many discussions, as well as Xian Chen and Dong Subo.
We are indebted with Andrew Taylor, Stefan Ohm and Rolf B{\"u}hler for their
input and in general to the THAT group of DESY for an extended visit in which
part of this work was done during 2020-2021. We thank Jeremy Goodman and Jill
Knapp to find the origin of the approximation used in the estimation of the
density of Bohdan Paczy{\'n}ski. Jakob Nordin pointed us to the Zwicky
Transient Facility observational data that seems to match the conceptual idea
we have presented. Kostas Tzanavaris suggested to use an expansion in powers to
solve the integral to derive the coalescence time, which has a faster
convergence as compared to the incomplete beta function.  This work was
supported by the 111 Project under Grant No. B20063 and the National Key R\&D
Program of China (2016YFA0400702) and the National Science Foundation of China
(11721303).

\section*{Appendix 1: Analytical solution of the integral associated to $T_{\rm gas}$}

\noindent
The integral \ref{eq.RHSMonsterIntegral} to be computed is the following.

\begin{equation}
I(\tau) = \int_{0}^{T_{\rm mrg,\,m}} e^{\,a\,\tau^2} \left(1 + b\,\tau\right)^{-3}\,d\tau.
\end{equation}

\noindent
We change now the notation, $\tau=t$, $x= T_{\rm mrg,\,m}$, so that

\begin{equation}
	I(x) = \int_0^x \frac{e^{\,ct^2}}{(1+bt)^3}dt.
\end{equation}

\noindent
Expand the exponential as a power series of $t$.
\begin{equation}
	I(x) = \sum_{n=0}^\infty \frac{c^{\,n}}{n!}\int_0^x \frac{t^{2n}}{(1+bt)^3}dt.
\end{equation}

\noindent
We now reparametrize the variable $x$ in such a way that the limits of the integral are $0$ and $1$. 
\begin{equation}
	t=xs, \;\; s = t/x, \;\; dt = xds,
\end{equation}
\begin{equation}
	I(x) = \sum_{n=0}^\infty \frac{c^{\,n} x^{\,2n+1}}{n!} \int_0^1 \frac{s^{\,2n}}{(1+bxs)^3}ds = \sum_{n=0}^\infty \frac{c^{\,n} x^{\,2n+1}}{n!} I_n(x),
\end{equation}

\noindent
where

\begin{equation}
	I_n(x) = \int_0^1 \frac{s^{\,2n}}{(1+bxs)^3}ds.
\end{equation}

\noindent
At this step, we compute the integral $I_0$.

\begin{align}
	I_0(x) 	& = \int_0^1 \frac{1}{(1+bxs)^3}ds = \frac{1}{bx} \int_0^1\frac{(1+bxs)'}{(1+bxs)^3} ds \nonumber\\
			& = -\frac{1}{2bx} \int_0^1  \left[ \frac{1}{(1+bxs)^2} \right]' ds\nonumber\\
			& = \frac{1}{2bx}\left[1 - \frac{1}{(1+bx)^2}\right]
\end{align}

\noindent
We simplify the integral $I_n$ for $n\geq 1$ by reducing the power of the
denominator, using the method of integration by parts.

\begin{align}
	I_n(x)	& = \int_0^1 \frac{s^{\,2n}}{(1+bxs)^3} ds = -\frac{1}{2bx} \int_0^1 s^{\,2n}\left[\frac{1}{(1+bxs)^2}\right]'ds \nonumber\\
			& = -\frac{1}{2bx}\left[\frac{s^{\,2n}}{(1+bxs)^2}\right]_{s=0}^{s=1} + \frac{n}{bx} \int_0^1 \frac{s^{\,2n-1}}{(1+bxs)^2}ds \nonumber\\
			& = -\frac{1}{2bx}\frac{1}{(1+bx)^2} - \frac{n}{bx} \int_0^1 s^{\,2n-1}\left[\frac{1}{1+bxs}\right]'ds \nonumber\\
			& = -\frac{1}{2bx}\frac{1}{(1+bx)^2} - \frac{n}{(bx)^2}\left[\frac{s^{\,2n-1}}{1+bxs}\right]_{s=0}^{s=1} \nonumber\\
			& + \frac{n(2n-1)}{(bx)^2}\int_0^1 \frac{s^{\,2n-2}}{1+bxs}ds \nonumber\\
			& = -\frac{1}{2bx}\frac{1}{(1+bx)^2} - \frac{n}{(bx)^2}\frac{1}{1+bxs} \nonumber\\
            & + \frac{n(2n-1)}{(bx)^2}\int_0^1 \frac{s^{\,2n-2}}{1+bxs}ds.
\end{align}

\noindent
So, we have to compute the integral

\begin{equation}
	f_n(x) = \int_0^1 \frac{s^{\,2n-2}}{1+bxs}ds, \;\; n\geq 1.
\end{equation}

\begin{enumerate}
	\item The first thing to do is to simplify the denominator, by using the reparametrization
	\begin{equation}
		z = 1+bxs, \;\; dz = bxds, \;\; s = \frac{1}{bx}(z-1),
	\end{equation}
	which gives us
	\begin{equation}
		f_n(x) = \frac{1}{(bx)^{2n-1}} \int_1^{1+bx} \frac{1}{z}(z-1)^{2n-2}dz.
	\end{equation}
	
	\item Next, we use the binomial theorem to expand the polynomial inside the integral. 
	\begin{align}
		(z-1)^{2n-2} & =  \sum_{k=0}^{2n-2} {2n-2 \choose k}(-1)^{\,k} z^{\,k} \nonumber \\
                     & = 1 + \sum_{k=1}^{2n-2} {2n-2 \choose k}(-1)^{\,k} z^{\,k}.
	\end{align}
	
	\item We substitute and have that
	\begin{equation}
		\int_1^{1+bx} \frac{1}{z}(z-1)^{\,2n-2}dz = \ln(1+bx) + F_n(x),
	\end{equation}
	where we have introduced the polynomial
	\begin{equation}
		F_n(x) = 
		\begin{cases} 
      		0 & n=1 \\
     		\sum_{k=1}^{2n-2} {2n-2 \choose k} \frac{(-1)^{\,k}}{k}\left[(1+bx)^{\,k} - 1\right] & n > 1 .
   		\end{cases}
	\end{equation}
	
	\item The integral $f_n$ can be expressed via those functions
	\begin{equation}
		f_n(x) = \frac{1}{(bx)^{2n-1}}\Big[\ln(1+bx) + F_n(x)\Big].
	\end{equation}
\end{enumerate}

\noindent
We now substitute and have that
\begin{align}
	I_n(x) & = \frac{n(2n-1)}{(bx)^{2n+1}}\Big[\ln(1+bx) + F_n(x)\Big] \nonumber \\
           & -\frac{1}{2bx}\frac{1}{(1+bx)^2} - \frac{n}{(bx)^2}\frac{1}{1+bx}
\end{align}

Finally, we combine (4), (6) and (15). Note that all terms apart from the one containing 
the polynomial $F_n$ yield elementary functions.

\begin{align}
	\sum_{n=1}^\infty & \frac{c^{\,n} x^{\,2n+1}}{n!}\left[-\frac{1}{2bx}\frac{1}{(1+bx)^2}\right] \nonumber \\
                      & = - \frac{1}{2bx}\frac{1}{(1+bx)^2}\sum_{n=1}^\infty \frac{c^{\,n} x^{\,2n+1}}{n!}		\nonumber\\
					  & = - \frac{1}{2bx}\frac{e^{\,cx^2}-1}{(1+bx)^2},
\end{align}
\begin{align}
	\sum_{n=1}^\infty &  \frac{c^{\,n} x^{\,2n+1}}{n!} \left[-\frac{n}{(bx)^2}\frac{1}{1+bx}\right]	\nonumber \\
                      & = -\frac{1}{(bx)^2}\frac{1}{1+bx} \sum_{n=1}^\infty \frac{c^{\,n} x^{\,2n+1}}{(n-1)!}	\nonumber\\
					  & = -\frac{1}{(bx)^2}\frac{1}{1+bx} \sum_{n=0}^\infty \frac{c^{\,n+1} x^{\,2n+3}}{n!}	\nonumber\\
					  & = -\frac{cx}{b^2}\frac{e^{\,cx^2}}{1+bx},
\end{align}

\begin{align}
	\sum_{n=1}^\infty & \frac{c^{\,n} x^{\,2n+1}}{n!}\frac{n(2n-1)}{(bx)^{2n+1}} \nonumber\\ 
                      & = \frac{1}{b}\sum_{n=1}^\infty \frac{2n-1}{(n-1)!}\left(\frac{c}{b^2}\right)^n = \frac{1}{b}\sum_{n=0}\frac{2n+1}{n!}\left(\frac{c}{b^2}\right)^{n+1}	\nonumber\\
																			& = \frac{c}{b^3}e^{\,c/b^3} + \frac{2c}{b^3}\sum_{n=0}^\infty \frac{n}{n!}\left(\frac{c}{b^2}\right)^n													\nonumber\\
																			& = \frac{c}{b^3}e^{\,c/b^3} + \frac{2c}{b^3}\sum_{n=1}^\infty \frac{n}{n!}\left(\frac{c}{b^2}\right)^n													\nonumber\\
																			& = \frac{c}{b^3}e^{\,c/b^3} + \frac{2c}{b^3}\sum_{n=1}^\infty \frac{1}{(n-1)!}\left(\frac{c}{b^2}\right)^n												\nonumber\\
																			& = \left(\frac{c}{b^3} + \frac{2c^2}{b^5}\right)e^{\,c/b^2}.
\end{align}
Thus:
\begin{align}
	I(x)	& = \frac{1}{2b}\left[1-\frac{1}{(1+bx)^2}\right] + \left(\frac{c}{b^3} + \frac{2c^2}{b^5}\right)e^{\,c/b^2}\ln(1+bx)																										\nonumber\\
			& - \frac{1}{2bx}\frac{e^{\,cx^2}-1}{(1+bx)^2} -\frac{cx}{b^2}\frac{e^{\,cx^2}}{1+bx}																																		\nonumber\\
			& + \sum_{n=1}^\infty \frac{n(2n-1)c^{\,n}}{n!} F_n(x).
\end{align}

\section*{Appendix 2: A scheme for the gravitational-wave analysis}

When comparing the polarizations in the evolving gaseous cloud and vacuum,
because of the big difference in $T_{\rm mrg,\,m}$ and $\Lambda(\tau)$, the two
polarizations diverge from the beginning. This leads to a significant mismatch
of the waveforms.

\begin{figure*}
\resizebox{\hsize}{!}
          {\includegraphics[scale=1,clip]{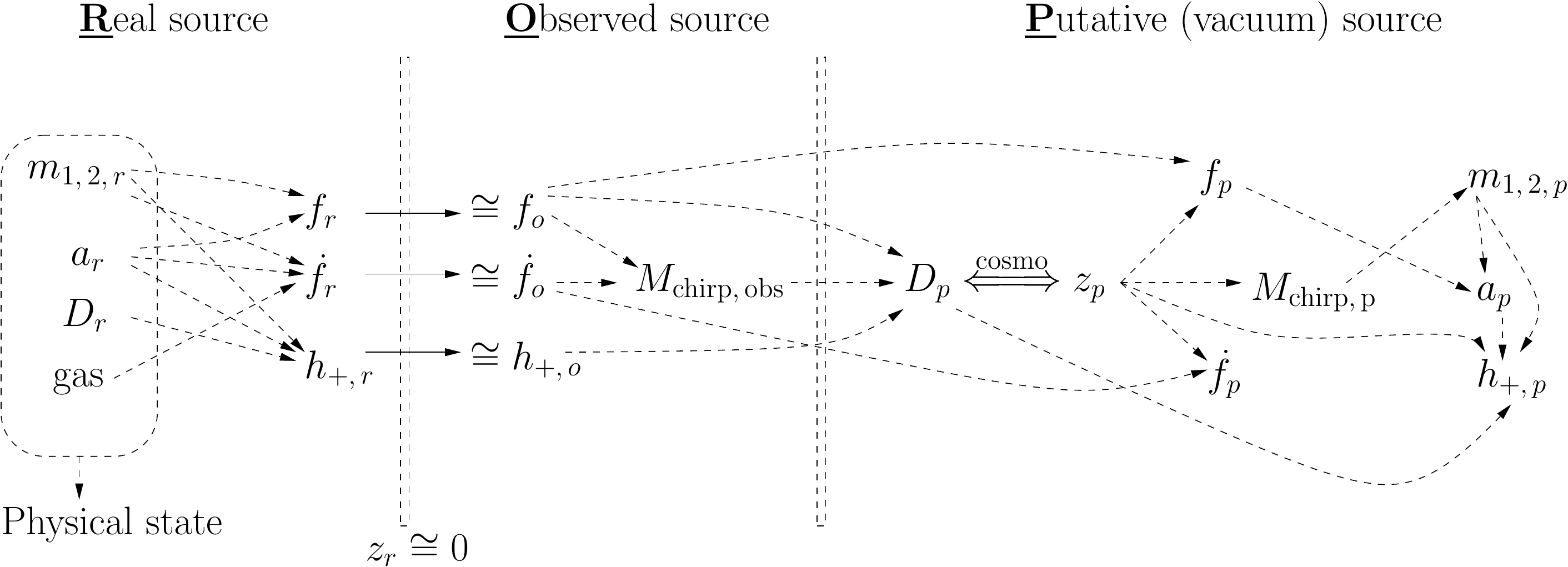}}
\caption
   {
Scheme to derive the various quantities required to compare
the observed source and the putative one.
   }
\label{fig.mismatch_scheme}
\end{figure*}

The reality is more complex. On the one hand, we have a real, physical source
which is producing the gravitational radiation. We will refer to this source
from now as the ``real'' source and will use the subscript ``r'' for it. On the
other hand, detectors will receive data for a source which we describe as the
``observed'' source for obvious reasons, and use the subscript ``o'' for it.
Finally, in
order to extract parameters from the observed source, data analysts will use a
theoretical model which assumes that the source is in vacuum.  This is our
``putative'' source, and we will use the subscript ``p'' for it. The connection
between these three different sources is displayed in
Fig.~(\ref{fig.mismatch_scheme}). 

\section*{Appendix 3: Neglection of the constant of integration to derive the
density of the H-fusing shell}

When trying to define the constants of integration of Eq.~(\ref{eq.dPdPrad})
and Eq.~(\ref{eq.dTdr}) we came accross the unpublished notes of Bohdan
Paczy{\'n}ski, where he explains that 

\begin{quote}
\it{
The constant (...) can be calculated from the matching conditions between the
radiative zone and the outer convective envelope, and it is very important near
the radiative - convective boundary. However, deep inside the radiative zone
the other two terms in the equation (...) become much larger than the constant,
and (it) may be neglected.
}
\end{quote}

We found this explanation in the notes of Jill Knapp in Princeton, who told us
it was not her work and after looking for the origin, she found out that the
link to the original notes written by Paczy{\'n}ski was Jeremy Goodman. In his
turn, he explained that ``He (Bohdan Paczy{\'n}ski) taught a class in stellar
structure to graduate students for many years, which I had the privilege of
helping him with in later years.'' Unfortunately, Jeremy could not find a
published version of this derivation by Paczy{\'n}ski, so that we acknowledge
here the origin of what has led us to the neglection of the constant of
integration, crucial in defining the analytical expression for the density of
the H-fusing shell.

\label{LastPage}
\end{document}